\documentclass[12pt,a4paper,twoside]{book}

\usepackage{graphicx,subfigure,epstopdf}
\usepackage{float}
\usepackage{amssymb,amsmath}
\usepackage{mathrsfs}
\usepackage{multirow}
\usepackage{appendix}
\usepackage{fancyhdr,braket}
\usepackage{time}
\usepackage{xspace}
\usepackage{slashed}

\usepackage{bookmark}

\hypersetup{}

\usepackage{xcolor}
\usepackage{rotating}
\usepackage{setspace}
\usepackage{type1cm}
\usepackage{enumitem}
\usepackage{chngcntr}
\usepackage{afterpage}
\usepackage{threeparttable}
\usepackage{color} 
\usepackage{enumitem}
\usepackage{amsfonts}
\graphicspath{{./figures/}}
\setlist[itemize,1]{label=$\times$}
\setlist[itemize,2]{label=$\checkmark$}
\setlist[itemize,3]{label=$\diamond$}
\setlist[itemize,4]{label=$\bullet$}
\definecolor{darkgreen}{rgb}{0.0,0.5,0.0}
\definecolor{greenblue}{rgb}{0.0,.1,.4}
\definecolor{brickred}{rgb}{0.8, 0.25, 0.33}
\definecolor{brass}{rgb}{0.71, 0.65, 0.26}
\definecolor{darkorchid}{rgb}{0.6, 0.2, 0.8}

\usepackage{amsmath}

\usepackage{soul}
\usepackage[normalem]{ulem}

\usepackage{amsmath}
\usepackage{makeidx}
\usepackage{amssymb}
\usepackage{graphicx}
\usepackage{bmpsize}
\usepackage{epstopdf}

\newcommand{\chushi}[1]{ }
\let\calccommentout\iffalse 
\let\calcshow\iftrue 

\newcommand{\p}{\partial}

\newcommand {\mathsym}[1]{{}}
\newcommand {\unicode}[1]{{}}

\labelformat{chapter}{Chapter #1} 
\labelformat{section}{Section #1} 
\labelformat{subsection}{Section #1} 
\labelformat{subsubsection}{Section #1}
\labelformat{subsubsubsection}{Section #1}
\labelformat{equation}{Eq.~(#1)} 
\labelformat{figure}{Fig.~#1} 
\labelformat{subfigure}{Fig.~\thefigure#1} 
\labelformat{table}{Tab.~#1} 
\labelformat{appendix}{Appendix #1}

\def\be{\begin{equation}}
\def\ee{\end{equation}}

\DeclareUnicodeCharacter{2212}{-}
\begin{document}

\pagestyle{empty}  
\begin{titlepage}

\begin{center}
\vspace*{0.25in}

   \textbf{\Large{\MakeUppercase{ Some aspects of quantum correlations and decoherence in the cosmological spacetimes}}}\\

\vspace*{0.9in}

 
\textbf{\Large{ Ph.D Thesis}}

\vspace*{0.4in}



\textbf{\Large {by}} \vspace*{0.15in}

\textbf{\Large{\MakeUppercase{ Nitin Joshi}}}\\


 

\vfill

\begin{figure}[h!]
\begin{center}
        \includegraphics[width=5cm]{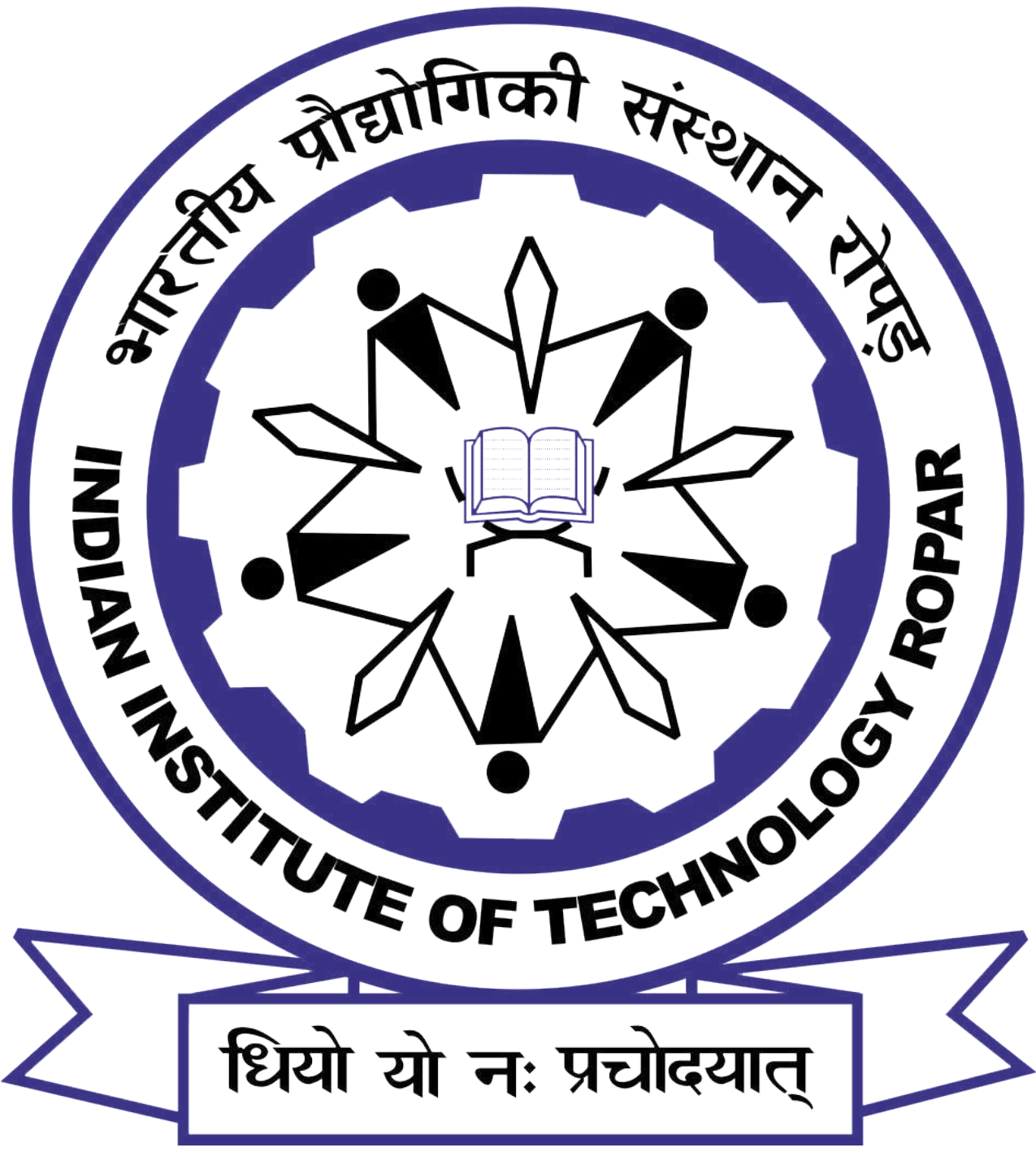}
    \end{center}
\end{figure}

\textbf{\Large{\MakeUppercase{ Department of Physics}}}\\ \vspace*{0.1in}
\textbf{\Large{\MakeUppercase{ Indian Institute of Technology Ropar}}}\\

\Large{\textbf{2023}}

\end{center}
\end{titlepage}

\begin{titlepage}

\begin{center}
\vspace*{0.25in}

   \textbf{\Large {\bf \MakeUppercase {Some aspects of quantum correlations and decoherence in the cosmological spacetimes}}}\\

\vspace*{0.5in}



\textbf{By}\\ \vspace*{0.1in}

\textbf{\Large {\bf Nitin Joshi}}\\

\vspace*{0.5in}

 {\it Submitted \\ in fulfillment of the requirements for the degree \\ of}\\
 \vspace*{0.25in}
 
{\Large \bf Doctor of Philosophy}

\vfill

\begin{figure}[h!]
\begin{center}
        \includegraphics[width=4cm]{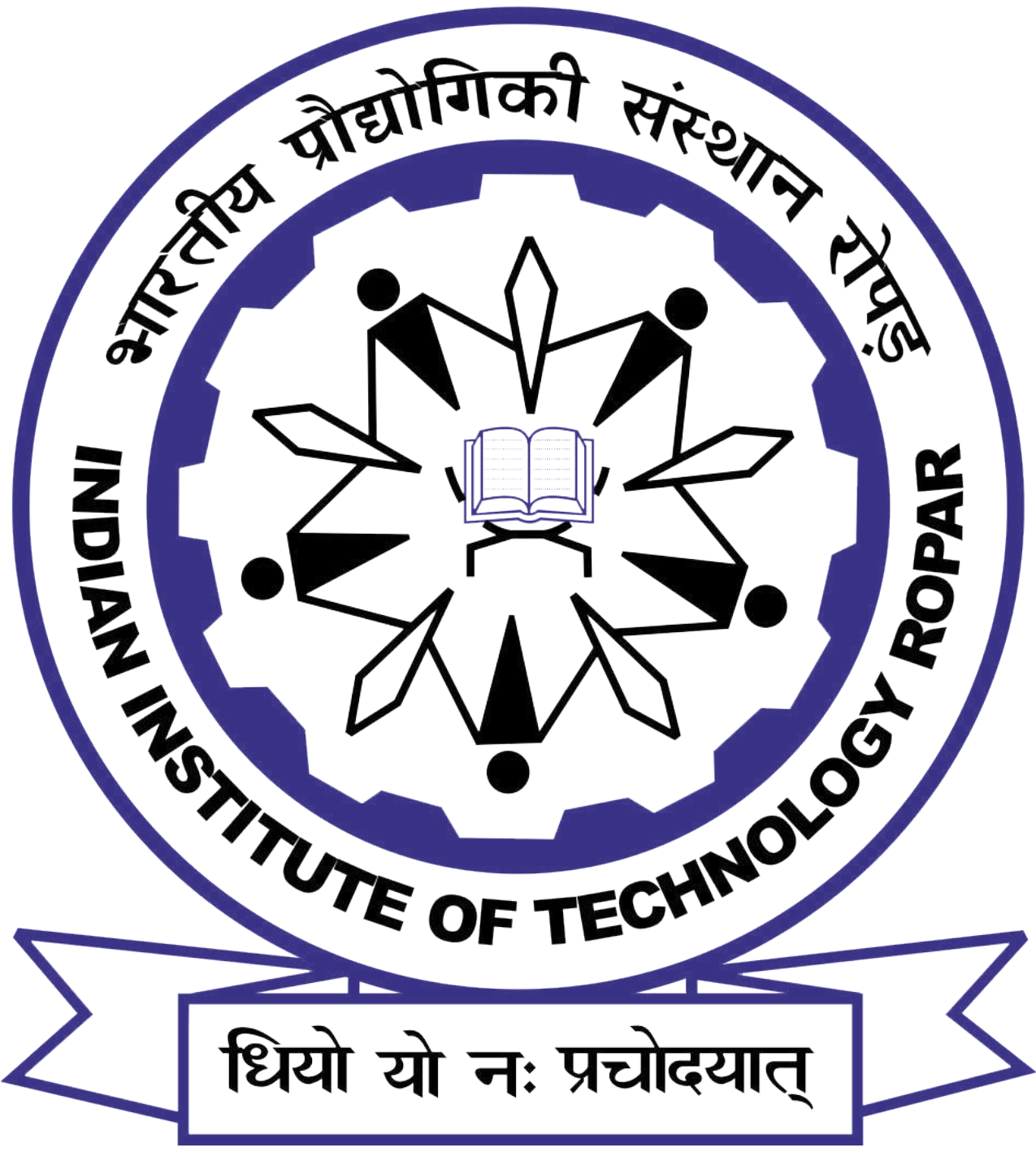}
    \end{center}
\end{figure}

{\large {\bf Department of Physics}}\\
{\large {\bf Indian Institute of Technology Ropar}}\\
{\large {\bf September 2023}}

\end{center}
\end{titlepage}

\thispagestyle{empty}
\vspace*{\fill}
\begin{center}
\end{center}
\vspace*{\fill}

\newpage

\thispagestyle{empty}
\vspace*{\fill}
\begin{center}
\large{\textcopyright Indian Institute of Technology Ropar}\\
\large All rights reserved.
\end{center}
\vspace*{\fill}

\cleardoublepage
\thispagestyle{empty}

\chapter*{}
\thispagestyle{empty}

\vspace*{\fill}
\begin{center}
 \large{This thesis is dedicated to}\\
 \textit{My late grandfather, my father, and to all the people who are struggling out there.}
\end{center}
\vspace*{\fill}
\thispagestyle{empty}

\chapter*{Certificate}
\thispagestyle{empty}
It is certified that the work contained in this thesis entitled ``{\bf Some aspects of quantum correlations and
decoherence in the cosmological spacetimes}" by {\bf Mr. Nitin Joshi}, a research scholar in the Department of Physics, Indian Institute of Technology Ropar, for the award of degree of {\textbf{Doctor of Philosophy}} has been carried out under our supervision and has not been submitted elsewhere for a degree.

\vspace{2in}

\begin{minipage}{\linewidth}
              
              
\end{minipage}
\vspace{-1.5in}
\begin{minipage}[t]{0.5\textwidth}
\begin{flushleft}
\vspace{2mm}
Dr. Sourav Bhattacharya\\
Associate Professor \\
Department of Physics\\
Jadavpur University\\
Kolkata 700 032, India\\
\vspace{0.5in}
\textbf{September 2023}
\end{flushleft}
\end{minipage}
\begin{minipage}[t]{0.45\textwidth}
\begin{flushright}
\textbf{} \\
\vspace{-3mm}
Dr. Rajesh Kumar Gupta\\
Assistant Professor \\
Department of Physics\\
Indian Institute of Technology Ropar \\
Rupnagar 140 001, Punjab, India
\end{flushright}
\end{minipage}


\chapter*{Declaration}
\thispagestyle{empty}
I hereby declare that the work presented in the thesis entitled ``{\bf Some aspects of quantum correlations and
decoherence in the cosmological spacetimes}" submitted for the degree of \textbf{Doctor of Philosophy} in Physics by me to Indian Institute of Technology Ropar has been carried out under the supervision of \textbf{Dr. Sourav Bhattacharya} and \textbf{Dr. Rajesh Kumar Gupta}. This work is original and has not been submitted in part or full by me elsewhere for a degree.
\vspace{1in}

\begin{minipage}[t]{0.5\textwidth}
\begin{flushleft}
\vspace{2.45in}
\textbf{September 2023}
\end{flushleft}
\end{minipage}
\begin{minipage}[t]{0.45\textwidth}
\begin{flushright}
\textbf{}
\vspace{1.05in}

Nitin Joshi\\
Ph.D Research Scholar \\
Department of Physics\\
Indian Institute of Technology Ropar \\
Rupnagar 140 001, Punjab, India
\end{flushright}
\end{minipage}

\cleardoublepage



\newpage
%

\begin{center}
{
    \fontencoding{OT1}
    \fontfamily{ppl}
    \fontseries{b}
    \fontshape{n}
    \fontsize{17}{40}
    \selectfont
    Acknowledgements\\
  \vspace*{0.2in}
}

\vspace*{0.1in}

{
    \selectfont
    }

\end{center}

\vspace*{0.1in}

\begin{spacing}{1.2}

{\noindent Almost five years ago, as I embarked on my Ph.D journey, I had no idea of the profound experiences that awaited me. Pursuing my dream and passion of a Ph.D in theoretical high energy physics, time seemed to fly by as I delved deeper into the realm of academia. I am sincerely thankful to Indian Institute of Technology Ropar for granting me the opportunity and financial support to pursue my Ph.D.

I would like to express my sincere gratitude to my Ph.D supervisors, Dr. Sourav Bhattacharya and Dr. Rajesh Kumar Gupta, for graciously imparting their vast academic expertise and playing a pivotal role in nurturing my academic growth. Their guidance has not only greatly contributed to my publications but has also unlocked opportunities I had never imagined. Specially, I would like to express my sincere appreciation to Dr. Sourav Bhattacharya for his exemplary professionalism and kind demeanor.

I extend my heartfelt gratitude to my Doctoral Committee members, Dr. Asoka Biswas, Dr. Shubhrangshu Dasgupta, Dr. Shankhadeep Chakrabortty, and Dr. Sudipta Kumar Sinha, for their invaluable insights and discussions. 

I wish to express my appreciation to Dr. Hironori Hoshino, Dr. Karunava Sil, Dr. Md. Sabir Ali, and Dr. Deepak Tomar for their invaluable contributions in assisting me. Additionally, I extend my appreciation to all the past and present members of the {\it Gravity and String Group}, Shagun, Sanjay, Arpit, Meenu, Sudesh, Pronoy, and Siddant. Their contributions have not only enhanced the quality of our discussions but also fostered a delightful and harmonious atmosphere within our group. I would also like to acknowledge the exceptional collaboration I had with Dr. Gopal Yadav, whom I found to be the most enjoyable colleague to work with. I would like to express my appreciation to the following individuals who have contributed significantly to my academic journey: Kusum for her invaluable assistance with analytical and numerical calculations, Raghvendra, Vasu Dev, Sahil, Mukesh, Rakhi, Damanpreet, and Pardeep for their engaging discussions and help throughout the coursework. Additionally, I would like to thank Ravi, Jaswant, Rahul, Prerna, Faizan bhai, Ajit and everyone mentioned above for all the entertainment and joy they brought to this amazing journey. 

I would also like to thank Mr. Brij Mohan Mamgain, my physics teacher during $11^{th}$ and $12^{th}$ standard at {\it GIC Kanwaghati}, for igniting my curiosity in physics. His engaging teaching style and enthusiasm turned each class into a captivating exploration, making the complex concepts of physics a thrilling journey of discovery.

It is a great pleasure to express my heartfelt gratitude to my childhood football club, {\it Kishanpuri FC}, and my lifelong football companions, who have consistently been a immense source of joy to my life. Additionally, I want to express my love for the beautiful game of football, which has instilled in me qualities such as honesty, courage, adventurous spirit, and unwavering confidence. All of my life's lessons have been learned with a football at my feet.

Most importantly, I want to convey my profound gratitude to my family. Firstly, I would like to remember my late grandfather, a VDO, Mr. Tika Ram Joshi, whose cherished memories continue to resonate with me. I fondly recall those childhood hours you dedicated in teaching Atul and me, along with the countless stories and jokes that never failed to make us laugh. The valuable lessons you have bestowed will remain forever with us. My grandmother, Mrs. Madhavi Joshi, your boundless love has been a constant source of warmth and affection. My father, Mr. Brij Mohan Joshi, your tireless dedication and steadfast commitment to providing for all of us have made an immeasurable impact on our lives. My mother, Mrs. Neelam Joshi, who has gone above and beyond to ensure we receive a quality education, your efforts have paved the way for our success. My genius beloved brother, an Engineer, Mr. Atul Joshi, your support has been a pillar of strength in my life, and words will never suffice to convey the depth of my gratitude. My heartfelt appreciation to my sister-in-law, an Engineer, Mrs. Megha Joshi, whom I hold dear. Your presence brings immense joy to our family, and you have been a beacon of happiness and a solution to all the problems. I am especially thankful to both Atul and Megha for your assistance during the initial stage of my Ph.D, particularly in coding and creating plots. I trust that you all recognize and take pride in the substantial role you have played in shaping my life's trajectory.

I also want to express my gratitude to all the incredible individuals I have encountered during my travels. Their captivating journeys deeply impressed me, imparting the essence of traveling and life, and fueling my yearning to explore the world even more.

Finally, I thank all the people who has been part of this journey, including those I may have inadvertently omitted.

    \begin{flushright}
Nitin Joshi\\Indian Institute of Technology Ropar\\ India-140001\\ September 2023
    \end{flushright}
}

\newpage

\begin{center}
{\LARGE \textbf{List of Publications}}
\end{center}
{\vspace{0.5 cm}}

{\large\textbf {Included in this thesis}\footnote{For the list of publications included in this thesis, the authors' names are presented in alphabetical order because this is the accepted convention in the field of theoretical high-energy physics.}}

\begin{enumerate}

\item S.~Bhattacharya and \textbf{N.~Joshi},
 {\it Decoherence and entropy generation at one loop in the inflationary de Sitter spacetime for Yukawa interaction}, under communication [arXiv:2307.13443 [hep-th]]. 

\item  S.~Bhattacharya, \textbf{N.~Joshi} and S.~Kaushal,
{\it Decoherence and entropy generation in an open quantum scalar-fermion system with Yukawa interaction},
Eur. Phys. J. C \textbf{83}, 208 (2023)
[arXiv:2206.15045 [hep-th]].

\item S.~Bhattacharya and \textbf{N.~Joshi},
{\it Entanglement degradation in multi-event horizon spacetimes},
Phys. Rev. D \textbf{105}, 065007 (2022)
[arXiv:2105.02026 [hep-th]].

\item S.~Bhattacharya, H.~Gaur and \textbf{N.~Joshi},
{\it Some measures for fermionic entanglement in the cosmological de Sitter spacetime},
Phys. Rev. D \textbf{102}, 045017 (2020)
[arXiv:2006.14212 [hep-th]].

\end{enumerate}
\vspace{2mm}

{\large \textbf{Not included in this thesis}}
{\vspace{0.50 mm}}
\begin{enumerate}
\item S.~Bhattacharya and \textbf{N.~Joshi},
{\it Non-perturbative analysis for a massless minimal quantum scalar with V(\ensuremath{\phi}) = \ensuremath{\lambda}\ensuremath{\phi} $^{4}$/4! + \ensuremath{\beta}\ensuremath{\phi} $^{3}$/3! in the inflationary de~Sitter spacetime},
JCAP \textbf{03}, 058 (2023)
[arXiv:2211.12027 [hep-th]].

\item G.~Yadav and \textbf{N.~Joshi},
{\it Cosmological and black hole islands in multi-event horizon spacetimes},
Phys. Rev. D \textbf{107}, 026009 (2023)
[arXiv:2210.00331 [hep-th]].

\item S. Bhattacharya, \textbf{N. Joshi} and K. Roy, {\it Resummation of local and non-local scalar self energies via the Schwinger-Dyson equation in de Sitter spacetime}, under communication [arXiv:2310.19436 [hep-th]].

\item S. Bhattacharya, \textbf{N. Joshi} and S. Kumar, {\it Perturbations in stochastic inflation for a massless minimal coupled quantum scalar field with asymmetric interaction in the quasi-de Sitter spacetime},
{manuscript in preparation}.

\end{enumerate}

\end{spacing}





%
%


\frontmatter
\pagestyle{plain}

\begin{center}
{
    \fontencoding{OT1}
    \fontfamily{ppl}
    \fontseries{b}
    \fontshape{n}
    \fontsize{20}{40}
    \selectfont
    Abstract
}

\vspace*{0.1in}

{
    \selectfont
    }

\end{center}

\vspace*{0.1in}

\begin{spacing}{1.2}

{\noindent This thesis presents a theoretical investigation into the quantum field theoretic aspects of quantum correlations and decoherence in the cosmological spacetimes. We shall focus on the inflationary or dark energy dominated phase of the universe, and we shall take the spacetime background to be de Sitter. The aim of this thesis is to investigate the physics of the very early universe and to gain insight into the interesting interplay among quantum correlations, entanglement and decoherence which can affect the evolution of our universe.

We begin with an introduction on the foundational motivation and outline the scope of this thesis. A comprehensive introduction to quantum correlations and decoherence within the framework of quantum field theory is provided. Furthermore, we provide a concise overview of diverse correlation quantification methods, setting the stage for further detailed investigations in the ensuing chapters. As the first objective of this thesis, we chiefly explore two measures of quantum correlations and entanglement, mainly, the violation of the Bell-Mermin-Klyshko (BMK) inequalities and the quantum discord, in the cosmological de Sitter background for the Dirac fermions. Specifically, we focus on the two and four-mode squeezed states, constructed from the initial  Bunch-Davies vacuum. We have demonstrated the maximum violation of the BMK inequalities. We also have analyzed the quantum discord for a maximally entangled initial states. We have compared our results with that of a couple of other coordinatisations of the de Sitter. The reason behind this is, such different coordinatisations used different time like coordinates, and hence the vacuum states corresponding to them are different. Hopefully, these computations provide us some insights into the inter-relationship between quantum correlations, entanglement, and quantum fields living in the very early universe.

Next, we focus on the computation of decoherence and entropy generation in the Minkowski and inflationary de Sitter spacetime for the Yukawa interaction. We use the correlator approach proposed a few years ago. The scalar field is treated as the system, and the fermions are considered as the environment in both cases. We have taken the Minkowski spacetime as a preliminary model to understand decoherence in the presence of fermions, before we move into the more complex scenario of the de Sitter spacetime. We have constructed the two loop and 2-particle irreducible effective action for the theory and have derived the renormalized Kadanoff-Baym equations, in the Schwinger-Keldysh or the in-in or the closed time path formalism. The Kadanoff-Baym equations are the equations of motion satisfied by the propagators in the in-in formalism, and they are causal. They contain corrections due to the self-energy, and can be thought of as the generalisation of the Schwinger-Dyson equations of the standard quantum field theory. Using these equations, we compute the loop corrected statistical propagator, the phase space area of the theory and finally the von Neumann entropy. We analyze the variation of this entropy with different relevant parameters and compare the result with scenario involving scalar fields as both the system and the environment. Our result in the Minkowski spacetime is non perturbative. However for the de Sitter background we have been able to find out a perturbative result at one loop level. Also in particular, this latter result shows qualitative similarity with an earlier one obtained using the  Feynman-Vernon influence functional technique. 

As the last objective of this thesis, we wish to investigate the degradation or survival of quantum entanglement in a cosmological back hole background, by studying the mutual information and the logarithmic negativity for maximally entangled, bipartite states for massless minimal scalar fields. We take the spacetime to be Schwarzschild-de Sitter and for simplicity we restrict ourself to (1+1)-dimensions. This spacetime basically represent a static and spherically symmetric black hole sitting in the de Sitter universe. Interestingly, this spacetime is endowed with a black hole as well as a cosmological event horizon, giving rise to particle creation at two different temperatures. This makes this multi-event horizon spacetime qualitatively different from black holes located in flat or anti-de Sitter background. We first show that the entanglement or correlation degrades with increasing Hawking temperatures, in agreement with the results found earlier. However, while treating both the horizons in an equal footing in order to define a total Bekenstein-Hawking entropy for this spacetime and an effective equilibrium temperature, we also aim to show that unlike the usual cases, the particle creation does not occur here in causally disconnected spacetime wedges but instead in a single region. Using the associated quantum state, we show that in this scenario the entanglement never degrades but increases with increasing black hole temperature and holds true no matter how hot the black hole becomes or how small the cosmological constant is. We argue that this phenomenon can have no analogue in the asymptotically flat/anti-de Sitter black hole spacetimes.

Finally, we summarize our results presented in the preceding main chapters of the thesis. We also mention here some future directions that might be important to understand the very early universe and the study of quantum correlations and decoherence in such backgrounds.



}
\end{spacing}

\newpage
\begin{center}
\Large\textbf{Notations}    
\end{center}

\hspace{-06mm}$S$ : Action\\
$\mathcal{L}$ : Lagrangian density\\
$\Lambda$ : Cosmological constant\\
$H$ : Hubble rate\\
$G$ : Newton constant\\
$t$ : Cosmological time\\
$\eta$ : Conformal time\\
$\eta_{\mu\nu}$ : Flat spacetime metric\\
$g_{\mu\nu}$ : Curved spacetime metric\\
$\Gamma^{\lambda} _{\mu \nu}$ : Christoffel connections\\
$\nabla_\mu$ : Covariant derivative\\
$D_\mu$ : Spin covariant derivative\\
$\psi$ : Fermionic field\\
$\phi$ : Scalar field\\
$\otimes$ : Tensor product between quantum states\\
$||\hspace{1.5mm} ||$ : Trace norm\\
$\ln$ : Natural logarithms\\
$\log$ : $\log_2$\\
$|A,B\rangle$ : $|A\rangle \otimes |B\rangle$

\pagestyle{plain}
\include{abr}
\pagestyle{plain}
\tableofcontents
\listoffigures
\pagestyle{plain}
\pagestyle{empty}



\pagestyle{fancy} 
\renewcommand{\chaptermark}[1]{\markboth{\textbf{\thechapter}\ \emph{#1}}{}}
\renewcommand{\sectionmark}[1]{\markright{\thesection.\ #1}}

\fancyhf{} 
\fancyhead[LE,RO]{\textbf{\thepage}} 
\fancyhead[RE]{\nouppercase{\rightmark}} 
\fancyfoot[RO]{\bfseries{\leftmark}}
\renewcommand{\headrulewidth}{0.5pt} 
\renewcommand{\footrulewidth}{0.5pt} 




\mainmatter

\chapter{Motivation and Overview}
\label{Motivation and Overview}
The standard big bang model, along with the inclusion of dark matter and dark energy, is a widely accepted cosmological framework that describes the origin and evolution of our universe. This model, coupled with the concept of various cosmological epochs, provides a comprehensive framework for understanding the history and evolution of our vast cosmos. We refer our reader to \cite{Weinberg:2008zzc, Mukhanov:2005sc, Guth:1997nq, Kolb:1990vq, Ryden:2016edk, Rindler, Dicke} for extensive and pedagogical reviews on these outlooks. According to this model, the universe began with a singularity, a point of infinite density and temperature, around 13.8 billion years ago. The subsequent epochs can be classified into several distinct phases. The first epoch, known as the Planck epoch, spans from the initial singularity until approximately $10^{-43}$ seconds after the big bang, where quantum effects probably including that of non-perturbative quantum gravity dominated and the fundamental forces are unified. Following the Planck epoch, the universe enters the primordial inflationary era, lasting approximately from $10^{-36}$ to $10^{-32}$ seconds, characterized by an exponential accelerated expansion. Inflation not only offers compelling solutions to cosmological puzzles such as spatial flatness, the horizon problem, and the scarcity of relics like the magnetic monopoles, but also furnishes an elegant mechanism for generating primordial cosmological density perturbations in the very early universe, as a seed to the large scale cosmic structures we observe today in the sky \cite{Weinberg:2008zzc, Mukhanov:2005sc, Guth:1997nq, Kolb:1990vq}. In order to drive the accelerated expansion of the spacetime, one usually expects that the universe has to be dominated by some sort of exotic matter with positive energy density but negative and isotropic pressure. This is called the dark energy, the simplest and phenomenologically very successful form of which is just a positive constant, $\Lambda$, known as the cosmological constant. We shall be discussing more on this in the subsequent Chapters. After the inflation ends, the universe transitions into the electroweak epoch, encompassing the next $10^{-12}$ seconds, during which the electromagnetic and weak nuclear forces become distinct. The subsequent quark epoch, lasting until about $10^{-6}$ seconds, witnesses the emergence of quarks and gluons, the fundamental particles that compose protons and neutrons. As the universe cools further, the quarks combine to form protons and neutrons during the hadron epoch, which lasts until about 1 second. Following this, the universe enters the lepton epoch, characterized by the dominance of leptons such as electrons and neutrinos, until approximately 10 seconds. Subsequently, the radiation-dominated epoch commences, lasting until about 50,000 years, where photons and other forms of radiation dominate the energy density of the universe. The matter dominated epoch follows then, with protons, neutrons, and cold dark matter becoming dominant. During this matter domination era, large scale structure formations was chiefly encouraged. We note that, the present era is also believed to be dominated by the dark energy or cosmological constant. However, the present dark energy density is much small compared to what it was during the primordial inflationary era. This leads to the hitherto elusive fine tuning issues, known as the cosmic coincidence puzzle, e.g. \cite{Weinberg:2008zzc, Mukhanov:2005sc, Grande:2008re, Zheng:2021uee, Davis:2007na, Riess, Perlmutter} and references therein. In this thesis, we shall be interested to probe the primordial inflationary period of our universe. 

It is well understood that the physics operating at quantum scales hold tremendous significance in shaping the large scale structure of the universe, as exemplified by the cosmic microwave background (CMB). The CMB, a relic radiation from the early universe presents us with a remarkable snapshot of a time when the universe was merely 380,000 years old \cite{Kosowsky:1996yc, Peebles:1993ud, Hu:2002rs, Planck:2018jrf, Durrer:2008aa}. Through careful analysis and observation, we can glean invaluable insights into the origin and evolution of the cosmos. According to the prevailing theoretical paradigm, a comprehensive understanding of the quantum physics underlying inflationary perturbations is crucial for comprehending the temperature fluctuations observed across the expanse of the CMB sky today. By examining these fluctuations, we can get information about the fundamental properties of the early universe, such as its density variations and the nature of the primordial gravitational waves generated during inflation. Moreover, a profound understanding of nuclear fusion and the pivotal role played by stars in this process empowers us to calculate the abundances of light elements that were generated a mere three minutes after the colossal event of the big bang. This process, known as the big bang nucleosynthesis, sheds light on the production of elements like hydrogen, helium, and lithium, offering crucial evidence in support of our current cosmological models.

In the realm of theoretical cosmology, the exploration of the early universe and the formation of cosmic structures intertwines with the profound concepts of quantum correlations and decoherence. The interplay between quantum physics and the large scale structure of the universe becomes apparent through the examination of CMB \cite{Parker:2009uva}. Furthermore, the study of decoherence and quantum correlations becomes particularly relevant when investigating the geometry of very early universe and the formation of large scale structures. Quantum correlations, such as entanglement, hold potential insights into the universe's quantum origin and cosmic structure. The process of decoherence, driven by interactions between these quantum systems and their surrounding environment, leads to the loss of quantum coherence and the emergence of classical behavior on cosmological scales that eventually give rise to galaxies, clusters of galaxies, and other cosmic structures we observe today in the sky. Understanding the interplay between quantum correlations and decoherence is crucial for comprehending the intricate mechanisms that govern the evolution of the universe.

Entanglement which can be regarded as a non-local form of quantum correlations, arise when two or more quantum systems become intertwined in such a way that their states cannot be factorised into two different Hilbert spaces \cite{Nielsen:2000}. We shall be discussing more in depth on quantum entanglement in \ref{Quantum correlations}. The relativistic sector, where particle pair creation may occur, is always of particular interest in this context because these pairs are found to be entangled. One of the most extensively studied cases of quantum entanglement in this context involves the Rindler left-right wedges or the maximally extended near-horizon geometry of a non-extremal black hole, as discussed in~\cite{MartinMartinez:2010ar, Qiang, Wang, Yao, Xiao:2018cxg, Alsing:2003es, alsing:2006, Friis:2011fy, A.dutta, wang, Brown:2012iz} (and references therein). Due to thermal pair creation, the initial entanglement or quantum correlation between two maximally entangled Bell pairs degrades in the Rindler frame, as observed in~\cite{MartinMartinez:2010ar}. See e.g. \cite{Bhattacharya:2020sjr} for a discussion on entanglement in the context of Schwinger pair creation. We further refer our reader to~\cite{Tomaras:2019sjq} and references therein for a discussion on entanglement in the soft sector of quantum electrodynamics. Moreover, the presence of entangled quantum states in the early universe was substantiated through an examination of photon pairs from certain high-redshift quasars, wherein the violation of Bell inequalities was observed, discussed in \cite{Rauch}, indicating the existence of entangled quantum states in the very early universe. We also refer to \cite{Morse} (and references therein) for the exploration on signature of Bell violation in the CMB and its observational constraints.

In this thesis, we first shall be studying quantum correlations or entanglement in the context of inflationary de Sitter spacetime for the Dirac fermions. Understanding the role that entanglement played in the early universe could provide valuable insights into the initial state as well as the geometry of the same. We would also be studying the degradation or survival of quantum entanglement in a cosmological black hole background. We take the spacetime to be the Schwarzschild-de Sitter, representing a static and spherically symmetric black hole located in the de Sitter universe. The primary qualitative distinction between these black holes and those with $\Lambda\leq 0$ is due to the presence of a cosmological event horizon. This cosmological event horizon serves as an outer causal boundary of our universe, restricting the length scale of the universe an observer can see. Since this spacetime features two event horizons, the black hole and the cosmological, a two-temperature thermodynamics framework is applicable here, making it markedly distinct from the spacetimes with a single event horizon, see e.g.~\cite{Gibbons, Shankaranarayanan:2000qv, Shankaranarayanan:2003ya, Robson:2019yzx}. The study of entanglement in the relativistic sector usually involves computation of measures like entanglement entropy, logarithmic negativity, mutual information, quantum discord etc., for various bosonic and fermionic fields. People are also often interested to investigate the quantum decoherence of cosmological perturbations and their possible observational consequences. For these aspects, we refer our reader to e.g.~\cite{SHN:2020, SN:2021, Choudhury:2016cso, Choudhury:2017bou, Fuentes:2010dt, Wu:2023pge, Maldacena:2015bha, Lim, Vennin, Liu:2016aaf, Kanno:2016gas,  Chen:2017cgw, Feng:2018ebt, Matsumura:2020uyg, Maldacena:2012xp, kanno:2015, Albrecht:2018prr,  Bhattacharya:2018yhm, dePutter:2019xxv, Bhattacharya:2019zno, Brahma:2020zpk, Menezes:2017oeb, Menezes:2015veo} and references therein.

Decoherence is a process that causes a quantum system to lose its coherence and become more classical in nature. This occurs when a quantum system interacts with its environment which is classical, creating entanglement between them. In the context of cosmology, decoherence is of interest since it may provide a way to explain how the classical world emerges from the underlying quantum world. Furthermore, decoherence can help to explain the emergence of a classical spacetime, which is essential for the formulation of general relativity. One aspect of quantum correlations and decoherence in cosmological spacetimes that has been studied extensively is the role they play in the problem of time in canonical quantum gravity, see e.g. \cite{Hu:2012rg, Rovelli:2004tv, Oriti:2009wn}. This problem arises due to the fact that time is treated as an external parameter in quantum mechanics, whereas in general relativity, time is a dynamical variable. The study of quantum correlations and decoherence in cosmological spacetimes may provide a way to solve this puzzle and give us a consistent framework for quantum gravity.

It is also widely recognized that massless, conformally non-invariant quantum fields, such as gravitons and a massless minimally coupled scalar, exhibit pronounced infrared temporal growth during the late times in inflationary backgrounds. This phenomenon is known as the {\it secular effect}. This effects is characterized by the powers of logarithm of the scale factor and originates from the existence of sufficiently long lived virtual particles residing within loops in the deep infrared or super-Hubble regime, indicating a breakdown of perturbation theory at late times \cite{nitin, Bhattacharya:2022aqi, Friedrich:2019hev, Onemli:2002hr, Tsamis:2005hd, Cabrer:2007xm, Glavan:2021adm, Akhmedov:2013xka, Akhmedov:2014doa, Akhmedov:2015xwa, Akhmedov:2019cfd, Kaplanek:2020iay, Hu:2018nxy}. Quantum fields that exhibit such secular effects are commonly referred to as {\it spectator} fields, such as a massless and minimally coupled scalar and gravitons. When a {\it spectator} field interacts with a conventional matter field, quantum effects can become strongly entwined, potentially imparting similar effects to the latter. As highlighted in \cite{Friedrich:2019hev}, this interplay alters the correlator and, consequently, the extent of decoherence and classicalisation in a non-trivial manner. Considering these elements collectively, investigating the phenomenon of decoherence or the quantum-to-classical transition in an inflationary background, particularly in the presence of these spectator fields, presents an interesting research endeavor.

The interface of quantum field theory and quantum decoherence is rapidly gaining interest among the research community, for e.g. we refer our reader to \cite{Parker:2009uva, Birrell:1982ix, Fulling:1989nb, Bertlmann:2000da, Peskin:2018, Weinberg, Sean, QFTCS1, Wald:1994rg, Hollands:2014eia} and also references therein. However, doing quantum field theory in a time dependent background such as the cosmological ones inherently embodies a non-equilibrium framework. Various strategies have been proposed in the literature to quantify decoherence in the context of non-equilibrium quantum field theory. Typically, these approaches involve calculating some measure of decoherence using the master equation, after tracing out the environmental degrees of freedom. In this thesis, we shall instead be interested in approach based upon the 2-particle irreducible effective action and the correlation functions, in order to compute the decoherence in terms of the von Neumann entropy generation. This approach, as usual, divides the entire theory into a system we observe and an environment. The information about the system is characterised by the correlation functions. We refer our reader to \cite{Calzetta Hu, Breuer:2002, Weiss:2012, Caldeira:1983, Hu:1992, Paz:1993, Breuer:2016, Rivas:2014, deVega:2017, Breuer:2009, Weiss:2008} for some relevant discussion. We shall be discussing more on these aspects in the due course.

The rest of this Chapter is organized as follows. In \ref{Quantum correlations}, we provide a basic understanding of the key concepts of quantum correlations from the point of view of entanglement, along with the various measures that will be used to quantify the same. This lays the foundation for their utility in the subsequent Chapters. Next, in \ref{deco}, we review the phenomenon of quantum decoherence. We also discuss the method used to quantify decoherence and the approach we will be using in this thesis. In \ref{Quantum field theory in curved spacetime}, we outline some fundamental ingredients of quantum field theory in curved spacetime. In \ref{The de Sitter spacetime}, we briefly review the geometry of the de Sitter spacetime. Finally, in \ref{A synopsis of the thesis}, we offer a brief overview of the remaining Chapters of this thesis.

We shall work with the mostly positive signature of the metric. If not otherwise stated our spacetime dimension will be four. We will set $c=\hbar=k_B=G=1$ throughout. In our numerical calculations, we interpret $\log$ as $\log_2$ and $\ln$ as the natural logarithm.

\section{Quantum entanglement}
\label{Quantum correlations}

Entanglement, a special kind of non-local quantum correlation, is one of the fundamental aspects of the quantum world \cite{Nielsen:2000, Einstein, bell:1964, clauser:1969, cirel:1980, Aspect1, Aspect2, Zyczkowski:1998yd, Vidal:2002zz, chen:2,  mermin:1990, belinski, Gisin:1998, nagata:2002, yu:2003, uffink:2002, Zurek, Dakic, Huang, aharon:2011}. In classical physics, two systems can be described independently of each other, but in the quantum realm, entanglement allows for a deeper and more intricate connection between particles or systems. When two or more quantum particles become entangled, their properties become interdependent, regardless of the spatial separation. This means that measuring the state of one of the entangled particles instantaneously affects the state of the other, no matter how far apart they are, and seemingly violates the principle of locality. This phenomenon, famously referred to as ``spooky action at a distance" by Albert Einstein, is a striking departure from classical intuitions. Entanglement also hold significant practical implications across various domain such as quantum computing. Moreover, entangled particles assume a pivotal role in quantum cryptography, furnishing secure techniques for the transmission and encryption of sensitive information e.g. \cite{Liu:2021xyz, Guhne:2022abc, Luo:2023def, Li:2023ghi, Gharibian:2023jkl}.

Our objective in this thesis is to investigate quantum correlations in the inflationary de Sitter spacetime. In de Sitter spacetime, which models an expanding universe with a positive cosmological constant, quantum correlations undergo alterations due to the curvature and the presence of a cosmological horizon. The interplay between quantum correlations and the expansion of the universe in the de Sitter spacetime leads to an interesting phenomena. The universe's expansion introduces an effective temperature linked to the de Sitter vacuum, and this thermal radiation or Hawking temperature has a notable impact on the entanglement characteristics of quantum systems. The expansion can stretch and dilute the entanglement between particles, resulting in the decay of correlations over time, a process known as entanglement evaporation/degradation, for e.g. see \cite{MartinMartinez:2010ar, Choudhury:2016cso, Choudhury:2017bou, Fuentes:2010dt, Wu:2023pge, Maldacena:2015bha, Bhattacharya:2018yhm, FuentesSchuller:2004xp, Torres-Arenas:2018vei, Wang:2010qq, Wu:2021pja, Dong:2019jhs, Hwang:2001etg, Agullo:2009zza, Antoniadis:1986sb, Martin:2015qta, EE, bell:2017, vaccum EE for fermions, QC in deSitter} and references therein.

Now, a natural question to begin with will be, how to detect or quantify entanglement? Various measures of entanglement have been defined to address this questions, and studied extensively in \cite{Zyczkowski:1998yd, Vidal:2002zz, Anastopoulos:2022owu, Preskill, RevModPhys, Jordan:2011ci, Giddings:2012bm, Calabrese:2004eu, Wu:2022glj, Vidal:1998re, measure, Monogamy, Horodecki:2009zz} and also references therein. Let us first try to understand, how to find out whether a state is entangled or not.

We consider two quantum states $|\psi\rangle_A$ on the Hilbert space of system $A$, and similarly, $|\psi\rangle_B$ on the Hilbert space of system $B$. Hence, we can express the total state as

\begin{equation}\label{cij}
|\psi\rangle_{AB} = |\psi\rangle_A \otimes |\psi\rangle_B 
 = \sum_{i,j} c_{ij} |\psi_i\rangle_A \otimes |\psi_j\rangle_B    
\end{equation}

where we have made the expansion in terms of the base kets, $|\psi_{A}\rangle = \sum_{i} c^A_{i}|\psi_{i}\rangle$ and $|\psi_{B}\rangle = \sum_{j} c^{B}_{j}|\psi_{j}\rangle$, where $c_{ij} = c^A_{i}c^B_{j}$ in \ref{cij}. Thus the composite system $|\psi\rangle_{AB}$ can be factorised into subparts $A$ and $B$, and in such instances $|\psi\rangle_{AB}$ is regarded as separable. However, there can be scenarios where one can construct $|\psi\rangle_{AB}$ in such a manner that, $c_{ij} \neq c^A_{i}c^B_{j}$. In such cases $|\psi\rangle_{AB}$ is not separable and we regard it as entangled state, indicating that one part of it cannot be characterized independently without the knowledge of the other part. The most popular of examples would be the four Bell states constructed for the spin-1/2 particles. For example one of which reads,
$$|\psi\rangle_{AB} = \frac{1}{\sqrt{2}}\Big[|+\rangle_A |+\rangle_B + |-\rangle_A |-\rangle_B  \Big] 
$$
which is manifestly non separable. Let us now briefly review some measures to quantify quantum entanglement in the following subsections. 
\subsection{Measures of quantum entanglement}
\label{Measures of correlations}
Measures of quantum correlations are mathematical tools used to quantify and characterize the extent of entanglement in a quantum system. Various measures have been developed based on different aspects of quantum correlations, such as entanglement entropy, mutual information, concurrence, entanglement witnesses, quantum discord, entanglement of formations, distillable entanglement and many more for which the reader can refer to \cite{ Vidal:2002zz, Plenio:2005, Calabrese:2012nk}. Different measures of quantum correlations are useful in different contexts, depending on the nature of the quantum system and the information of interest. In this thesis, we shall be working with the following measures. 

\subsubsection{Entanglement entropy}
\label{Entanglement entropy}
\noindent Entanglement entropy quantifies the amount of non-local correlations or information shared between two quantum systems that are entangled with each other. Consider a bipartite system comprising two subsystems, $A$ and $B$, such that the Hilbert space can be decomposed as ${\mathscr{H}}_{AB}$ $ = {\mathscr{H}}_{A}$ $ \otimes $ ${\mathscr{H}}_{B}$. Let $\rho_{AB}$ be the density matrix of states on ${\mathscr{H}}_{AB}$. The entanglement entropy of subsystem $A$ is defined as the von Neumann entropy,
\begin{eqnarray}
S(\rho_A) = -\mathrm{Tr}_A \left(\rho_A \log \rho_A\right)    
\end{eqnarray}
where reduced density matrix of subsystem $A$, $\rho_A = \text{Tr}_B \rho_{AB}$. The partial trace $\text{Tr}_B$ is taken over the Hilbert space $\mathscr{H}_B$. One can define $S(\rho_B)$ in a likewise manner. If $\rho_{AB}$ corresponds to a pure state, the entanglement entropy $S(\rho_A) = S(\rho_B)$. Also, when $\rho_{AB}$ is separable, the entanglement entropy is vanishing. The entanglement entropy also satisfy a subadditivity property: $S(\rho_{AB}) \leq S(\rho_A) + S(\rho_B)$, where $S(\rho_{AB})$ represents the von Neumann entropy corresponding to density matrix $\rho_{AB}$. This inequality holds true, if and only if $\rho_{AB}$ is separable. For further elaboration we refer our reader to \cite{Nielsen:2000}.

\subsubsection{Logarithmic negativity}
\label{Entanglement negativity and logarithmic negativity}
\noindent
The logarithmic negativity is calculated by taking the logarithm of the trace norm of the partially transposed density matrix of the system. In order to define the logarithmic negativity let us first consider,
$$ \mathcal{N}(\rho_{AB})
= \frac{1}{2} \left(\lvert \lvert \rho_{AB}^{T_A} \rvert \rvert_1-1\right)$$
where $\rho_{AB}^{\text{T}_A}$ is the partial transpose of $\rho_{AB}$ w.r.t the subspace $A$, $\left( \lvert i\rangle_{\! A} \hspace{-0.2ex} \langle n \rvert \otimes \lvert j \rangle_{\! B} \hspace{-0.2ex} \langle \ell \rvert \right)^{\text{T}_A}: = \lvert n \rangle_{\! A} \hspace{-0.2ex} \langle i \rvert \otimes \lvert j \rangle_{\! B} \hspace{-0.2ex} \langle \ell \rvert$.
Here, $\lvert \lvert \rho_{AB}^{T_A} \rvert\rvert_{1}$ is the trace norm, $\lvert \lvert\rho_{AB}^{T_A}\rvert \rvert_{1} = \sum_{i=1}^{\text{all}} \lvert \mu_i \rvert$, where $\mu_i$ is the $i$-th eigenvalue of $\rho_{AB}^{T_A}$. The logarithmic negativity represents the logarithm of the absolute value of the trace norm of the partially transposed density matrix $\rho_{AB}^{T_A}$
\begin{eqnarray}
L_{N} (\rho_{AB}) = \log (1 + 2 \mathcal{N}(\rho_{AB}) )    
\end{eqnarray}
The quantities mentioned above quantify the violation of positive partial transpose (PPT) in the density matrix $\rho_{AB}$. The PPT criterion states that if $\rho_{AB}$ is separable, then the eigenvalues of $\rho_{AB}^{T_A}$ are non-negative. Therefore, if $\mathcal{N} \neq 0$ ($L_\mathcal{N} \neq 0$), it indicates that $\rho_{AB}$ is an entangled state. On the contrary, if $\mathcal{N} = 0$ ($L_\mathcal{N} = 0$), we cannot determine the existence of entanglement solely based on this measure, as there are PPT states that can still be entangled. Further in depth discussions on this topic can be found in \cite{Horodecki:2009zz}.

\subsubsection{Mutual Information}
\label{Mutual Information}
\noindent
Mutual information is a measure of the amount of information shared between two random variables. In the context of quantum correlations, mutual information quantifies the amount of correlations between two subsystems of a larger quantum system. It captures both classical and quantum correlations, between the subsystems $A$ and $B$, defined as,
\begin{eqnarray}
I(A,B) = S(\rho_{A})+ S(\rho_{B}) -	S(\rho_{AB})
\end{eqnarray}

The subadditivity of the entanglement entropy directly implies that $I(A,B) \geq 0$. This bound is achieved only when the joint state of the subsystems, $\rho_{AB}$, is separable, i.e., when it can be written as a tensor product of the individual subsystem states, $\rho_A \otimes \rho_B$. For further pedagogical discussion we refer our reader to ~\cite{Nielsen:2000}.

\subsubsection{Quantum discord} \label{discord} 

Quantum discord is a measure of the quantum correlation between two quantum systems, which extends beyond the classical correlation described by mutual information. Quantum discord is defined as the difference between two measures of mutual information, one calculated using the full joint quantum state of the two systems, and the other using a restricted set of measurements on one of the systems. This difference is nonzero when the correlation between the two systems is quantum in nature. Quantum discord has found many applications in quantum information theory, including in quantum cryptography, quantum computing, and quantum metrology. It has also been used as a tool to investigate the role of quantum correlations in quantum thermodynamics, where it is relevant for understanding the efficiency of heat engines and the emergence of thermodynamic behavior in small quantum systems.

Let us now define the quantum discord~\cite{Zurek, Dakic}. In classical information theory, the mutual information between two random variables $X$ and $Y$ is defined as
\begin{eqnarray}
{\cal I}(X,Y)=H(X)+H(Y)-H(X,Y),
\label{d1}
\end{eqnarray}
where $H(X)=-\sum_{X} P(X)\log P(X)$ and $H(Y)=-\sum_{Y} P(Y)\log  P(Y))$ are the Shannon entropies with probabilities $P(X)$ and $P(Y)$, respectively, and $H(X,Y)\\=-\sum_{X,Y} P(X,Y)\log  P(X,Y)$ is the joint Shannon entropy with joint probability $P(X,Y)$ for both variables $X$ and $Y$. The joint probability $P(X,Y)$ can be related to the conditional probability $P(Y|X)$ as 
\begin{eqnarray}
P(X,Y)=P(X)P(Y|X),
\label{d2}
\end{eqnarray}
where $P(Y|X)$ is the probability of $Y$ if $X$ is given with probability $P(X)$. Thus, the joint entropy $H(X,Y)$ can be rewritten as
\begin{eqnarray}
H(X,Y)=-\sum_{X,Y} P(X,Y)\left[\log  P(X)+\log P(Y|X) \right].
\label{d3}
\end{eqnarray}  
Therefore, \ref{d1} can be rewritten as
\begin{eqnarray}
{\cal I}(X,Y)=H(Y)-H(Y|X),
\label{d4}
\end{eqnarray}
where 
$H(Y|X)=-\sum_{X,Y} P(X)P(Y|X)\log P(Y|X)$ is the conditional entropy, representing the average over $X$ of the Shannon entropy of $Y$ given $X$.

The above construction is purely classical. For a quantum system, however, the Shannon entropy is replaced by the von Neumann entropy, $S(\rho)=-{\rm Tr}\rho\log \rho$, where $\rho$ is the density operator. Also, $P(X,Y)$, $P(X)$, and $P(Y)$ are replaced, respectively, by the density operator of the entire system $\rho_{X,Y}$, the reduced density operator of subsystem $X$ $(\rho_{\small X}={\rm Tr}_{\footnotesize Y}\rho_{X,Y})$, and the reduced density operator of subsystem $Y$ $(\rho_{Y}=\operatorname{Tr_{X}\rho_{X,Y}})$.

In quantum mechanics, the concept of conditional probability $P(Y|X)$ requires projective measurements using a complete set of projection operators $\Pi_{i}=|\psi_{i}\rangle\langle\psi_{i}|$ for all $i$. The density operator of $Y$ after measuring $X$ is expressed as
\begin{eqnarray}
\rho_{Y|i}=\frac{{\rm Tr}_{X}\left(\Pi_i\rho_{X,Y}\Pi_i \right)}{p_{i}}, \quad \text{where}\quad p_{i}=\operatorname{Tr}_{X,Y}(\Pi_{i}\rho_{X,Y}\Pi_{i}).
\label{d5}
\end{eqnarray}
Using the density operator defined in \ref{d5}, we can define a quantum analogue of the conditional entropy as
\begin{eqnarray}
S(Y|X)=\min_{\Pi_{i}}\sum_{i}p_{i}S(\rho_{Y|i}),
\label{d6}
\end{eqnarray}
where $S(\rho_{Y|i})$ is the von Neumann entropy. The term `min' in the above expression corresponds to measurements that disturb the system the least, minimizing the influence of projectors. We can write the quantum analogues of \ref{d1} and \ref{d4} as
\begin{eqnarray}
{\cal I}_{X,Y}=S(\rho_{X})+S(\rho_{Y})-S(\rho_{X,Y}), \quad 
{\cal J}_{X,Y}=S(\rho_{Y})-S(X|Y),
\label{d7}
\end{eqnarray}
respectively. However, unlike in the classical scenario, the equivalent expressions \ref{d1} and \ref{d4} for the mutual information need not necessarily be equal in the quantum case. This is because they involve different measurement procedures, and a quantum measurement on one subsystem can affect the other. Therefore, the quantum discord is defined as,~\cite{Zurek, Dakic},
\begin{eqnarray}
\mathscr{D}_{X,Y} ={\mathcal I}_{X,Y}-{\mathcal J}_{X,Y}=S(\rho_{X})-S(\rho_{X,Y})+S(Y|X)
\label{d8}
\end{eqnarray}

\subsubsection{The violation of the  BMK inequalities}\label{Bell1}

Bell's inequality violation occurs when the predictions of quantum mechanics contradict classical physics. It demonstrates that entangled particles can exhibit correlations that cannot be explained by local hidden variables, challenging the idea of locality and realism. By conducting carefully designed experiments \cite{Aspect1, Aspect2}, it has been confirmed that the predictions of quantum mechanics hold true, revealing the complete non-classical nature of the quantum world and providing evidence for the existence of quantum entanglement.

Consider two sets of non-commuting observables defined over the Hilbert spaces $\mathscr{H}_{X}$ and $\mathscr{H}_{Y}$, denoted by $\{X, X^{\prime} \in \mathscr{H}_{X}\}$ and $\{Y, Y^{\prime} \in \mathscr{H}_{Y} \}$. We assume that these observables represent spin-$1/2$ operators along specific directions, such as $X = n_i \sigma_i$ and $X^{\prime} = n'_i \sigma_i$, where $\sigma_i$'s are the Pauli matrices and $n_i$, $n'_i$ are unit vectors in three-dimensional Euclidean space. The eigenvalues of these operators are $\pm 1$. To investigate the correlations between these observables, we introduce the Bell operator ${\cal B} \in \mathscr{H}_{X}\otimes \mathscr{H}_{Y}$, which is defined as
\begin{equation}
{\cal B}=\frac12\left(X \otimes Y+X^{\prime} \otimes Y +X \otimes Y^{\prime} -X^{\prime} \otimes Y^{\prime}\right)=\frac{1}{2}X\otimes(Y+{Y^\prime})+\frac{1}{2}{{X^\prime}}\otimes(Y-{Y^\prime})
\label{b1}
\end{equation}
In theories incorporating local classical hidden variables, Bell's inequality, $\langle{\cal B}^2\rangle\leq 1$ and $|\langle{\cal B}\rangle|\leq 1$ ~\cite{clauser:1969}, holds. However, this inequality is violated in the realm of quantum mechanics. In fact, considering \ref{b1} (omitting the tensor product sign),
\begin{eqnarray}
{\cal B}^2={\bf I} - \frac{1}{4} [X,X^\prime]\,[Y,Y^\prime],
\label{b2}
\end{eqnarray}
where ${\bf I}$ is the identity operator. By applying the commutation relations for the Pauli matrices, it can be deduced that $\langle{\cal B}^2\rangle\leq2$ or $|\langle{\cal B}\rangle|\leq\sqrt2$. This violation of Bell's inequality, commonly referred to as the {\it maximum violation} ~\cite{cirel:1980}.

The above construction can be extended to multipartite systems, leading to the Mermin-Klyshko inequalities, which are also collectively known as the Bell-Mermin-Klyshko inequalities or the Clauser-Horne-Shimony-Holt inequality. These inequalities are defined by a relevant operator that can be recursively constructed as
\begin{eqnarray}
{\cal B}_{n}=\frac{1}{2}{ \cal B}_{n-1}({ \cal O}_{n}+{\cal O}_{n}^\prime)+\frac{1}{2}{\cal B}_{n-1}^\prime({\cal O}_{n}-{\cal O}_{n}^\prime), \qquad n=2,3,4,\dots,
\label{b3}
\end{eqnarray}
where we have defined ${\cal B}_{1} ={\cal O}_{1}$, ${\cal B}^{\prime}_{1} ={\cal O}_{1}^{\prime}$. As earlier, all the operators correspond to spin-1/2 systems. In classical hidden variable theories we have ${\cal O}_{n}=\pm  {\cal O}_{n}^\prime$, yielding  the  Mermin-Klyshko inequalities
\begin{eqnarray}
|\langle{\cal B}_{n}\rangle|\leq 1, \qquad n=1,2,3, \dots 
\end{eqnarray}
whereas in quantum mechanics one has~\cite{Gisin:1998, nagata:2002, yu:2003},
\begin{eqnarray}
|\langle{\cal B}_{n}\rangle|\leq 2^{\frac{n-1}{2}}, \qquad n=1,2,3,\dots
\label{b4}
\end{eqnarray}
Thus the BMK inequality will also be violated for multipartite states, $n\geq 2$. We note that the construction of the Bell and the Bell-Mermin-Klyshko (BMK) operators for fermions is similar to that of the scalar field theory~\cite{bell:2017} (also references therein).

\section{Decoherence}\label{deco}

The superposition principle is a fundamental aspect of quantum mechanics, allowing for coherence and interference effects in ideal, isolated quantum systems. However, in realistic scenarios, quantum systems are not completely isolated and interact with their environment, leading to the phenomenon of decoherence. This arises out of the interaction and possibly entanglement between the system and its surrounding. The presence of this entanglement can have an impact on our local observations and measurements of the system, even from a classical perspective. Quantum decoherence plays a crucial role in the transition from quantum to classical behavior, ensuring the consistency between quantum predictions and the classical observations of the system. We refer our reader to \cite{Zurek:2003zz, Schlosshauer:2007zz, Giulini:2002zz, Breuer:2002rev, Paz:2001zza, Joos:2003} and references therein for an extensive review.

Quantum decoherence has gained significant attention within the research community, particularly in the context of interacting quantum field theories~\cite{Calzetta Hu, Calzetta Hu1}. It is closely associated with the loss or lack of information in an {\it open quantum system}, typically described within a system + environment framework. Various correlations, including mutual information, discord, and entanglement entropy, serve as useful characterisations of decoherence. The study of decoherence encompasses diverse scenarios. For instance, a comprehensive model for decoherence involving a non-relativistic quantum particle interacting with weak stochastic gravitational perturbations has been examined in~\cite{gravity, gravity1}. The generation of decoherence resulting from an accelerated time-delay source, as seen by an inertial observer, has been investigated in~\cite{acceleration}. In the context of dark matter, decoherence induced by gravitational interactions with ordinary matter in the environment has been analyzed in~\cite{darkmatter}. Notably, the decoherence mechanism may have interesting implications for the classicalisation of primordial inflationary quantum field theoretic perturbations, ultimately contributing to the formation of the large scale structures we observe in the sky today. For further discussions on these topics, we refer our reader to~\cite{Hollowood:2017bil, Bhattacharya:2022wpe, Bhattacharya:2023twz, cosmology, cosmology1, DC, DC1, DC2, DC3, Janssen:2007ht, Friedrich, Markkanen:2016vrp, Hu:1992xp, Hu:1990cr, Stargen:2016cft} and references therein.

Such interactions between the system and the surrounding are usually out of equilibrium phenomenon. Accordingly one uses the framework of non-equilibrium quantum field theory in order to quantify the decoherence in various ways, see e.g. \cite{Calzetta Hu, NEQFT, DTLN, noise, buyanovsky, FCL} and references therein. For example, such quantification is often characterised by the generation of the von Neumann entropy for the system at the late times \cite{buyanovsky, buyanovsky1, FA, entropy, entropy1, Berges, Schmidt}. Most popularly, the decoherence problem is addressed by tracing out the environmental degrees of freedom which are not accessible, resulting in a mixed density matrix for the system and hence an effective loss of information for the same. Also, decoherence is often characterized by the decay of the off-diagonal entries of the reduced density matrix in observer's basis. While any Hermitian matrix can be diagonalized by selecting appropriate basis, it is crucial to recognize that decoherence relies on a dynamical process that eliminates at late times these off-diagonal matrix elements in a specific basis, known as the pointer basis. A diagonal (reduced) density matrix is easy to interpret, as each entry corresponds to one of the classical outcomes of a measurement. In reality, a system does not fully decohere, and the von Neumann entropy is used to quantify the amount of decoherence that has occurred. Apart from the overall increase in entropy, there is also interest in determining the decoherence time, which is the characteristic timescale at which decoherence becomes effective. If this timescale is sufficiently long naturally, the system can maintain coherence over an extended period.

Let us examine a simple example of a spin-1/2 system with a superposition state
\begin{equation}
\vert \psi \rangle_S = \alpha \vert \uparrow \rangle + \beta \vert \downarrow \rangle, 
\end{equation}

where $|\alpha|^2 + |\beta|^2 = 1$. The corresponding density matrix is obtained as 

\begin{equation}\label{den}
{\rho}_S = \vert \psi \rangle_S \langle \psi \vert = 
\begin{pmatrix} 
|\alpha|^2 & \alpha \beta^* \\ 
\alpha^* \beta & |\beta|^2
\end{pmatrix}
\end{equation}

\ref{den} describes a pure state and has zero entropy. Let us consider another spin-1/2 particles with state $|E_{\uparrow}\rangle$ and $|E_{\downarrow}\rangle$. Let us also suppose that this latter particle serves as the environment. Then one of states which can represent a coupling between $\psi_S$ and the environment will be $
|\psi\rangle_{S+E} = \alpha| \uparrow \rangle|E_{\uparrow}\rangle + \beta|\downarrow\rangle|E_{\downarrow}\rangle
$. The coupling between the system and the environment produces entanglement, as evidenced by the non-separability of the state. The total density operator is given by

\begin{multline}\label{enden}
{\rho}_{S+E} = |\alpha|^2 |\uparrow \rangle \langle \uparrow | \otimes |E_{\uparrow}\rangle \langle E_{\uparrow}| + \alpha \beta^* |\uparrow \rangle \langle \downarrow | \otimes |E_{\uparrow}\rangle \langle E_{\downarrow}| \\
+ \alpha^* \beta |\downarrow \rangle \langle \uparrow | \otimes |E_{\downarrow}\rangle \langle E_{\uparrow}| + |\beta|^2 |\downarrow \rangle \langle \downarrow | \otimes |E_{\downarrow}\rangle \langle E_{\downarrow}|    
\end{multline}

We trace over the environmental degrees of freedom from \ref{enden}, yielding the reduced density matrix for the system,

\begin{equation}
{\rho}_{\rm red} = \text{Tr}_E({\rho}_{S+E}) = \begin{pmatrix} |\alpha|^2 & 0 \\ 0 & |\beta|^2 \end{pmatrix}
\end{equation}
We note that, ${\rho}_{\rm red}$ is mixed. The resulting change in the von Neumen entropy is given by
\begin{equation}
\Delta S = S[{\rho}_{\rm red}] - S[{\rho}] 
= -| \alpha |^2 \log| \alpha |^2 - | \beta |^2 \log| \beta |^2 > 0
\end{equation}
The above process of tracing out certain degrees of freedom may introduce complexities in some instances, \cite{Calzetta Hu}. After such tracing the unitary von Neumann equation undergoes a transformation into a non-unitary master equation for ${\rho}_{red}$, \cite{Hollowood:2017bil, Shaisultanov:1995cf, mastereq1, master3}. Due to the non-unitarity of the system, energy conservation is may not be preserved, which hinders the conventional method of verifying the numerical evolution of the reduced density matrix. So, the current model for decoherence may offer complications in some cases. For example:


\noindent 1. \textbf{Non-unitary evolution}: It is concerning that the reduced density matrix evolves non-unitarily, despite the underlying quantum theory being unitary. This raises the need for thorough examination to ensure the implications of this non-unitary evolution are properly understood and justified. \\

\noindent 2. \textbf{Complexity of perturbative master equation}: The perturbative master equation is highly intricate, making it difficult to address fundamental field theoretical questions. The existing treatments for incorporating perturbative corrections into the reduced density matrix lack a well-established framework. Additionally, the issue of renormalization for the reduced density matrix remains unresolved, further hindering progress in this area.

The decoherence phenomenon within the quantum field theory framework poses significant challenges, primarily because it involves intricate computations in out-of-equilibrium conditions, finite-temperature setting, interactions within the quantum field, renormalisation, and resummations etc. In order to address the issues mentioned above, we shall instead adopt a different perspective by employing the correlator approach introduced in  \cite{JFKTPMGS, koksma, kok} a few years ago. We note that the complete information about the system in principle we obtained by an idealized {\it perfect observer} resides in the $n$-point correlators which can be generated using the $n$-particle irreducible effective action, capturing the information of interaction between the system and its surrounding. However, the real world observers are constrained by the limitations of their measurement apparatus, preventing them from directly accessing correlation functions of an arbitrary order. Consequently, the negligence of the information stored in higher-order correlators for both system and surrounding which are inaccessible from observer's perspective contribute to an increase in the system's entropy. It is also essential to emphasize that this approach does not necessitate the use of a non-unitary process involving the tracing out of any environmental degrees of freedom.

We will now provide a concise overview of the essential components required for calculating the von Neumann entropy in both Minkowski and de Sitter spacetime via correlator approach, to be addressed in \ref{chapter3}. For example, in a simplest case we encounter the two point correlators, some of which for a scalar field can be written as \cite{JFKTPMGS}
\begin{eqnarray}
\langle\phi(x)\phi(x^\prime)\rangle &=&  \mathrm{Tr}\left[{\rho}(t)
\phi(x)\phi(x^\prime) \right]
\nonumber\\
\langle \pi(x)\pi(x^\prime)\rangle &=&  \mathrm{Tr}\left[\rho(t)
\pi(x)\pi(x^\prime) \right]
\nonumber\\
\frac{1}{2}\langle[\phi(x),\pi(x^\prime)]_+\rangle &=&  \frac{1}{2}\mathrm{Tr}[{\rho}(t)
[\phi(x),\pi(x^\prime)]_+]:=\partial_{t'}F_{\phi}(x,x')
\label{p}
\end{eqnarray}
where $\rho(t)$ is some density matrix. In this thesis, we shall take $\rho(t)$ to be pure and it correspond to some initial vacuum state. Also the first line appearing in the above equations represents the Wightman function, $\pi(x)=\dot{\phi}(x)$ is the canonically conjugate momentum of the scalar field and the quantity  $F_{\phi}(x,x')$ is called the statistical propagator defined as
\begin{equation}\label{statisticalpropagator}
F_{\phi}(x,x') := \frac{1}{2} \mathrm{Tr} \left(
\rho(t)[\phi(x), \phi(x')]_+ \right)=
\frac{1}{2} \mathrm{Tr} \left[ \rho(t) (
\phi(x)\phi(x') + \phi(x')\phi(x)) \right]
\end{equation}
Thus the statistical propagator is basically the Wightman functions symmetrised in $x$ and $x'$. For a given density operator $\rho(t)$ (pure or mixed), the statistical propagator tell us about how the states are occupied. One can also relate this to the average particle density. In other words, the statistical propagator is a mathematical representation of the probability distribution of states, and it describes how the probability density evolves with time. The connection between the statistical propagator and the phase space area arises from the conservation of the phase space volume (i.e., the  Liouville theorem) \cite{JFKTPMGS}. Liouville's theorem states that the phase space volume occupied by a system remains constant during its evolution if the system is isolated. As the system evolves, the statistical propagator governs how the probability density changes, effectively redistributing the same in the phase space. For an open quantum system however, the phase space volume does not remain constant due to the system-environment interactions. We shall focus only on the above two point correlators and their quantum corrections, as the simplest practical scenario. In the spatial momentum space, the statistical propagator reads,
\begin{equation}\label{statpropagatorFourier}
F(|\vec{k}|,t,t')=\int d^3\Delta\vec{x} \, F_{\phi}(t,\vec{x},t',\vec{x'})
e^{-\imath \vec{k}\cdot\Delta \vec{x}}
\end{equation}
One also defines the phase space area for each spatial momentum mode as the Fourier transform of the quantity 
$$4\Big[\langle\phi(x)\phi(x^\prime)\rangle \langle\pi(x)\pi(x^\prime)\rangle-\left(\frac{1}{2}\langle[\phi(x),\pi(x^\prime)]_+\rangle\right)^2 \Big]_{t=t'}$$
The phase space area given by the Gaussian invariant
\begin{equation}\label{deltaareainphasespace}
\Xi_{|\vec{k}|}^{2}(t)=4 \left[
F(|\vec{k}|,t,t')\partial_{t}\partial_{t'}F(|\vec{k}|,t,t') -
(\partial_{t}F(|\vec{k}|,t,t'))^{2} \right]_{t=t'}
\end{equation} 
where we have used the equal time limits of \ref{p} and \ref{statpropagatorFourier}. In order to see the analogy of this construction with that of the ordinary quantum mechanics, let us recall the generalised uncertainty relation (with $\hbar=1$),
\begin{eqnarray} \label{un}
\left\langle q^{2}\right\rangle\left\langle p^{2}\right\rangle-\left[\left\langle\frac{1}{2}[q, p]_+\right\rangle\right]^{2}=\frac{1}{4}\qquad  \text { (pure state) }  \nonumber\\
\left\langle{q}^{2}\right\rangle\left\langle p^{2}\right\rangle-\left[\left\langle\frac{1}{2}[q, p]_+\right\rangle\right]^{2}>\frac{1}{4} \qquad  \text { (mixed state)}
\end{eqnarray}
Combining the two above, we write
\begin{eqnarray} \label{unc}
\left\langle q^{2}\right\rangle\left\langle p^{2}\right\rangle-\left[\left\langle\frac{1}{2}[q,p]_+\right\rangle\right]^{2}=\frac{\Xi^2}{4}
\end{eqnarray}
Thus  the quantity $\Xi \geq 1$, interpreted as the phase space area, can be thought of as a measure of the impurity of the quantum state.  One plausible way to understand the increase in $\Xi$ is the transfer of momentum between the system and the environment, thereby increasing the momentum uncertainty.  Putting these all in together, the loss or ignorance of information due to the inaccessibility of all the correlations in the system, environment and between them, is characterised via the von Neumann entropy     for our field theoretic continuum system~\cite{Koksma:2010zi},
\begin{equation}\label{entropy}
S_{|\vec{k}|}(t) = \frac{ \Xi_{|\vec{k}|}(t)+1}{2}
\ln\left(\frac{\Xi_{|\vec{k}|}(t)+1}{2}\right) - \frac{
\Xi_{|\vec{k}|}(t)-1}{2} \ln\left(\frac{\Xi_{|\vec{k}|}(t)-1}{2}\right) 
\end{equation}
One can also relate the phase space area with the statistical particle number density per mode as
\begin{equation}\label{particlenumber}
n_{|\vec{k}|}(t) = \frac{ \Xi_{|\vec{k}|}(t)-1}{2}
\end{equation}
$\Xi_{|\vec k|}(t)$ becomes identity in the absence of the interaction. For a detailed derivation of \ref{entropy} based on \cite{Koksma:2010zi}, see \ref{Derivation of phase space area and entropy}. 

As our chief objective is to investigate the decoherence within the context of the inflationary de Sitter spacetime, it is pertinent to find the expressions for phase space area and entropy within this spacetime. This task involves an extension of the principles established above in the flat spacetime scenario, \cite{JFKTPMGS, koksma}, to the curved spacetimes such as \cite{ Friedrich:2019hev} for the interacting scalar field theory in the de Sitter spacetime. In this thesis, we shall be interested to explore the Yukawa interaction.

The Gaussian invariant in the 3-momentum space in the de Sitter spacetime corresponding to the scalar is given by \cite{Friedrich:2019hev, JFKTPMGS},
\begin{eqnarray}\label{gaussian invariant}
\frac{\Xi_{ \phi }^2(\eta, k)}{4a^4} =  \left[F_{\phi}(\eta, \eta^{\prime},k) \partial_{\eta}\partial_{\eta^{\prime}} F_{\phi}(\eta, \eta^{\prime},k) - \left( \partial_{\eta^{\prime}} F_{\phi}(\eta, \eta^{\prime},k) \right)^2 \right]_{\eta=\eta^{\prime} }   
\end{eqnarray}

where we have abbreviated $k = |\vec{k}|$, which we will retain for our calculations within the inflationary de Sitter spacetime. Note that $\Xi_{ \phi }^2(\eta, k)$  is dimensionless. The Gaussian part of the von Neumann entropy in the de Sitter spacetime looks formally the same as \ref{entropy}, \cite{ Friedrich:2019hev}. Taking the first order variation of  \ref{gaussian invariant} we get, 
\begin{multline}\label{phase}
\delta \left(\frac{ \Xi^2_{\phi}}{4 a^4} \right) = \delta \Big[ F_{\phi}(\eta, \eta) \partial_{\eta} \partial_{\eta^{\prime}} F_{\phi}(\eta ,\eta^{\prime}) -  \big[ \partial_{\eta^{\prime}}  F_{\phi}(\eta ,\eta^{\prime})\big]^2 \Big]_{\eta = \eta^{\prime}}
\\ = \Big[  F_{\phi}(\eta, \eta, k) \partial_{\eta} \partial_{\eta^{\prime}} \delta F_{\phi}(\eta ,\eta^{\prime}, k) + \delta F_{\phi}(\eta, \eta, k) \partial_{\eta} \partial_{\eta^{\prime}}  F_{\phi}(\eta ,\eta^{\prime}, k) \\ \qquad \qquad \qquad \qquad \qquad \qquad \qquad \qquad -2  \partial_{\eta^{\prime}}   F_{\phi}(\eta ,\eta^{\prime}, k)\partial_{\eta^{\prime}}  \delta F_{\phi}(\eta ,\eta^{\prime}, k)\Big]_{\eta=\eta^{\prime}} \\
=   \frac{H^2}{2 k} \Big[\frac{(1 + k^2 \eta^2)}{k^2} \partial_{\eta} \partial_{\eta^{\prime}} \delta F_{\phi}(\eta ,\eta^{\prime}, k) + k^2 \eta^2 \delta F_{\phi}(\eta, \eta, k) - 2   \eta\partial_{\eta^{\prime}}   \delta F_{\phi}(\eta ,\eta^{\prime}, k)\Big]_{ \eta=\eta^{\prime}}\,
\end{multline}

We shall be using the above equation to compute the perturbative correction to entropy generation due to the Yukawa interaction in the inflationary de Sitter spacetime in \ref{chapter3}. However, prior to that we also have computed the entropy generation in the flat spacetime non-perturbatively.


\section{Quantum field theory in curved spacetime}
\label{Quantum field theory in curved spacetime}
Quantum field theory in curved spacetime is a branch of theoretical physics that explores the behavior of quantum fields in the presence of gravitational fields and curved geometries. It combines the principles of quantum mechanics and general relativity to provide a framework for understanding the quantum nature of particles and their interactions in classical curved spacetime backgrounds. In this formalism, particles are represented by quantum fields, and their dynamics are governed by field equations that incorporate the effects of gravity. The concept of particle pair creation arises naturally in curved spacetime due to the presence of gravitational fields, leading to phenomena such as the Hawking radiation from black holes. Quantum field theory in curved spacetime has found applications in various areas, including cosmology, astrophysics, and black holes, where it offers valuable insights into the quantum aspects. It allows us to investigate the behavior of particles and their interactions in extreme gravitational environments, providing a deeper understanding of the fundamental nature of matter and spacetime. The relevant Einstein-Hilbert action reads \cite{Weinberg, Sean},

\begin{equation}
    \label{eq:EHA}
    S = \int d^4x \sqrt{-g} \Big[  \frac{1}{16\pi G}  (R - 2\Lambda) + {\cal{L}}_M \Big]
\end{equation}

where $\Lambda$ is the cosmological constant, $G$ is the Newton constant and $g$ is the determinant of spacetime metric $g_{\mu\nu}(x)$. The metric $g_{\mu\nu}$ describes the curved spacetime background. ${\cal{L}}_M$ is the Lagrangian density of any relevant matter field and $R$ is the Ricci curvature scalar given by $R = R_{\mu\nu}g^{\mu\nu}$ where Ricci tensor $R_{\mu\nu}$ is given as

$$
R_{\mu\nu} = \partial_{\lambda}\Gamma^{\lambda}_{\mu\nu} -\partial_{\nu}\Gamma^{\lambda}_{\mu\lambda} + \Gamma^{\lambda}_{\mu\nu}\Gamma^{\rho}_{\lambda \rho} - \Gamma^{\lambda}_{\mu\rho}\Gamma^{\rho}_{\nu\lambda}
$$
where $\Gamma^{\lambda}_{\mu\nu}$ are the Christoffel connections given by
$$
\Gamma^{\lambda}_{\mu\nu} = \frac{1}{2} g^{\lambda\sigma} \left( \partial_\mu g_{\sigma\nu} + \partial_\nu g_{\mu\sigma} - \partial_\sigma g_{\mu\nu} \right)
$$

The Einstein field equations, $ R_{\mu\nu}-\frac{1}{2}R_{\mu\nu}+\Lambda g_{\mu\nu}=8\pi G T_{\mu\nu}$, can be obtained by extremising \ref{eq:EHA} with respect to $g_{\mu\nu}$. Where $T_{\mu\nu}$ is the energy-momentum tensor or stress-energy tensor defined as $T_{\mu\nu}=-\frac{2}{\sqrt{-g}}\frac{\delta (\sqrt{-g}{\cal{L}}_M)}{\delta g^{\mu\nu}}$. In the case of quantum fields, $T_{\mu\nu}$ is replaced by its expectation value and computed with respect to a some suitable vacuum state. These modified field equations along with perhaps some suitable initial state provide us a description of how quantum fields propagate and evolve in a curved spacetime.

We recall that the simplest Poincare invariant action for a real scalar field $\phi(x)$ in the flat spacetime is given by,
\begin{equation}
    \label{actionM}
    S= - \frac{1}{2}\int d^4x \Big(\eta^{\mu\nu}\partial_\mu\phi\partial_\nu \phi+ m^2 \phi^2\Big)
\end{equation}
where $\eta^{\mu\nu}$ is the inverse Minkowski metric. To generalize this action to a curved spacetime, we replace $\eta_{\mu\nu}$ with $g_{\mu\nu}$ and the ordinary derivative with the covariant derivative. We also replace $d^4x$ with the invariant volume element $d^4x\sqrt{-g}$. This results in the following action for a scalar field coupled to gravity minimally
\begin{equation}
    \label{actionC}
    S= - \frac{1}{2}\int d^4x\sqrt{-g} \Big(g^{\mu\nu}\nabla_\mu\phi\nabla_\nu \phi+ m^2 \phi^2\Big)
\end{equation}
In a more general case, a scalar curvature term is also added of the form $\sqrt{-g}\xi R \phi^2$, where $\xi$ is a coupling constant. By varying \ref{actionC} with respect to $\phi$, we obtain the Klein-Gordon equation in curved spacetime
\begin{equation}
    \label{KGC}
    (\nabla_\mu \nabla^\mu -m^2)\phi=\Big[\frac{1}{\sqrt{-g}}\partial_\mu(\sqrt{-g}g^{\mu\nu}\partial_\nu)-m^2\Big]\phi=0
\end{equation}

Similar analysis can be done for fermionic field. The Dirac field action in the Minkowski spacetime is given by
\begin{equation}
    \label{DactionM}
    S=\int d^4x \;\Bar{\psi}(x)(\imath \gamma^{\mu} \partial_\mu-m)\psi(x)
\end{equation}
here $\gamma^{\mu}$'s are the flat space gamma matrices, $\psi(x)$ is the fermionic field and $\Bar{\psi}(x)=\psi^{\dagger}(x)\gamma^0$ is the adjoint. Since we are working with mostly positive signature, these flat space gamma matrices satisfy the anti-commutation relation
\begin{eqnarray}\label{gamma}
\left[\gamma^\mu, \gamma^\nu\right]_+ = - 2\eta^{\mu \nu}\, {\bf I_{4\times 4}}     
\end{eqnarray}
In curved spacetime the above action modifies as
\begin{equation}
    \label{DactionC}
     S=\int d^4x \;\sqrt{-g}\Bar{\psi}(x)(\imath \gamma^{\mu} D_\mu-m)\psi(x)
\end{equation}
where $D_\mu$ is the spin covariant derivative and the curved space gamma matrices satisfy
\begin{eqnarray}\label{gamma1}
\left[\gamma^\mu, \gamma^\nu\right]_+ = - 2g^{\mu\nu}\, {\bf I_{4\times 4}}     
\end{eqnarray}

One can relate flat and curved spacetime gamma metrics using the tetrads $e_{a}^{\mu}$, where the Greek and Latin indices respectively stand for the general frame and the local Lorentz frame \cite{Parker:2009uva, Peskin:2018}, satisfying the relations $ g^{\mu\nu}(x)=e^{\mu}_a(x) e^{\nu}_b(x) \eta^{ab}$ and $\gamma^\mu=e^{\mu}_a \gamma^{a}$. The spin covariant derivative $D_{\mu}$ is given as ~\cite{Parker:2009uva},
	\begin{eqnarray}
	D_\mu:=\partial_\mu+\frac{1}{2}\omega_{\mu a b}\Sigma^{ab},
	\end{eqnarray}
where $\Sigma^{ab}=\left[\gamma^a,~\gamma^b\right]/4$ and the Ricci rotation coefficients $\omega$'s are given by 
 $$\omega_\mu{}^a{}_b = - e_b{}^\nu \left(\partial_\mu e^a{}_\nu - \Gamma^\lambda_{\mu \nu} e^a{}_\lambda \right)$$

From \ref{DactionC}, we obtain the Dirac equation as
\begin{equation}
    \label{DQ}
    (\imath \gamma^{\mu} D_\mu -m)\psi(x)=0
\end{equation}

In this thesis, we are interested in studying quantum field theory in cosmological background. We shall solve the Dirac equation in the inflationary de Sitter spacetime in order to look into some aspects of entanglement.

\section{A brief review of the de Sitter spacetime}
\label{The de Sitter spacetime}

The de Sitter spacetime is a solution to Einstein's field equations with a positive cosmological constant and without any other backreacting matter field. It provides a model for an expanding universe that is dominated by vacuum energy. In this spacetime, the geometry is characterized by a constant positive curvature, similar to that of the sphere. This de Sitter is the simplest example where one can realise the accelerated expansion of the spacetime.

The de Sitter spacetime has several notable properties. First, it is maximally symmetric, that is, it has the maximum number of Killing vectors possible for any given spacetime dimensions. For example in four spacetime dimensions the number is 10 analogous to that of the Minkowski spacetime. Second, it has a cosmological event horizon beyond which events are causally disconnected from an observer located within. The de Sitter spacetime serves as an important background in various areas of modern gravitation, cosmology and quantum field theory. This spacetime is physically very well motivated and is very popular among the researchers, including early inflationary cosmology, the current cosmological epoch as well as quantum gravity. It provides insights into the expansion history of our universe, the generation of primordial density fluctuations, and the origin of the large scale structures.

The cosmological spacetimes are modeled by the Friedmann-Lemaitre-Robertson-Walker (FLRW) universe \cite{Weinberg, Sean}, generically written as 
\begin{equation}
    \label{FRW}
    ds^2=-dt^2+a^2(t)\Big[\frac{dr^2}{1-Kr^2}+r^2(d\theta^2+\sin{\theta}^2d\phi^2)\Big]
\end{equation}
The function $a(t)$ is known as the scale factor and $K$ is a constant. The above ansatz for the metric is based upon the assumption of spatial homogeneity and isotropy of the universe at very large scales, confirmed observationally with excellent accuracy for length scale larger than 300 million lightyears.
The parameter $K$ corresponds to the curvature of the three space, and it can take three possible values. $K=0$ for a spatially flat universe, $K=1$, for a spatially closed universe, $K=-1$, for a spatially open universe. Each value of $K$ corresponds to a distinct spatial geometry, respectively, the flat, three sphere and three hyperboloid, 
\begin{equation}
\label{RW}
    ds^2=-dt^2+a^2(t)\begin{cases}
    d\psi^2+\sin^2\psi (d\theta^2+\sin^2\theta d\phi^2)\\
    dx^2+dy^2+dz^2\\
    d\psi^2+\sinh^2\psi (d\theta^2+\sin^2\theta d\phi^2)
    \end{cases}
\end{equation}
 
The physically most well accepted geometry is the spatially flat one \cite{Weinberg:2008zzc, Mukhanov:2005sc}, in which the de Sitter spacetime in particular reads
\begin{equation}
    \label{dS}
    ds^2=-dt^2+a^2(t) \Big[dx^2+dy^2+dz^2\Big]
\end{equation}
where $a(t) = e^{Ht}$, with $H=\sqrt{\Lambda/3}$, where $\Lambda$ is the positive cosmological constant. Also, we have the temporal range  $-\infty < t < \infty$. We note that the above metric can also be written in the conformally flat form as

\begin{equation}\label{desitterspacetime}
ds^2=a^2(\eta)(-d\eta^2+dx^2+dy^2+dz^2)
\end{equation}

where $a(\eta)=-1/H\eta$ and $\eta = e^{-Ht}/H$. Thus we have the temporal range $ -\infty < \eta <0$. The timelike coordinate $t$ in \ref{dS} is the cosmological time whereas $\eta$ in \ref{desitterspacetime} is known as the conformal time.
\begin{figure}
    \centering
\includegraphics[scale=.32]{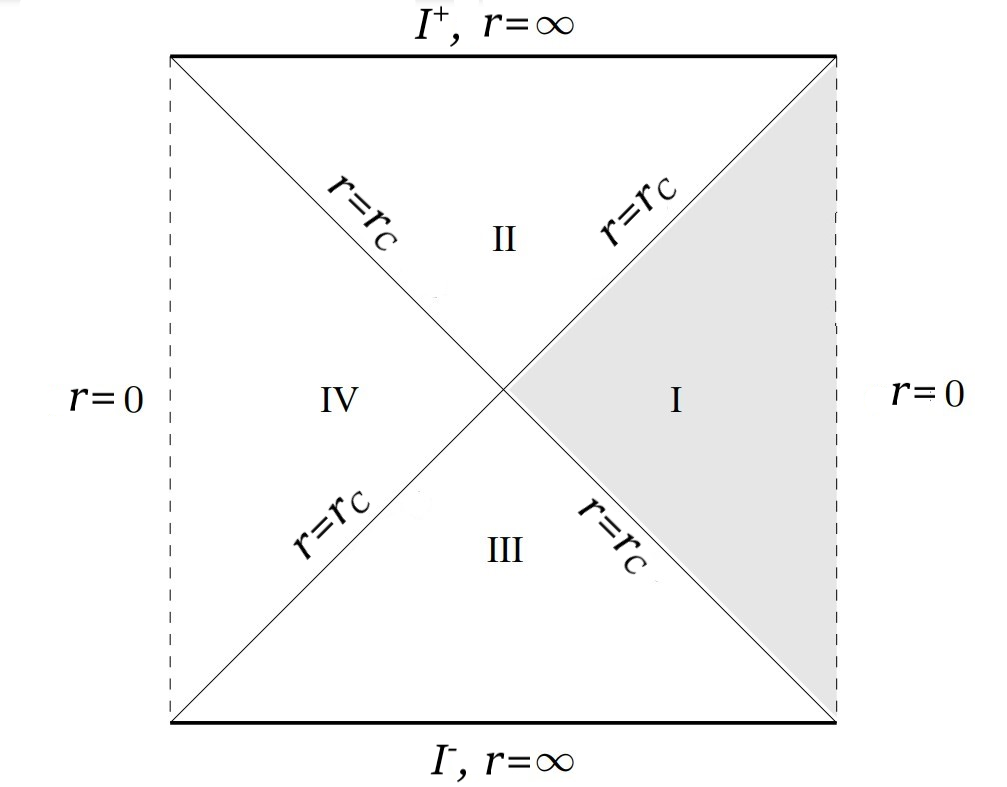}
    \caption{\small \it The Penrose-Carter diagram of the de Sitter spacetime. The shaded region is the static patch covered by the coordinates $(t_s, r, \theta, \phi)$ of \ref{1.48}. Region IV is also covered by the same with the direction of time reversed. $r_C = H^{-1}$ denotes the cosmological event horizon (CEH). Due to the time reversal symmetry of \ref{1.48} we have both the future and past segments of CEH and hence four causally disconnected spacetime wadges. Owing to the rapid accelerated expansion of the spacetime, such horizon is obtained and accordingly the infinities are spacelike here.}
    \label{deSitter-penrose}
\end{figure}

The de Sitter spacetime can also be written in a manifestly static form as
\begin{equation}\label{1.48}
ds^2 = -\Big(1 - H^2 r^2 \Big) dt_s^2 + \Big(1 - H^2 r^2 \Big)^{-1} dr^2 + r^2 (d\theta^2 + \sin^2\theta d\phi^2
)    
\end{equation}
We note that the static form holds in the region $0\le r<H^{-1}$, beyond which the timelike and the radial coordinates are not well defined.  $r=H^{-1}$ is a null hypersurface, known as the cosmological event horizon. \ref{deSitter-penrose} shows the maximally extended Penrose-Carter diagram for \ref{1.48}. For further detail on the static patch of the de Sitter spacetime we refer our reader to \cite{Parker:2009uva, Gibbons, Birrell:1982ix, Castineiras:2003hc} also references therein. In \ref{chapter4} of this thesis we shall consider a static and spherically symmetric black hole sitting in the de Sitter universe for the sake of studying entanglement, mainly Schwarzschild-de Sitter spacetime. This spacetime, natually has a black hole event horizon surrounded by the cosmological horizon. For a comprehensive analysis and original works on de Sitter spacetime pertaining to bosonic and fermionic fields in different coordinatisations, we refer our reader to \cite{Bhattacharya:2018yhm, EE, bell:2017, vaccum EE for fermions, QC in deSitter}, and the references therein. 


\section{Highlights of the thesis}
\label{A synopsis of the thesis}

The chief objective of this thesis is to comprehensively explore the analytical and numerical aspects of quantum correlations and decoherence in the cosmological spacetimes. We focus on exploration of the physics of the early inflationary de Sitter universe with the motivation to gain some insights into the interesting interplay among quantum correlations and decoherence during the evolution of the universe.\\ 

In \ref{chapter2}, we explore two measures of quantum correlations and entanglement, specifically the violation of Bell-Mermin-Klyshko (BMK) inequalities and the quantum discord, in the context of Dirac fermions in the cosmological de Sitter background. We focus on the two and four mode squeezed states to examine the extent of BMK violation, demonstrating the maximum violation is achievable. Additionally, we investigate the quantum discord for a maximally entangled initial states. We discuss both the qualitative similarities and differences between our findings to those obtained from different coordinatisations of de Sitter. \\

In \ref{chapter3}, we wish to compute the decoherence and entropy generation in the Minkowski and inflationary de Sitter spacetime for the Yukawa interaction, using the correlator approach reviewed earlier. The scalar field is treated as the system, and fermions are considered as the environment in both spacetimes. The scalar field is assumed to be massive in Minkowski spacetime and massless in the de Sitter spacetime, while the fermions are assumed to be massless in both cases. The Minkowski spacetime is used as a preliminary study to understand decoherence before delving into the more complex de Sitter spacetime, where perturbative results are obtained at one loop level. We wish to construct the renormalized Kadanoff-Baym equation accounting for self energy corrections. Using these equations we aim to compute the statistical propagator, phase space area and the von Neumann entropy. We analyze how the entropy varies with relevant parameters and compare the results with scenarios involving scalar fields serving as both the system and the environment. Additionally, this study seeks to identify qualitative similarities with the Yukawa theory results obtained using the Feynman-Vernon influence functional technique, \cite{Boyanovsky:2018soy}, in the de Sitter spacetime. \\

In \ref{chapter4}, we examine the degradation or survival of quantum entanglement in the Schwarzschild-de Sitter black hole spacetime. Specifically, we investigate the mutual information and logarithmic negativity for maximally entangled bipartite states associated with massless minimal scalar fields. 
For simplicity, we restrict ourselves to (1+1)-dimensions. As we have mentioned earlier, this spacetime consists of both a black hole and a cosmological event horizon, resulting in particle creation at distinct temperatures. We analyze two different perspectives regarding thermodynamics and particle creation within this background. The first approach considers thermal equilibrium for an observer associated with either of the horizons. Our findings reveal that akin to asymptotically flat/anti-de Sitter black holes, the entanglement or correlation degrades as the Hawking temperature(s) increase. The second approach combines both horizons to define a total Bekenstein-Hawking entropy and an effective equilibrium temperature. We provide a field theoretic derivation of this effective temperature and demonstrate that unlike conventional cases, particle creation does not occur in causally disconnected spacetime wedges but rather in a single region, in this case. By employing the corresponding the associated quantum states, we establish that in this scenario, entanglement never degrades but instead increases with rising black hole temperature. Remarkably, this holds true regardless of the black hole's temperature or the magnitude of the cosmological constant. We have argued that this phenomenon cannot happen in asymptotically flat/anti-de Sitter black hole spacetimes.

In the concluding \ref{Summary&discussion}, a brief summary of this thesis is provided, encompassing the key results and insights discussed in the preceding Chapters of this thesis. Furthermore, potential avenues for future research are highlighted, aiming to further enhance our understanding of quantum correlations and decoherence in the cosmological backgrounds.

\chapter{Some measures for fermionic entanglement in the cosmological de Sitter spacetime}
\label{chapter2}

In this Chapter, we wish to investigate two measures for quantum correlations namely the violation of the BMK ~\cite{bell:1964, clauser:1969,  mermin:1990, belinski, Gisin:1998} and the quantum discord~\cite{Zurek, Dakic} for Dirac fermions in the (1+3)-dimensional cosmological de Sitter background. Motivation for this study and these measures have already been reviewed in \ref{Motivation and Overview}, \ref{Quantum correlations}, \ref{discord} and \ref{Bell1}.


The Bell inequality~\cite{bell:1964} (see also~\cite{Nielsen:2000} and references therein) is a measure of non-locality for a two-partite quantum system. Later such inequality was extended to multipartite systems~\cite{clauser:1969,  mermin:1990, belinski, Gisin:1998}, altogether  regarded as  the Bell-Mermin-Klyshko (BMK) inequalities. In the nonlocal regime of quantum mechanics BMK inequalities may be violated, thereby clearly distinguishing quantum effects from that of any local classical hidden variables. As the number of partite is increased in a system, the upper bound of the BMK violation also increases, e.g.~\cite{nagata:2002, yu:2003}. Given two subsystems, on the other hand, quantum discord is a suitable measure of all correlations including entanglement between them~\cite{Zurek, Dakic}. Accordingly, even if there is no entanglement for a mixed state, the  quantum discord can be non-vanishing. The key ingredient of the computation of discord is the quantum mutual information between the subsystems. One also needs to optimise over all possible measurements performed on one of the subsystems. We refer our reader to~\cite{Qiang, Wang, Yao, Xiao:2018cxg, Choudhury:2016cso, Choudhury:2017bou, Maldacena:2015bha, Lim, Vennin, Kanno:2016gas,  Chen:2017cgw,  Feng:2018ebt, Matsumura:2020uyg} and references therein for discussions on the BMK violation and quantum discord in both non-inertial and inflationary scenarios.   

The basic computational tools we shall use in this paper can be seen in~\cite{Kanno:2016gas, bell:2017} and references therein. In~\cite{Kanno:2016gas}, the quantum discord corresponding to a maximally entangled state for two scalar fields was investigated in the hyperbolic de Sitter background. In~\cite{bell:2017}, the infinite BMK violation was demonstrated for a massless scalar field in a cosmological background which is de Sitter and radiation dominated respectively in the past and future. We shall compute these two measures for massive Dirac fermions in the cosmological de Sitter background in order to see how much similar or dissimilar the result is, with the already existing ones.
 
The rest of the Chapter is organised as follows. In the next section, we construct the relevant two and four mode squeezed states. We compute the BMK violation for the two and four mode squeezed states in \ref{Bell}. Computation of the discord can be seen in \ref{discord1}. Finally we discuss the results and conclude in \ref{con}.

\section{Fermionic squeezed states} \label{sqz}
Based upon the discussion of \ref{The Dirac mode functions and Bogoliubov coefficients}, we shall construct below the two- and four-mode fermionic squeezed states, to be useful for our purpose. Corresponding to the field quantisations \ref{f1}, \ref{f2}, we define the `in' and `out' vacua as,
$$
a_{\rm in}(\vec{k},s) |0_{\rm in}\rangle=0= b_{\rm in}(\vec{k},s) |0_{\rm in}\rangle,            \qquad {\rm and } \qquad a_{\rm out}(\vec{k},s) |0_{\rm out}\rangle=0= b_{\rm out}(\vec{k},s) |0_{\rm out}\rangle
$$
The Bogoliubov relations of  \ref{bglv3} show that the `in' vacuum can be expressed as a squeezed state over all the `out' states,
\begin{eqnarray}
|0_{\rm in}\rangle \sim \exp\left[-\sum_{\vec{k}, s} \frac{\beta_{k}} {\alpha_{k}^\ast} a^{\dagger}_{\rm out}(\vec{k},s) b^{\dagger}_{\rm out}(-\vec{k},s)     \right] |0_{\vec{k},s,\,{\rm out}}\rangle \otimes|0_{-\vec{k},s,\,{\rm out}}\rangle,
\label{sq1}
\end{eqnarray}
where $|0_{\pm \vec{k},s,\,{\rm out}}\rangle$ respectively represent a particle and an antiparticle vacuum. 

We shall work here with a specific value of the spatial momentum, $\vec{k}$, e.g.~\cite{Kanno:2016gas}.
We also note from \ref{bglv3} that the helicities do not mix in the Bogoliubov transformations. Thus due to the various anti-commutation relations, the squeezed state expansion corresponding to different $s$ values in \ref{sq1} will just factor out.  This permits us to go for another simplification -- to restrict ourselves to a specific $s$ value as well. In other words, we shall work with a subspace of $|0_{\rm in}\rangle$ corresponding to specific $\vec{k}$ and $s$. Thus instead of \ref{sq1}, we work with (after normalisation)
\begin{eqnarray}
|0_{{\vec{k}}, \rm in}\rangle= |\alpha_k| \left[ |0_{\vec{k},\,{\rm out}}\rangle \otimes|0_{-\vec{k},\,{\rm out}}\rangle - \frac{\beta_{k}} {\alpha_{k}^\ast} |1_{\vec{k},\,{\rm out}}\rangle \otimes|1_{-\vec{k},\,{\rm out}}\rangle\right],
\label{sq2}
\end{eqnarray}
where we have suppressed the index $s$, since we are restricting ourselves to   any single value of it. We shall further comment on the more general helicity summed state at the end of \ref{Bell2}.  The above is called a two-mode squeezed state.\\

\noindent
The notion of the two-mode squeezed state can easily be extended if we include more than one fermionic fields, say $\psi_1(x), \, \psi_2(x), \dots$, each quantised in a way described in \ref{The Dirac mode functions and Bogoliubov coefficients} and further mix these particle species via some Bogoliubov transformations (see \cite{bell:2017} for discussions on scalar field theory, also~\cite{Blasone, Blasone2}).

Let us consider two fermionic fields, $\psi_1$ and $\psi_2$ with their `in' vacuum $|0_{{\rm in}}\rangle_1$ and $|0_{{\rm in}}\rangle_2$ respectively,
\begin{eqnarray}
a_{{\rm in},i}(\vec{k},s) |0_{{\rm in}}\rangle_i=0= b_{{\rm in}, i}(\vec{k},s) |0_{{\rm in}}\rangle_i \qquad (i=1,2,\,\,\,{\rm no~sum~on}~i),           
\label{sq3}
\end{eqnarray}
where $a_{{\rm in},1},\,b_{{\rm in},1}$ and $a_{{\rm in},2},\,b_{{\rm in},2}$ are the annihilation operators corresponding to $\psi_1$ and $\psi_2$ respectively.
The combined  `in' vacuum for these two field system is then $|0_{{\rm in}}\rangle_1 \otimes |0_{{\rm in}}\rangle_2  $. 
Let us suppose that these two fields are correlated via a simple mixing transformation as,
\begin{multline}
\overline{a}^{(1)}_{{\rm in}}(\vec{k},s)=\Gamma_{k} \,a_{{\rm in},1}(\vec{k},s)+\Delta_{k}\,b^{\dagger}_{{\rm in}, 2}(-\vec{k},s), \qquad  \overline{a}^{(2)}_{{\rm in}}(\vec{k},s)=\Gamma_{k}\, a_{{\rm in},2}(\vec{k},s)+\Delta_{k}\,b^{\dagger}_{{\rm in}, 1}(-\vec{k},s)  \\
\overline{b}^{(1)}_{{\rm in}}(\vec{k},s)=\Gamma_{k} \,b_{{\rm in},1}(\vec{k},s)-\Delta_{k}\,a^{\dagger}_{{\rm in}, 2}(-\vec{k},s), \qquad  \overline{b}^{(2)}_{{\rm in}}(\vec{k},s)=\Gamma_{k}\, b_{{\rm in},2}(\vec{k},s)-\Delta_{k}\,a^{\dagger}_{{\rm in}, 1}(-\vec{k},s)
\label{sq4}
\end{multline}
where $\Gamma_{k}$ and $\Delta_{k}$ are Bogoliubov coefficients satisfying,  $|\Gamma_{k}|^2+|\Delta_{k}|^2=1$. It is easy to check that the operators defined above  satisfy the canonical anti-commutation relations. We assume that such squeezing  between different field species is weak, i.e.,
$$\left\vert\frac{\Delta_{k}} {\Gamma_{k}}\right \vert \ll 1$$
Let us denote the vacuum state corresponding to the new operators in \ref{sq4} by $|\overline{0}\rangle$, 
\begin{eqnarray}
\overline{a}^{(i)}_{{\rm in}}(\vec{k},s)|\overline{0}\rangle=0=\overline{b}^{(i)}_{{\rm in}}(\vec{k},s)|\overline{0}\rangle \qquad (i=1,2)
\label{sq5}
\end{eqnarray}
From \ref{sq4}, $|\overline{0}\rangle $ can then be expanded as,
\begin {eqnarray}
|\overline{0}\rangle \sim e^{-\sum_{\vec{k}, s} \frac{\Delta_{k}} {\Gamma_{k}^\ast} \left[a^{\dagger}_{{\rm in},1}(\vec{k},s)b^{\dagger}_{{\rm in},2}(-\vec{k},s)+a^{\dagger}_{{\rm in},2}(\vec{k},s)b^{\dagger}_{{\rm in},1}(-\vec{k},s)\right]} |0_{{\rm in}}\rangle_1 \otimes |0_{{\rm in}}\rangle_2 
\label{sq6}
\end{eqnarray}
We focus as earlier on a specific $\vec{k}$ and $s$ value. Suppressing the index $s$, making the expansion of the exponential   only up to the second order owing to the weak squuezing, and after normalising, \ref{sq6} takes the form
\begin{eqnarray}
|\overline{0}_{\vec k}\rangle=A_{k}|0_{{\rm in},\,\vec{k}}\rangle_1 \otimes |0_{{\rm in},\, \vec{k}}\rangle_2  +\frac{B_{k}}{\sqrt2}\left(|1_{{\rm in},\,\vec{k}}\rangle_1  \otimes |1_{{\rm in},\,-\vec{k}}\rangle_2+|1_{{\rm in},\,-\vec{k}}\rangle_1  \otimes |1_{{\rm in},\,\vec{k}}\rangle_2\right),
\label{sq7}
\end{eqnarray}
where $A_k$ and $B_k$  depend upon $\Gamma_k$ and $\Delta_k$ and   $|A_{k}|^2+|B_{k}|^2=1$.

The `in' states appearing on the right hand side of \ref{sq7} can further be expanded according to \ref{sq2}. If we take the rest mass of both the fields to be the same, the Bogoliubov coefficients ($\alpha_k,\,\beta_k$) corresponding to these two fields are then same as well, cf. \ref{The Dirac mode functions and Bogoliubov coefficients}. This yields,
\begin{equation}
\begin{split}    
|\overline{0}_{\vec k}\rangle=A_{k}|\alpha_k|^2 & \Bigg[ |0_{\vec{k},\,{\rm out}}\rangle_1 \otimes|0_{-\vec{k},\,{\rm out}}\rangle_1 - \frac{\beta_{k}} {\alpha_{k}^\ast} |1_{\vec{k},\,{\rm out}}\rangle_1 \otimes|1_{-\vec{k},\,{\rm out}}\rangle_1\Bigg] \\ & \otimes  \Bigg[ |0_{\vec{k},\,{\rm out}}\rangle_2 \otimes|0_{-\vec{k},\,{\rm out}}\rangle_{2} - \frac{\beta_{k}} {\alpha_{k}^\ast} |1_{\vec{k},\,{\rm out}}\rangle_2 \otimes|1_{-\vec{k},\,{\rm out}}\rangle_2\Bigg] \\ &+\frac{B_{k} |\alpha_k|^2}{\sqrt{2} (\alpha_k^{\ast})^2 }\Bigg[{|1_{{\vec{k}},\,{\rm out}}\rangle_1\otimes|0_{{-\vec{k}},\,{\rm out}}\rangle_1}\otimes{|0_{{\vec{k}},\,{\rm out}}\rangle_2\otimes|1_{-{\vec{k}},\,{\rm out}}\rangle_2} \\& +{|0_{{\vec{k}},\,{\rm out}}\rangle_1\otimes|1_{-{\vec{k}},\,{\rm out}}\rangle_1}\otimes{|1_{{\vec{k}},\,{\rm out}}\rangle_2\otimes|0_{-{\vec{k}},\,{\rm out}}\rangle_2}\Bigg],
\label{sq8}
\end{split}
\end{equation}
known as the four mode squeezed state.
Note that the above construction is similar but not exactly the same as the de Sitter $\alpha$-vacua (e.g.~\cite{Collins:2004wj}). This is because the latter mixes the modes of a single quantum field whereas here we have mixed two different fields themselves. It is also clear that the construction of such squeezed states goes beyond two fields. For example, with three fermionic fields one can construct a six mode squeezed state. 

Having constructed the necessary states, we are now ready to go into computing the BMK violation and the quantum discord.

\section{The violation of the  BMK inequalities}\label{Bell}
The construction of the Bell-Mermin-Klyshko (BMK) operators is reviewed in \ref{Bell1}. Let us now compute the violation.
\subsection{Computing the  violation } \label{Bell2}

\begin{figure}[h!]
  \includegraphics[width=8.4cm]{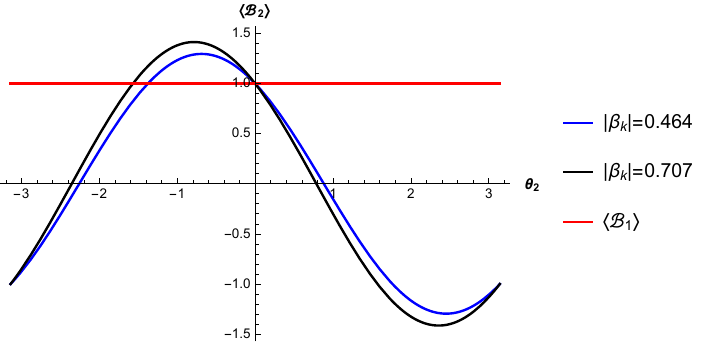}
  \includegraphics[width=8.4cm]{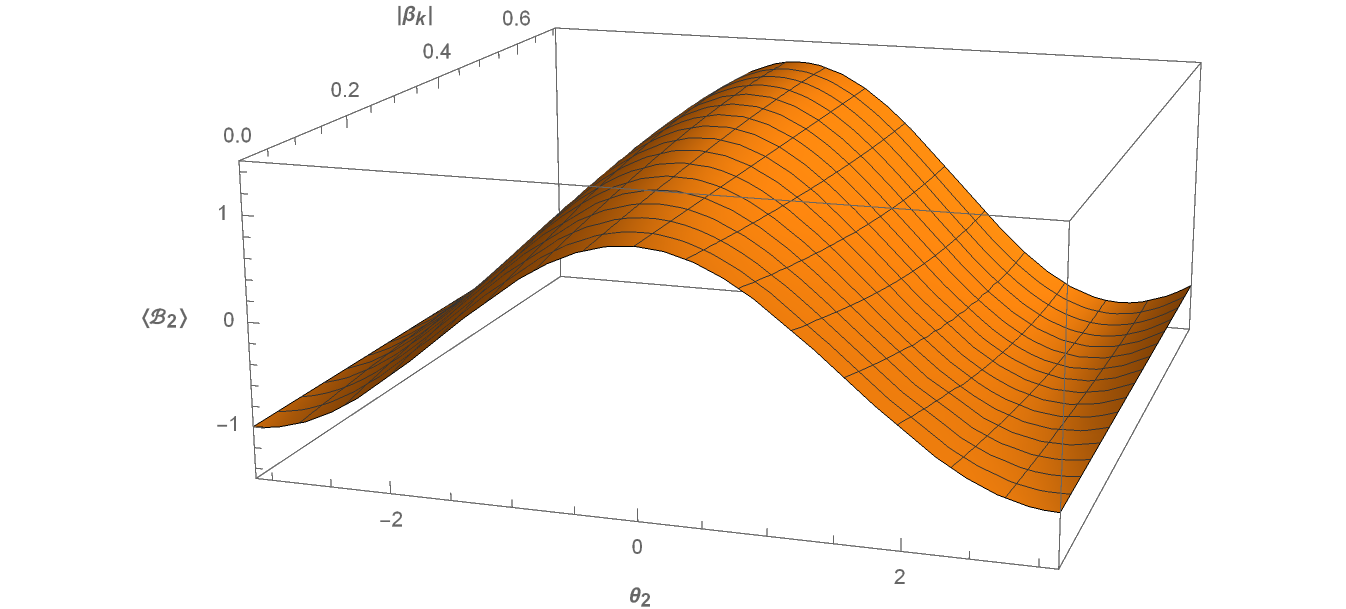}
  \caption{\small \it 2-d and 3-d plots for the violation of BMK inequalities, \ref{bv3}, for two-mode squeezed state with different values of the Bogoliubov coefficients.   In the 2-d plot, the red line, $\langle{\cal B}_{1}\rangle$, stands for the classical upper bound. $\langle{\cal B}_{2}\rangle \sim \sqrt{2}$ indicates the maximum violation. Note also from \ref{bglv3} that the variation of the Bogoliubov coefficient correspond to the variation of the rest mass  only. Thus for a given particle species, the value of the Bell violation does not change with respect to the momentum $\vec{k}$.   }
  \label{fig:bell1}
\end{figure}

Let us first  compute  the BMK violation for the two-mode squeezed state defined in~\ref{sqz}. Following \cite{chen:2, bell:2017}, we introduce a pseudospin operator $S$, 
\begin{eqnarray}
\textbf{n}.\textbf{S}:=S_{z}\cos{\theta}+\sin{\theta}(e^{i\phi}S_{-}+e^{-i\phi}S_{+})
\label{bv0}
\end{eqnarray} 
where $\textbf{n}\equiv (\sin{\theta} \cos{\phi},\sin{\theta}\sin{\phi},\cos{\theta})$ is a  spatial unit vector, $S_{\pm}$ are the ladder operators and $\textbf{n}.\textbf{S}$ has eigenvalues $\pm1$. We have the operations over the orthonormal states $|0\rangle$, $|1\rangle$ (corresponding to a fix value of the helicity $s$), 
\begin{eqnarray}
\begin{split}
&S_{z}|0\rangle=-|0\rangle,\,\, S_{z}|1\rangle=|1\rangle; \qquad 
S_{+}|0\rangle=|1\rangle,\,\, S_{+}|1\rangle=0; \qquad
S_{-}|0\rangle=0,\,\, S_{-}|1\rangle=|0\rangle
\label{bv1}
\end{split}
\end{eqnarray}
Since \ref{bv0} is defined on an Euclidean plane, we can use its rotational invariance to set  $\phi=0$ in \ref{bv0}, so that the expectation value of the Bell operator, \ref{b3}, in the two-mode squeezed state 
%
%
(\ref{sq2}) becomes
\begin{eqnarray}
\langle0_{\vec{k},{\rm in}}|{\cal B}_{2}|0_{\vec{k},{\rm in}}\rangle=\frac{1}{2}[F(\theta_{1},\theta_{2})+F(\theta_{1},\theta_{2}^{\prime})+F(\theta_{1}^{\prime},\theta_{2})-F(\theta_{1}^{\prime},\theta_{2}^{\prime})]
\label{exp}
\end{eqnarray}
where ${\cal O}_{1}=\textbf{n}_{1}.\textbf{S}$,  ${\cal O}_{2}=\textbf{n}_{2}.\textbf{S}$, ${\cal O}_{1}^\prime=\textbf{n}_{1}^\prime.\textbf{S}$, ${\cal O}_{2}^\prime=\textbf{n}_{2}^\prime.\textbf{S}$, in \ref{b3} and $F(\theta_{1},\theta_{2})$ is given by
\begin{eqnarray}
\begin{split} 
F(\theta_{1},\theta_{2})=&\langle0_{\vec{k},{\rm in}}| (S_{z}\cos{\theta_{1}}+S_{x}\sin{\theta_{1}})\otimes(S_{z}\cos{\theta_{2}}+S_{x}\sin{\theta_{2}})|0_{\vec{k},{\rm in}}\rangle ,\\&
=(\cos{\theta_{1}}\cos{\theta_{2}}- 2|\beta_{{k}} \alpha_{{k}}| \sin{\theta_{1}\sin{\theta_{2}}})
\end{split}
\end{eqnarray}
The other $F$'s can be found by replacing  $\theta_1$, $\theta_2$ suitably above. The violation of the BMK inequality can be examined for various values of the angles as well as the Bogoliubov coefficients. 
For example for $\theta_{1}=0$, $\theta_{1}^{\prime}=\pi/2$ and $\theta_{2}=-\theta_{2}^{\prime}$, we obtain
\begin{eqnarray}
\langle0_{\vec{k},{\rm in}}|{\cal B}_{2}|0_{\vec{k},{\rm in}}\rangle=\cos{\theta_{2}}-2|\beta_{{k}} \alpha_{{k}}|\sin{\theta_{2}},
\label{bv3}
\end{eqnarray}
which is plotted in \ref{fig:bell1}. In particular, the maximum violation  $ \langle{\cal B}_{2}\rangle \sim \sqrt{2} $, is achieved for the maximum value of the Bogoliubov coefficient, $|\beta_k| \sim 0.707 $ (cf., \ref{The Dirac mode functions and Bogoliubov coefficients}). 

\noindent
As \ref{b4} indicates,  the upper bound of the Bell violation can be increased  by going to four or higher mode squeezed states. We now wish to demonstrate such BMK violation for a four-mode squeezed state discussed in \ref{sqz}. Using \ref{b3}, the relevant BMK operator can be written as
\begin{eqnarray}\label{fm1}
\begin{split}
{\cal B}_{4}=\frac{1}{4}[&-{\cal O}_{1}\otimes {\cal O}_{2}\otimes {\cal O}_{3}\otimes {\cal O}_{4}-{\cal O}_{1}^\prime \otimes {\cal O}_{2}^\prime \otimes {\cal O}_{3}^\prime \otimes {\cal O}_{4}^\prime +{\cal O}_{1} \otimes {\cal O}_{2} \otimes {\cal O}_{3} \otimes {\cal O}_{4}^\prime
\\&+{\cal O}_{1} \otimes {\cal O}_{2} \otimes {\cal O}_{3}^\prime \otimes {\cal O}_{4}
+{\cal O}_{1} \otimes {\cal O}_{2}^\prime \otimes {\cal O}_{3} \otimes {\cal O}_{4}+{\cal O}_{1}^\prime \otimes {\cal O}_{2} \otimes {\cal O}_{3} \otimes {\cal O}_{4} \\& +{\cal O}_{1} \otimes {\cal O}_{2} \otimes {\cal O}_{3}^\prime \otimes {\cal O}_{4}^\prime+{\cal O}_{1} \otimes {\cal O}_{2}^\prime \otimes {\cal O}_{3} \otimes {\cal O}_{4}^\prime
+{\cal O}_{1}^\prime \otimes {\cal O}_{2} \otimes {\cal O}_{3} \otimes {\cal O}_{4}^\prime
\\&+{\cal O}_{1} \otimes {\cal O}_{2}^\prime \otimes {\cal O}_{3}^\prime \otimes {\cal O}_{4}+{\cal O}_{1}^\prime \otimes {\cal O}_{2} \otimes {\cal O}_{3}^\prime \otimes {\cal O}_{4}+{\cal O}_{1}^\prime \otimes {\cal O}_{2}^\prime \otimes {\cal O}_{3} \otimes {\cal O}_{4}\\&-{\cal O}_{1} \otimes {\cal O}_{2}^\prime \otimes {\cal O}_{3}^\prime \otimes {\cal O}_{4}^\prime
-{\cal O}_{1}^\prime \otimes {\cal O}_{2} \otimes {\cal O}_{3}^\prime \otimes {\cal O}_{4}^\prime -{\cal O}_{1}^\prime \otimes {\cal O}_{2}^\prime \otimes {\cal O}_{3} \otimes {\cal O}_{4}^\prime \\& -{\cal O}_{1}^\prime \otimes {\cal O}_{2}^\prime \otimes {\cal O}_{3}^\prime \otimes {\cal O}_{4}],
\end{split}
\end{eqnarray} 
where ${\cal O}_{i}=n_{i}\cdot\textbf{S}$ and ${\cal O}_{i}^{\prime}=n^{\prime}_{i}\cdot\textbf{S}$, 
for $i=1,2,3,4$ and $n_i$'s are unit spacelike vectors on the Euclidean $3$-plane as earlier. We shall compute the expectation value of the above operator in the state ~\ref{sq8}. Denoting  the first operator appearing within the square bracket on the right hand side of the above equation by $\mathscr{E}_{4}$, we find
\begin{equation}\label{f1'}
\begin{split}
&\langle \overline0_{\vec k}|\mathscr{E}_{4}|\overline0_{\vec k}\rangle=|A_{k}|^2  {_1}\langle0_{\vec{k},{\rm in}}|\otimes  {_2}\langle0_{\vec{k},{\rm in}}|\mathscr{E}_{4}| 0_{\vec{k},{\rm in}}\rangle_{1}\otimes|0_{\vec{k},{\rm in}}\rangle_{2}\\&+ \frac{A_{k}^\ast B_{k}}{\sqrt2}  {_1}\langle0_{\vec{k},{\rm in}}|\otimes  {_2}\langle0_{\vec{k},{\rm in}}|\mathscr{E}_{4}\left(|1_{\vec{k},{\rm in}}\rangle_1 \otimes |1_{-\vec{k},{\rm in}}\rangle_2+|1_{-\vec{k},{\rm in}}\rangle_1 \otimes |1_{\vec{k},{\rm in}}\rangle_2\right)\\&+ \frac{A_{k} B_{k}^\ast}{\sqrt2} \left(_1\langle1_{\vec{k},{\rm in}}| \otimes {_2}\langle1_{-\vec{k},{\rm in}}|+_1\langle1_{-\vec{k},{\rm in}}| \otimes {_2}\langle1_{\vec{k},{\rm in}}|\right)\mathscr{E}_{4}| 0_{\vec{k},{\rm in}}\rangle_{1}\otimes|0_{\vec{k}}^{\rm in}\rangle_{2}\\&+\frac{|B_{k}|^2}{2}\left(_1\langle1_{\vec{k},{\rm in}}| \otimes {_2}\langle1_{-\vec{k},{\rm in}}|+_1\langle1_{-\vec{k},{\rm in}}| \otimes {_2}\langle1_{\vec{k},{\rm in}}|\right)\mathscr{E}_{4}\left(|1_{\vec{k},{\rm in}}\rangle_1 \otimes |1_{-\vec{k},{\rm in}}\rangle_2+|1_{-\vec{k},{\rm in}}\rangle_1 \otimes |1_{\vec{k},{\rm in}}\rangle_2\right)\\&=F(\theta_{1},\theta_{2}, \theta_{3}, \theta_{4}, \phi_{1}, \phi_{2}, \phi_{3}, \phi_{4})~~~({\rm say}).
\end{split}
\end{equation}
We obtain after some algebra,
\begin{eqnarray}
\begin{split}
&F(\theta_{1},\theta_{2}, \theta_{3}, \theta_{4}, \phi_{1}, \phi_{2}, \phi_{3}, \phi_{4})=|A_{k}|^2 f(\theta_{1},\theta_{2}, \phi_{1}, \phi_{2})  f(\theta_{3},\theta_{4}, \phi_{3}, \phi_{4})\\&+ \frac{A_{k}^\ast B_{k}}{\sqrt2}[ g_{+}(\theta_{1},\theta_{2}, \phi_{1}, \phi_{2}) g_{-}(\theta_{3},\theta_{4}, \phi_{3}, \phi_{4})+g_{-}(\theta_{1},\theta_{2}, \phi_{1}, \phi_{2}) g_{+}(\theta_{3},\theta_{4}, \phi_{3}, \phi_{4})]\\&+ \frac{A_{k} B_{k}^\ast}{\sqrt2}[ g_{+}^\ast (\theta_{1},\theta_{2}, \phi_{1}, \phi_{2}) g_{-}^\ast (\theta_{3},\theta_{4}, \phi_{3}, \phi_{4})+g_{-}^\ast (\theta_{1},\theta_{2}, \phi_{1}, \phi_{2}) g_{+}^\ast (\theta_{3},\theta_{4}, \phi_{3}, \phi_{4})]\\&+\frac{|B_{k}|^2}{2}[ h_{++} (\theta_{1},\theta_{2}, \phi_{1}, \phi_{2}) h_{--} (\theta_{3},\theta_{4}, \phi_{3}, \phi_{4})+h_{--} (\theta_{1},\theta_{2}, \phi_{1}, \phi_{2}) h_{++} (\theta_{3},\theta_{4}, \phi_{3}, \phi_{4})\\&+ h_{+-} (\theta_{1},\theta_{2}, \phi_{1}, \phi_{2}) h_{-+} (\theta_{3},\theta_{4}, \phi_{3}, \phi_{4})+h_{-+} (\theta_{1},\theta_{2}, \phi_{1}, \phi_{2}) h_{+-} (\theta_{3},\theta_{4}, \phi_{3}, \phi_{4})]
\end{split}
\label{f2'}
\end{eqnarray}
where we have denoted for the sake of brevity,
\begin{multline}
 f(\theta_{1},\theta_{2},\phi_{1},\phi_{2})= {_1}\langle0_{\vec{k},{\rm in}}|(\textbf{n}_{1}.\textbf{S})\otimes (\textbf{n}_{2}.\textbf{S})| 0_{\vec{k},{\rm in}}\rangle_{1} = (\cos{\theta_{1}} \cos{\theta_{2}} \\\qquad \qquad \qquad \qquad \qquad \qquad \qquad \qquad \qquad \qquad  - 2|\beta_{k} \alpha_{k}| \cos(\phi_{1}+\phi_{2}) \sin{\theta_{1}}\sin{\theta_{2}})\\
g_{+}(\theta_{1},\theta_{2},\phi_{1},\phi_{2})={_1}\langle0_{\vec{k},{\rm in}}|(\textbf{n}_{1}.\textbf{S})\otimes (\textbf{n}_{2}.\textbf{S})| 1_{\vec{k},{\rm in}}\rangle_{1}
 =\frac{|\alpha_k|^2}{\alpha_k^\ast} \Big(-\sin{\theta_{1}} \cos{\theta_{2}}e^{i \phi_{1}} \\\qquad \qquad \qquad \qquad \qquad \qquad \qquad \qquad \qquad \qquad \qquad \qquad - \left\vert \frac{{\beta_{k}}}{ {\alpha_{k}}}\right \vert e^{-i\phi_{2}} \cos{\theta_{1}}\sin{\theta_{2}}\Big)  \\
g_{-}(\theta_{1},\theta_{2},\phi_{1},\phi_{2})={_1}\langle0_{\vec{k},{\rm in}}|(\textbf{n}_{1}.\textbf{S})\otimes (\textbf{n}_{2}.\textbf{S})| 1_{-\vec{k},{\rm in}}\rangle_{1}
 =\frac{|\alpha_k|^2}{\alpha_k^\ast} \Big(-\sin{\theta_{2}} \cos{\theta_{1}}e^{i\phi_{2}}  \\\qquad \qquad \qquad \qquad \qquad \qquad \qquad \qquad \qquad \qquad \qquad \qquad- \left\vert \frac{{\beta_{k}}}{ {\alpha_{k}}}\right \vert  e^{-i\phi_{1}}\cos{\theta_{2}}\sin{\theta_{1}}\Big)   \\
 h_{++}(\theta_{1},\theta_{2},\phi_{1},\phi_{2})={_1}\langle1_{\vec{k},{\rm in}}|(\textbf{n}_{1}.\textbf{S})\otimes (\textbf{n}_{2}.\textbf{S})| 1_{\vec{k},{\rm in}}\rangle_{1}
 = -\cos{\theta_{1}}\cos{\theta_{2}} \\\qquad \qquad \qquad \qquad \qquad \qquad \qquad \qquad \qquad \qquad \qquad \qquad \qquad= h_{--}(\theta_{1},\theta_{2},\phi_{1},\phi_{2}) \\
h_{+-}(\theta_{1},\theta_{2},\phi_{1},\phi_{2})={_1}\langle1_{\vec{k},{\rm in}}|(\textbf{n}_{1}.\textbf{S})\otimes (\textbf{n}_{2}.\textbf{S})| 1_{-\vec{k},{\rm in}}\rangle_{1}
=\sin{\theta_{1}}\sin{\theta_{2}}e^{-i(\phi_{1}-\phi_{2})} 
\\\qquad \qquad \qquad \qquad \qquad \qquad =h_{-+}(\theta_{1},\theta_{2},\phi_{1},\phi_{2})^\ast
\label{b4'}
\end{multline}

The  expectation values corresponding to the other operators in \ref{fm1} can be found by suitably permuting their angular arguments in \ref{f1'}, \ref{f2'} and \ref{b4'}. 

\begin{figure}[h!]
\begin{center}
  \includegraphics[width=10cm]{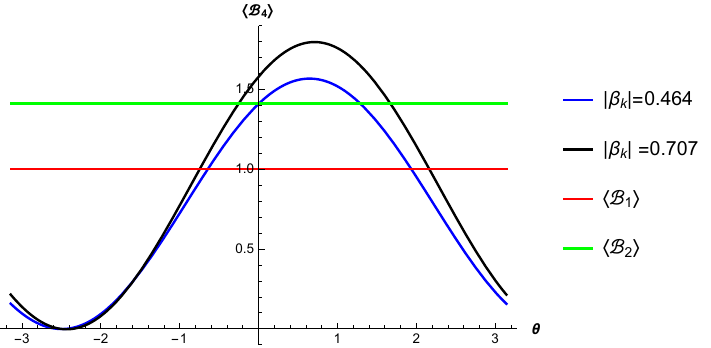}
  \caption{\small \it Variation of $\langle{\cal B}_{4}\rangle$ (c.f. \ref{fm1}), with respect to the angular variable $\theta$ and the  subsequent violation of the BMK inequality for four-mode squeezed state. The red line denotes the classical upper bound whereas the green line stands for the quantum upper bound for the two-mode squeezed state. The blue and black lines correspond to  two different values of $|\beta_{k}|$ with the fixed values of $A_{k}\sim 0.975$ and $B_{k}\sim 0.224$ in \ref{sq8}. Thus the maximum violation is larger here compared to the two-mode squeezed state, \ref{fig:bell1}.}
  \label{fig:bell3}
\end{center}
\end{figure}

\ref{fig:bell3} depicts the  expectation value of \ref{fm1} with respect to the angular variable $\theta$.\footnote{We have set all $\phi$'s to zero in \ref{b4'}. Also we have taken, $\theta_1= -0.31^{\circ},\,
\theta_2=40.8^{\circ},\,
\theta_3=-37.1^{\circ},\,
\theta_4=-37.3^{\circ},$
$\theta'_1=-90.5^{\circ},\,
\theta'_2=-40.4^{\circ},\,
\theta'_3=36.9^{\circ},\,
\theta'_4=37.1^{\circ}$.}
 We have taken $A_{k}\sim 0.975$ and $B_{k}\sim 0.224$ in the expression of the four-mode squeezed state, \ref{sq8}. Comparison with \ref{fig:bell1} shows that the BMK violation has increased, in agreement with \ref{b4}. 

Let us now recall that for a general many particle  quantum state which can be grouped into entangled and non-entangled parts, the  BMK violation  can also  be written in a more illuminating form~\cite{uffink:2002},
\begin{eqnarray}
\langle{\cal B}_{N}\rangle^2+\langle{\cal B}^{\prime}_{N}\rangle^2 \leq 2^Z
\label{b6}
\end{eqnarray}
where $N$ represents the total number of partite states and 
\begin{eqnarray}
Z=N-K_{1}-2L+1,
\label{b7}
\end{eqnarray}
where $K_{1}$ denotes the number of single partite states which are not entangled with other $(N-1)$-partite states. $L=\sum_{l=2}^M K_{l}$,  with $K_{l}$ being the number of groups consisting of ${l}$ entangled partite states and $M$ stands for the largest number of such states in a group.  It follows that $N=\sum_{l=1}^M lK_{l}$. \ref{b6} indicates that the upper bound of the BMK violation increases with the number of modes in the squeezed states discussed in \ref{sqz} as follows. For a two-mode squeezed state, we have $N=2,\,L=1,\, K_{1}=0$, so that $Z=1$. For a four mode squeezed state on the other hand, we have $N=4$, $L=1$, $K_1=0$ and hence $Z=3$. Following~\cite{bell:2017} as of the scalar field, it is then easy to argue that if we include more than one  $\vec{k}$ value into the state, for the two-mode squeezed state the BMK violation does not increase any further, but for the four mode squeezed state it increases without bound, eventually leading to an infinite violation of the BMK inequality. We shall not go into any detail of this.

Finally, we note that we have worked with states with a single value of the helicity  $s$,  \ref{sq2}, \ref{sq8}. It is easy to see that the above results will remain unchanged even if we work with a squeezed state where the $s$ values are summed over, as follows. As we have argued below \ref{sq1}, any squeezed state expansion that sums over  $s$, will factor out between two normalised sub-sectors corresponding to those $s$ values. On the other hand, since  the pseudospin operator  \ref{bv0} acts on states only with a specific $s$ value, it is clear that the  bra and ket for the squeezed state expansion corresponding to the other $s$ value will just give unity, while computing the expectation values like \ref{exp}.  

Finally, we note the qualitative similarity between the BMK violations for a scalar field~\cite{bell:2017} with that of the Dirac fermions. The BMK violation discussed above basically probes the vacuum state of the cosmological de Sitter spacetime. We also wish to discuss in this work some correlation properties (both classical and quantum) associated with the maximally entangled states. The quantum discord is one such suitable measure, which we investigate below using the two-mode squeezed state, \ref{sq2}.

 \section{Quantum discord}\label{discord1}
We wish to compute quantum discord, described in \ref{discord}, in the cosmological de Sitter background we are interested in, for a maximally entangled `in' state,
\begin{eqnarray}
|\psi\rangle=\frac{1}{\sqrt2}(|0\rangle_{\vec{l}}|0\rangle_{\vec{k}}+|1\rangle_{\vec{l}}|1\rangle_{\vec{k}})
\label{d9}
\end{eqnarray}
Using the expression for the two-mode squeezed state \ref{sq2}, the above state can be expanded in terms of the `out' states. We shall consider two cases below. In the first case the states denoted by the momentum $\vec{l}$ will be held intact. In another case we shall consider their squeezed state expansion as well. This will help us to probe the correlations between the `in-out' and the `out-out' sectors.

\begin{figure}[h!]
\centering
  \includegraphics[width=7cm]{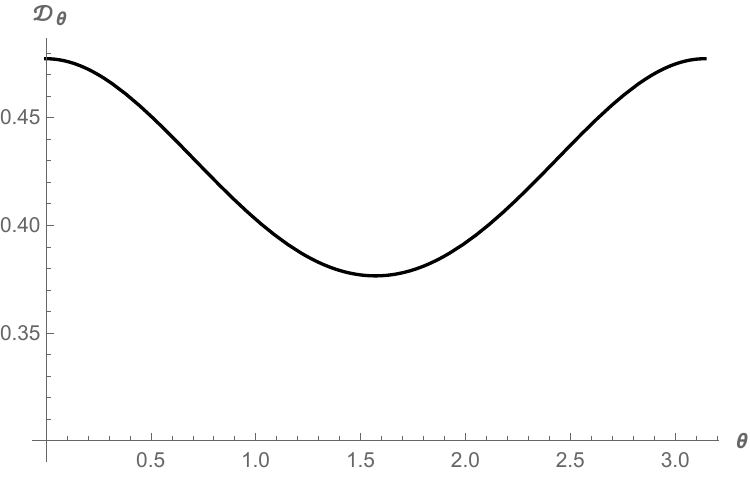} \hskip 1cm
  \includegraphics[width=7cm]{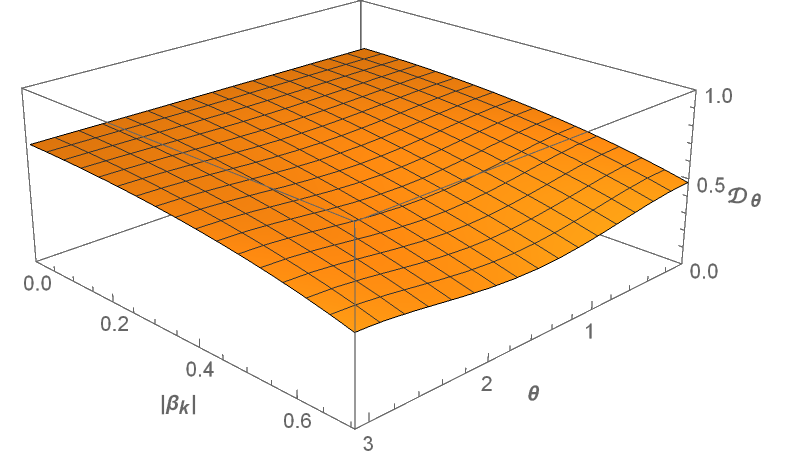}
 \caption{\small \it (Left) Plot for the quantum discord, \ref{d19}, with respect to the polar angle $\theta$ with $|\beta_k| \approx 0.707$. The minimisation occurs at $\theta =\pi/2$. This is valid for all possible values of the Bogoliubov coefficient. (Right) A 3-D plot for the same with the simultaneous variation of $\theta$ and $|\beta_{k}|$. As of \ref{bglv3}, the variation of the Bogoliubov coefficient correspond to the variation of the rest mass  only, and hence for a given species of field, $|\beta_k|$ or the quantum discord is fixed with respect to its spatial momentum ${\vec k}$.}
  \label{fig:discord1}
  \end{figure}

Accordingly as the first case, using \ref{sq2} for the modes $\vec{k}$, we rewrite \ref{d9} as
\begin{equation}\label{d11}
|\psi\rangle=\frac{1}{\sqrt2}\frac{|\alpha_{k}|}{\alpha_{k}^\ast}\left[|0\rangle_{\vec{l}}\otimes\left(\alpha_{k}^\ast|0_{\vec{k},{\rm out}}\rangle |0_{-\vec{k},{\rm out}}\rangle - {\beta_{k}} |1_{\vec{k},{\rm out}}\rangle |1_{-\vec{k},{\rm out}}\rangle \right)+|1\rangle_{\vec{l}}\otimes (|1_{\vec{k},{\rm out}}\rangle|0_{-\vec{k},{\rm out}}\rangle)\right]
\end{equation}
The reduced density matrix $\rho_{XY}$ ($X \equiv \vec{l},\,\, Y \equiv \vec{k}$), after tracing out over $-\vec{k}$ and suppressing the level `out' for the sake of brevity is given by,
\begin{eqnarray}
\begin{split}
\rho_{XY}=&\frac{1}{2}\left[|0\rangle_{\vec{l}} {_{\vec{l}}\langle0|}\otimes(|\alpha_{k}|^2 |0\rangle_{\vec{k}} {_{\vec{k}}\langle0|})+|0\rangle_{\vec{l}} {_{\vec{l}}\langle0|}\otimes (|\beta_{k}|^2 |1\rangle_{\vec{k}} {_{\vec {k}}\langle1|})\right. \\&\left.+|1\rangle_{\vec{l}} {_{\vec{l}}\langle0|}\otimes (\alpha_{{k}} |1\rangle_{\vec{k}} {_{\vec{k}}\langle0|}) +|0\rangle_{\vec{l}} {_{\vec{l}}\langle1|}\otimes (\alpha^\ast_{k} |0\rangle_{\vec{k}} {_{\vec{k}}\langle1|})+|1\rangle_{\vec{l}} {_{\vec{l}}\langle1|}\otimes|1\rangle_{\vec{k}} {_{\vec{k}}\langle1|}\right]
\label{d12}
\end{split}
\end{eqnarray}
We also have,
\begin{eqnarray}
\begin{split}
\rho_{X}&=\operatorname{Tr}_{Y} (\rho_{X,Y})
=\frac{1}{2}[(|0\rangle_{\vec{l}} {_{\vec{l}}\langle0|}+|1\rangle_{\vec{l}} {_{\vec{l}}\langle1|})],\\
\rho_{Y}&=\operatorname{Tr}_{X} (\rho_{X,Y})
=\frac{1}{2}[|\alpha_{k}|^2(|0\rangle_{\vec{k}} {_{\vec{k}}\langle0|}+(1+|\beta_{k}|^2)|1\rangle_{\vec{k}} {_{\vec{k}}\langle1|})]\\&
\label{d13}
\end{split}
\end{eqnarray}
Using \ref{d12} and \ref{d13}, we now compute the von Neumann entropies,
\begin{eqnarray}
\begin{split}
&S(\rho_{X})=\log 2\\&
S(\rho_{Y})=-\frac{1}{2}\left[|\alpha_{k}|^2\log\frac{|\alpha_{k}|^2}{2}+(1+|\beta_{k}|^2)\log\frac{(1+|\beta_{k}|^2)}{2}\right]\\&
S(\rho_{XY})=-\frac{1}{2}\left[(1+|\alpha_{k}|^2)\log\frac{(1+|\alpha_{k}|^2)}{2}+(1-|\alpha_{k}|^2)\log\frac{(1-|\alpha_{k}|^2)}{2}\right]
\end{split}
\label{d14}
\end{eqnarray}
In order to compute the conditional entropy $S(Y|X)$, \ref{d6}, which needs a minimisation over the projective measurements, we take the usual projection operators~\cite{A.dutta, Kanno:2016gas, Zurek},
\begin{eqnarray}
\begin{split}
\Pi_{\pm}  :&= \frac{1}{2}\left[(1\pm{\hat x}_{3})|0\rangle_{\vec{l}} {_{\vec{l}}}\langle0|+(1\mp{\hat x}_{3})|1\rangle_{\vec{l}} {_{\vec{l}}}\langle1|\pm({\hat x}_{1}- i{\hat x}_{2})|0\rangle_{\vec{l}} {_{\vec{l}}}\langle1|\pm({\hat x}_{1}+i{\hat x}_{2})|1\rangle_{\vec{l}} {_{\vec{l}}}\langle0|\right]
\label{d15}
\end{split}
\end{eqnarray}
where $ {\hat x}_{1}^2+{\hat x}_{2}^2+{\hat x}_{3}^2=1$, representing spatial unit vectors.
Note that the above operates only on the $\vec{l}$ sector of \ref{d12}. Since the relevant  Hilbert space is 2-dimensional, we have taken two projectors which are orthogonal to each other and follows the identity $\Pi^2_{\pm}=\Pi_{\pm}$. 
Using  now $\operatorname{Tr}(AB)= \operatorname{Tr}(BA)$, we get from \ref{d5} after some algebra,
\begin{eqnarray}
\begin{split}
\rho_{Y|\pm}=&\frac{1}{2} \left[{(1\pm {\hat x}_{3})(|\alpha_{k}|^2 |0\rangle_{\vec{k}} {_{\vec{k}}\langle0|}+ |\beta_{k}|^2 |1\rangle_{\vec{k}} {_{\vec{k}}\langle1|})}+(1\mp {\hat x}_{3})(|1\rangle_{\vec{k}}{_{\vec{k}}}\langle1|)\right.\\&\left.\pm(\hat{x}_{1}-i {\hat x}_{2})(\alpha_{k}|1\rangle_{\vec{k}}{_{\vec{k}}}\langle0|)\pm({\hat x}_{1}+i {\hat x}_{2})(\alpha^\ast_{k}|0\rangle_{\vec{k}}{_{\vec{k}}}\langle1|)\right]
\end{split}
\label{d17}
\end{eqnarray}
In the usual parametrisation,
$${\hat x}_{1}=\sin{\theta} \cos{\phi},\qquad {\hat x}_{2}=\sin{\theta}\sin{\phi}, \qquad {\hat x}_{3}=\cos{\theta},$$
the conditional entropy is given by \ref{d6} and is found to be independent of the azimuthal angle $\phi$,
\begin{equation}
\begin{split}
S(Y|X)&=-\frac{1}{2}\left[\left(\frac{1+\sqrt{1-(1+\cos{\theta})^2|\beta_{k}|^2|\alpha_{k}|^2}}{2}\right)\log\left(\frac{1+\sqrt{1-(1+\cos{\theta})^2|\beta_{k}|^2|\alpha_{k}|^2})}{2}\right)\right.\\&\left.+\left(\frac{1-\sqrt{1-(1+\cos{\theta})^2|\beta_{k}|^2|\alpha_{k}|^2})}{2}\right)\log\left(\frac{(1-\sqrt{1-(1+\cos{\theta})^2|\beta_{k}|^2|\alpha_{k}|^2})}{2}\right)\right.\\&\left.+\left(\frac{(1+\sqrt{1-(1-\cos{\theta})^2|\beta_{k}|^2|\alpha_{k}|^2})}{2}\right)\log\left(\frac{(1+\sqrt{1-(1-\cos{\theta})^2|\beta_{k}|^2|\alpha_{k}|^2})}{2}\right)\right.\\&\left.+\left(\frac{(1-\sqrt{1-(1-\cos{\theta})^2|\beta_{k}|^2|\alpha_{k}|^2})}{2}\right)\log\left(\frac{(1-\sqrt{1-(1-\cos{\theta})^2|\beta_{k}|^2|\alpha_{k}|^2})}{2}\right)\right]_{\rm min},
\end{split}
\label{d18}
\end{equation}
where the suffix `min' stands for the minimisation with respect to $\theta$. The quantum discord, \ref{d8}, is given by,
\begin{equation}
\begin{split}
{\mathscr{D}_{\theta}}
=&\log 2+\frac{1}{2}\left[(1+|\alpha_{k}|^2)\log\frac{(1+|\alpha_{k}|^2)}{2}+(1-|\alpha_{k}|^2)\log\frac{(1-|\alpha_{k}|^2)}{2}\right]\\&-\frac{1}{2}\left[\left(\frac{1+\sqrt{1-(1+\cos{\theta})^2|\beta_{k}|^2|\alpha_{k}|^2}}{2}\right)\log\left(\frac{1+\sqrt{1-(1+\cos{\theta})^2|\beta_{k}|^2|\alpha_{k}|^2})}{2}\right)\right.\\&\left.+\left(\frac{1-\sqrt{1-(1+\cos{\theta})^2|\beta_{k}|^2|\alpha_{k}|^2})}{2}\right)\log\left(\frac{(1-\sqrt{1-(1+\cos{\theta})^2|\beta_{k}|^2|\alpha_{k}|^2})}{2}\right)\right.\\&\left.+\left(\frac{(1+\sqrt{1-(1-\cos{\theta})^2|\beta_{k}|^2|\alpha_{k}|^2})}{2}\right)\log\left(\frac{(1+\sqrt{1-(1-\cos{\theta})^2|\beta_{k}|^2|\alpha_{k}|^2})}{2}\right)\right.\\&\left.+\left(\frac{(1-\sqrt{1-(1-\cos{\theta})^2|\beta_{k}|^2|\alpha_{k}|^2})}{2}\right)\log\left(\frac{(1-\sqrt{1-(1-\cos{\theta})^2|\beta_{k}|^2|\alpha_{k}|^2})}{2}\right)\right]_{\rm min}
\end{split}
\label{d19}
\end{equation}
For different values of the Bogoliubov coefficients,  ${\mathscr{D}_{\theta}}$ appearing above minimises  at $\theta=\pi/2$,  \ref{fig:discord1}.  For the maximum value $|\beta_{k}|\approx 0.707$ in the cosmological de Sitter, we have  ${\mathscr{D}_{\pi/2}}\approx 0.377$. Note also that ${\mathscr{D}_{\theta}}$ is never vanishing, not even for $|\beta_k| \to 0$, in which limit it equals $\log 2$.
\\

\noindent
As we mentioned earlier,  we shall now examine the second scenario where  both $l$ and $k$ sectors in \ref{d9} undergo the squeezed state expansion, so that
\begin{equation}
\begin{split}
|\psi\rangle&=\frac{|\alpha_{l}| |\alpha_{k}|}{\sqrt2 \alpha_{l}^\ast \alpha_{k}^\ast   }
\left[(\alpha_{l}^\ast |0_{\vec{l},{\rm out}}\rangle |0_{-\vec{l},{\rm out}}\rangle - {\beta_{l}} |1_{\vec{l},{\rm out}}\rangle |1_{-\vec{l},{\rm out}}\rangle)\otimes(\alpha_{k}^\ast|0_{\vec{k},{\rm out}}\rangle |0_{-\vec{k},{\rm out}}\rangle - {\beta_{k}} |1_{\vec{k},{\rm out}}\rangle |1_{-\vec{k},{\rm out}}\rangle)\right.\\& + \left. |1_{\vec{l},{\rm out}}\rangle |0_{-\vec{l},{\rm out}}\rangle \otimes |1_{\vec{k},{\rm out}}\rangle|0_{-\vec{k},{\rm out}}\rangle\right]
\end{split}
\label{d21}
\end{equation}
We define the reduced density operator  $\rho_{XY}$ $(X\equiv \vec{l},\, Y\equiv \vec{k})$ by tracing out over the $(-\vec{l}, -\vec{k})$ sector,
\begin{eqnarray}
\begin{split}
\rho_{XY}&=\frac{1}{2}\left[(|\alpha_{l}|^2|0\rangle_{\vec{l}} {_{\vec{l}}\langle0|})\otimes(|\alpha_{k}|^2 |0\rangle_{\vec{k}} {_{\vec{k}}\langle0|})+(|\alpha_{l}|^2|0\rangle_{\vec{l}} {_{\vec{l}}\langle0|})\otimes (|\beta_{k}|^2 |1\rangle_{\vec{k}} {_{\vec {k}}\langle1|})\right.\\&\left.+(|\beta_{{l}}|^2|1\rangle_{\vec{l}} {_{\vec{l}}\langle1|})\otimes (|\alpha_{{k}}|^2 |0\rangle_{\vec{k}} {_{\vec{k}}\langle0|}) +(\alpha^\ast_{l}|0\rangle_{\vec{l}} {_{\vec{l}}\langle1|})\otimes (\alpha^\ast_{k} |0\rangle_{\vec{k}} {_{\vec{k}}\langle1|})\right.\\&\left.+(\alpha_{l}|1\rangle_{\vec{l}} {_{\vec{l}}\langle0|})\otimes (\alpha_{k} |1\rangle_{\vec{k}} {_{\vec{k}}\langle0|})+(|\beta_{l}|^2|1\rangle_{\vec{l}} {_{\vec{l}}\langle1|})\otimes(|\beta_{k}|^2|1\rangle_{\vec{k}} {_{\vec{k}}\langle1|})+|1\rangle_{\vec{l}} {_{\vec{l}}\langle1|}\otimes|1\rangle_{\vec{k}} {_{\vec{k}}\langle1|}\right]
\label{d22}
\end{split}
\end{eqnarray}
where we have suppressed the level `out' as earlier for the sake of brevity. We find after some algebra, the following von Neumann entropies,
\begin{equation}
\begin{split}
&S(\rho_{X})=-\frac{1}{2}\left[|\alpha_{l}|^2\log\frac{|\alpha_{l}|^2}{2}+(1+|\beta_{l}|^2)\log\frac{(1+|\beta_{l}|^2)}{2}\right]\\&
S(\rho_{Y})=-\frac{1}{2}\left[|\alpha_{k}|^2\log\frac{|\alpha_{k}|^2}{2}+(1+|\beta_{k}|^2)\log\frac{(1+|\beta_{k}|^2)}{2}\right]\\&
S(\rho_{XY})=-\frac{1}{2}\left[|\beta_{l}|^2|\alpha_{k}|^2\log\frac{|\beta_{l}|^2|\alpha_{k}|^2}{2}+|\beta_{k}|^2|\alpha_{l}|^2\log\frac{|\beta_{k}|^2|\alpha_{l}|^2}{2}\right.\\&\left.+\left(\frac{(1+|\alpha_{l}|^2|\alpha_{k}|^2+|\beta_{l}|^2|\beta_{k}|^2)+\sqrt{(1+|\alpha_{l}|^2|\alpha_{k}|^2+|\beta_{l}|^2|\beta_{k}|^2)^2-4|\alpha_{l}|^2|\alpha_{k}|^2|\beta_{l}|^2|\beta_{k}|^2}}{2}\right)\right.\\&\left. \times \log\left(\frac{(1+|\alpha_{l}|^2|\alpha_{k}|^2+|\beta_{l}|^2|\beta_{k}|^2)+\sqrt{(1+|\alpha_{l}|^2|\alpha_{k}|^2+|\beta_{l}|^2|\beta_{k}|^2)^2-4|\alpha_{l}|^2|\alpha_{k}|^2|\beta_{s}|^2|\beta_{k}|^2}}{4}\right)\right.\\&\left.+\left(\frac{(1+|\alpha_{l}|^2|\alpha_{k}|^2+|\beta_{l}|^2|\beta_{k}|^2)-\sqrt{(1+|\alpha_{l}|^2|\alpha_{k}|^2+|\beta_{l}|^2|\beta_{k}|^2)^2-4|\alpha_{l}|^2|\alpha_{k}|^2|\beta_{l}|^2|\beta_{k}|^2}}{2}\right)\right. \\&\left.\times \log\left(\frac{(1+|\alpha_{l}|^2|\alpha_{k}|^2+|\beta_{l}|^2|\beta_{k}|^2)-\sqrt{(1+|\alpha_{l}|^2|\alpha_{k}|^2+|\beta_{l}|^2|\beta_{k}|^2)^2-4|\alpha_{l}|^2|\alpha_{k}|^2|\beta_{l}|^2|\beta_{k}|^2}}{4}\right)\right]
\end{split}
\label{d23}
\end{equation}
 Since the $\vec{l}$ sector of \ref{d22} is two dimensional by the virtue of the Pauli exclusion principle, the conditional entropy, \ref{d6}, for this system can still be computed as earlier using the projectors of~\ref{d15}.\footnote{For Bosons on the other hand, we would have obtained an infinite dimensional Hilbert space for the $\vec{l}$ sector, requiring the construction of an infinite number of projection operators to define the quantum discord.  One possible way to tackle this issue seems to be the   truncation of the relevant squeezed state expansion at some finite order by assuming  $|\beta_l|/|\alpha_l| \ll 1$. A full resolution of this situation, however, is not clear to us.} We find
\begin{eqnarray}
\begin{split}
S({Y|X})&=p_{+}S(\rho_{Y|+})+p_{-}S(\rho_{Y|-})\\&
=p_{+}\left[-\left(\frac{2p_{+}+\sqrt{(2p_{+})^2-C_{+}}}{4p_{+}}\right)\log\left(\frac{2p_{+}+\sqrt{(2p_{+})^2-C_{+}}}{4p_{+}}\right)\right.\\&\left.-\left(\frac{2p_{+}-\sqrt{(2p_{+})^2-C_{+}}}{4p_{+}}\right)\log\left(\frac{2p_{+}-\sqrt{(2p_{+})^2-C_{+}}}{4p_{+}}\right)\right] \\&+p_{-}\left[-\left(\frac{2p_{-}+\sqrt{(2p_{-})^2-C_{-}}}{4p_{-}}\right)\log\left(\frac{2p_{-}+\sqrt{(2p_{-})^2-C_{-}}}{4p_{-}}\right)\right.\\&\left.-\left(\frac{2p_{-}-\sqrt{(2p_{-})^2-C_{-}}}{4p_{-}}\right)\log\left(\frac{2p_{-}-\sqrt{(2p_{-})^2-C_{-}}}{4p_{-}}\right)\right]_{\rm min}
\end{split}
\label{d28}
\end{eqnarray}
where $p_{\pm}$ and  $C_{\pm}$ are given by
\begin{equation}
\begin{split}
{p_{\pm}}=&\operatorname{Tr}_{X,Y}\left(\Pi_{\pm}\rho_{X,Y}\Pi_{\pm}\right)=\frac{1}{2}\left(1\mp \cos\theta|\beta_{k}|^2\right),   \\
C_{\pm}=& {(1\pm \cos\theta)^2|\alpha_{l}|^4|\alpha_{k}|^2|\beta_{k}|^2}+2 |\beta_{l}|^2|\beta_{k}|^2|\alpha_{l}|^2|\alpha_{k}|^2\,\sin^2\theta \\& +(1\mp\cos\theta)^2 |\beta_{l}|^2|\alpha_{k}|^2(|\beta_{l}|^2|\beta_{k}|^2+1)
\end{split}
\label{d29}
\end{equation}
and the suffix `min' in \ref{d28} refers to minimisation with respect to the polar angle $\theta$. Note that alike \ref{d18}, the above is independent of the azimuthal angle $\phi$.

\begin{figure}[h!]
\centering
  \includegraphics[width=7cm]{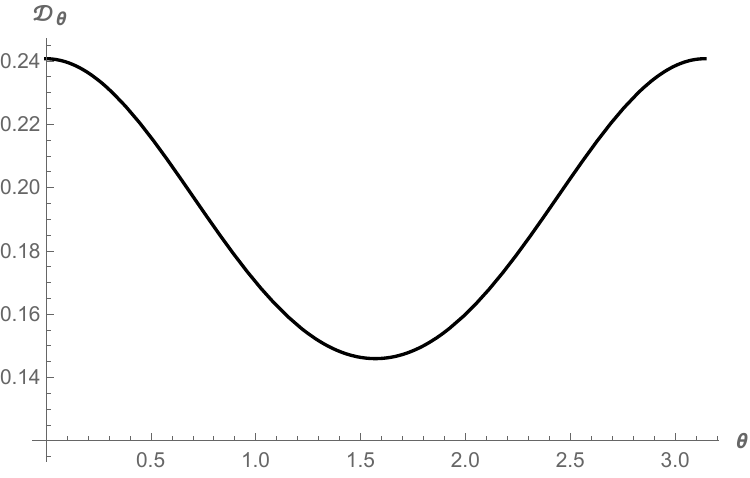} \hskip 1cm
   \includegraphics[width=7cm]{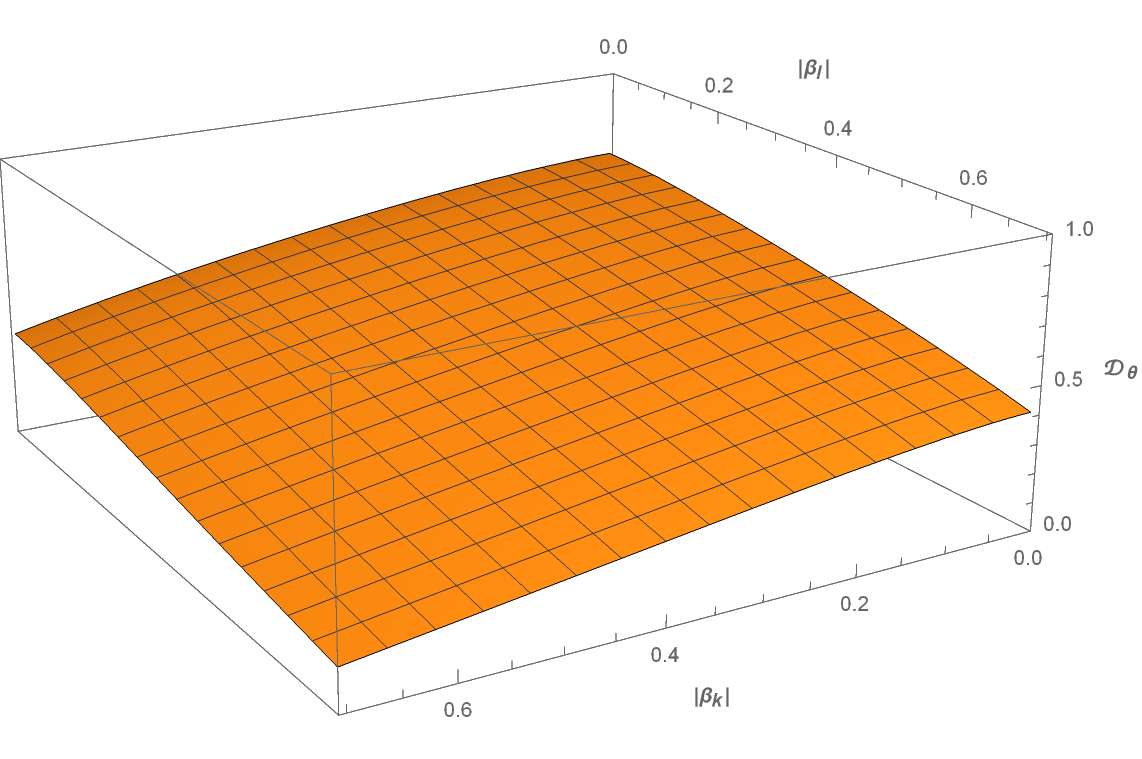}
  \caption{\small \it  (Left) Plot for the quantum discord characterising `out-out' correlations, with respect to the polar angle $\theta$ with $|\beta_l|=|\beta_k| \approx 0.707$. The minima occurs at $\theta =\pi/2$ for all possible values of the Bogoliubov coefficient, similar to the `in-out' sector, \ref{fig:discord1}. (Right) A 3-D plot of the same, depicting its variation with respect to the two Bogoliubov coefficients, $|\beta_{k}|$ and $|\beta_{l}|$. As of \ref{bglv3}, the variation of the Bogoliubov coefficients correspond to the variation of the rest mass only.}
  \label{fig:discord}
\end{figure}

Even though the $\theta$ dependence of \ref{d29} looks  different compared to \ref{d18}, the discord, ${\mathscr{D}_{\theta}}=S(\rho_{X})-S(\rho_{X,Y})+S({Y|X})$, still minimises at $\theta = \pi/2$, as depicted in \ref{fig:discord}. For the Bogoliubov coefficients close to their maximal values, $|\beta_{l}|=|\beta_{k}|\approx 0.707$, we have ${\mathscr{D}_{\pi/2}}\approx 0.146$, which is less compared to our previous case, \ref{fig:discord1}. This shows the degradation of correlations in the `out-out' states compared to the `in-out' states. We also have plotted the variation of the discord with respect to the two Bogoliubov coefficients in \ref{fig:discord}. As earlier, the discord is never vanishing.
%
%

The above analysis was done by tracing out the $(-\vec{k}, -\vec{l})$ subsectors in \ref{d21}.  One can also trace out the other subsectors and analogously compute the quantum discord between them. We expect qualitatively similar results to hold. We also note the qualitative similarity of \ref{fig:discord1}, \ref{fig:discord} with the earlier studies on scalar and fermion fields in the Rindler spacetime~\cite{A.dutta, wang, Brown:2012iz} and as well as on the scalar field in the hyperbolic de Sitter spacetime~\cite{Kanno:2016gas}.

Finally, we wish to mention very briefly the logarithmic negativity~\cite{Zyczkowski:1998yd, Vidal:2002zz} for this system. The log-negativity is a measure of pure quantum correlations,  useful in particular, for mixed ensembles. In order to compute this, one needs to take the partial transpose of the matrix representation of \ref{d12},
%
$$(\rho_{XY})^{\rm T}=\frac{1}{2} \begin{pmatrix} |\alpha_{k}|^2 & 0 & 0 & 0 \\   0 & |\beta_{k}|^2 &\alpha_{k}^\ast & 0 \\ 0 &\alpha_{k} & 0 & 0 \\ 0 & 0& 0&1  \end{pmatrix},$$
%
whose eigenvalues are $1/2,\,1/2,\, \pm |\alpha_{k}|^2/2$. The log-negativity is given by,
\begin{eqnarray}
{\cal{L}_N}=\log\left({1+{|\alpha_{k}|^2}}\right)
\label{d31}
\end{eqnarray}
Since we do not have any extreme squeezing limit $(|\beta_k| \to 1,\,|\alpha_k| \to 0)$ for our case, cf. \ref{The Dirac mode functions and Bogoliubov coefficients}, the above quantity is always non-vanishing.  Similar conclusion holds for the mixed density operator of \ref{d22} as well. This is qualitatively different for the scalar field theory  in the context of transition from the de Sitter to radiation dominated era~\cite{Kanno:2016gas}, where such extreme squeezing may happen for which the log-negativity vanishes, even though the quantum discord remains non-vanishing. For our case {\it both} log-negativity and the quantum discord are always non-vanishing. 

\section{Discussion}\label{con}

In this Chapter, we have computed the Bell-Mermin-Klyshko (BMK) violation and quantum discord for massive Dirac fermions in the cosmological de Sitter background, respectively in \ref{Bell} and \ref{discord1}. For the BMK violation, we have focused on the vacuum corresponding to the two- and four-mode squeezed states whereas for the latter we have used a maximally entangled Bell state as our `in' state. Our motivation was to see how the results differ subject to different coordinatisation of the de Sitter as well as the spin of the field. We thus first note the qualitative similarities of the variations with respect to parametrisation angles  (\ref{fig:bell1}, \ref{fig:bell3}, \ref{fig:discord1}, \ref{fig:discord}) with a scalar field theory discussed respectively in the context of transition from de Sitter to radiation dominated era~\cite{bell:2017} and the hyperbolic de Sitter spacetime~\cite{Kanno:2016gas}. However, the chief qualitative difference for the cosmological de Sitter from them or the non-inertial frame (e.g.~\cite{Alsing:2003es}), or even the static de Sitter spacetime (e.g.~\cite{Bhattacharya:2019zno}) is that the Bogoliubov coefficients for the `in' and `out' vacua in this case is independent of the spatial momentum or the total energy, \ref{bglv3}. Similar feature is seen for a scalar field theory in the global or the cosmological de Sitter spacetime, e.g.~\cite{Mottola:1984ar, Markkanen:2016aes}. Also in alternative coordinates like hyperbolic de Sitter spacetime \cite{Bhattacharya:2018yhm}, describing two causally disconnected expanding regions in the de Sitter spacetime, Bogoliubov coefficients decreases with higher frequency resulting diminished entanglement between causally disconnected regions. A naive explanation to this is, the Compton wavelength decreases with the increase in frequency, if wave length decrease than there is less non-locality or entanglement. Thus as we have indicated earlier (e.g.~\ref{fig:discord1}), for our case the variation of the Bogoliubov coefficients correspond to the variation of the rest mass only. In other words, unlike the standard R-L entanglement, for a given particle species, any of our results in the cosmological de Sitter spacetime is fixed.

We also note that there exists no extreme squeezing limit ($|\beta_k| \to 1$) for the states constructed in \ref{sqz} and we always have $|\beta_k| \lesssim 0.707$, as discussed at the end of \ref{The Dirac mode functions and Bogoliubov coefficients}. Due to this reason, as discussed at the end of the preceding section, the logarithmic negativity is never vanishing.  This is qualitatively different from the 
scenario reported in~\cite{Kanno:2016gas} for a scalar field theory where the log-negativity can indeed vanish, indicating the complete decay of quantum entanglement due to particle creation but the discord, being a measure of all correlations,  survive.   

Investigation of these results and also the decoherence properties in the presence of background primordial electric and magnetic fields seems interesting, due to the Schwinger pair creation mechanism. We hope to return to this issue in a future work. 
\chapter{Decoherence and scalar entropy generation with Yukawa interaction in the Minkowski and inflationary de Sitter spacetimes}
\label{chapter3}

In this Chapter, we wish to investigate the decoherence and entropy generation for a scalar field coupled to fermions via the Yukawa interaction. We shall use the correlator approach based on ~\cite{JFKTPMGS, koksma, kok} reviewed in \ref{deco}. Although our primary motivation is to investigate such phenomenon in the inflationary de Sitter spacetime, we would first like to investigate the same in the simpler case of the Minkowski background. The underlying foundation of decoherence is described in \ref{deco}. We treat scalar field as the system and the fermions as the surrounding. We shall assume that the scalar field is massive in the Minkowski spacetime and massless in the de Sitter spacetime, while the fermions are assumed to be massless in both cases. The scalar field is taken to be massless for the de Sitter background corresponds to the anticipation of the possible existence of the late time {\it secular effect} \cite{Kahya:2007cm, Kahya:2007bc}, mentioned in \ref{Motivation and Overview}, whereas the fermions are taken to be massless just in order to keep our computations simple. Some earlier analysis on open quantum systems with scalars and fermions can be found in~\cite{Nusseler:2019ghw, Enqvist:2004pr, Anirban, Lankinen:2019vgv}.

Precisely, there can be $n$-point correlators which can be generated using the $n$-particle irreducible effective action~\cite{Calzetta Hu, Eaction, Calzetta:1986cq}, capturing the information of interaction between the system and its surrounding. In a realistic scenario however, it is certainly impossible for an observer to quantify the correlators of all orders exactly.  Thus one needs to consider  a practical scenario allowing us to compute only some finite order correlators such as the $2$-point or $4$-point Gaussian ones. As we have mentioned in \ref{Motivation and Overview}, the ignorance of the higher order Gaussian and non-Gaussian correlators of the system and surrounding leads to the lack of information, generation  of quantum decoherence and hence  a non-zero entropy defined in some suitable manner. However, we also note that due to progresses in cold atom experiments, simulating quantum field theory models and measuring higher order correlations becomes an experimentally relevant problem. For example in \cite{prl}, multipoint correlation functions in both in and out of equilibrium quantum field theories have been experimentally studied to show the deviation from Gaussianity due to the presence of interaction. 

In both spacetimes, we shall restrict ourselves to the one loop self energy for the scalar. However for the Minkowski background, such one loop self energy diagram are resummable, as we shall see below. We note that the one loop scalar self energy in the Yukawa theory involves a closed fermion loop~\ref{figa}, and hence no secular effect  is possible at this level of perturbation theory in the inflationary de Sitter spacetime. However, we note also that we are interested to compute the correlator for the scalar field and not simply the self energy. The two point correlator corresponding  to the amputated diagram \ref{figa} involves adding external scalar lines, eventually indicating a secular growth for the scalar two point function and its subsequent effect in the decoherence. Previous literature has considered various types of interactions, such as self-interactions of the inflaton field \cite{Lombardo:2005iz, Martineau:2006ki, Nelson:2016kjm, Nelson:2017pmc}, interactions with gravitational waves \cite{Calzetta:1995ys, piao}, interactions with the system scalar with another scalar~\cite{ Liu:2016aaf, Rostami:2017akw, Martin:2018zbe, Martin:2018lin}, as well as interactions with massless and massive fermionic fields \cite{Duffy:2005ue, Prokopec:2003qd, Miao:2006pn, Toms:2018oal, Toms:2018wpy}. We also refer our reader to~\cite{Nusseler:2019ghw, Enqvist:2004pr, Anirban, Lankinen:2019vgv} for some earlier analysis on open quantum systems with scalars and fermions and \cite{Schaub:2023scu, Pethybridge:2021rwf} for the analytical aspects of correlators involving Dirac spinors in de Sitter spacetime with applications to cosmological perturbation theory.  Finally we refer to ~\cite{Boyanovsky:2018soy} for a study of entropy generation in the Yukawa theory in the de Sitter background via the influence functional technique.

In the following, as the simplest realistic scenario, we assume that the observer measures only the two point correlator for the scalar, that is our assumed system, and does not measure any other correlator for the system, the surrounding or the system and surrounding. We also assume that both the system and the surrounding are at zero temperature. A couple of clarifications are in order here. We note that the entire setup might seem to rest upon the implicit assumption that the observer cannot measure the correlations corresponding to the fermions. For the case when the fermions represent a thermal bath and the scalar is initially at zero temperature, such distinction between the system and the environment seems obvious. For the zero temperature case we have considered presently, however, such distinction might seem albeit arbitrary. For example, we do not have any explicit hierarchy of scales of physical quantities here which decides such partitioning. Thus for the zero temperature field theory, the above analysis is based upon the fact that the observer only measures the correlations for the scalar field, and the effect of fermions comes only as virtual particles inside a loop. Such implicit assumption is justified only when the system is much ‘small’ compared to the surrounding, i.e., when it can be considered as a ‘bath’ at zero temperature. This  seems to have some qualitative similarity with the standard formalism of tracing out the fermionic degrees of freedom and look into the effective scalar field dynamics. However, we would also like to emphasize that for inflationary background, observing the scalar field fluctuations/correlations seems to be very natural. Hence at least for that case we have natural way to distinguish between the system and the environment.

The rest of this chapter is structured into two primary sections, \ref{11} and \ref{22}. The first focuses on computations in the Minkowski spacetime, while the latter pertains to the computations in the inflationary de Sitter spacetime.

In \ref{section : The model} we have described the model we are considering in the Minkowski spacetime. In \ref{The Kadanoff-Baym equations} we have computed the 2-loop 2-particle irreducible (2PI) effective action using the Schwinger-Keldysh formalism and have found out the Kadanoff-Baym equations for our model in the Minkowski spacetime. In \ref{Renormalising the Kadanoff-Baym Equations}, we have renormalised the Kadanoff-Baym equations by renormalising the self-energy of the scalar field. Finally, in \ref{Phase space area and Entropy}, we have computed the statistical propagator, the phase space area and the entropy of the system, which can be thought of as a quantifier of the decoherence due to the surrounding and the subsequent ignorance  of the higher order correlators. The variation of the entropy and phase space area with respect to the rest mass of the system as well as the Yukawa coupling strength is obtained. 

Then in \ref{basic}, we provides a description of the model we are considering in the inflationary de Sitter spacetime. \ref{2PF} derives the Kadanoff-Baym equations in the de Sitter background, and as well as the statistical propagator. \ref{renorm} outlines the derivation of  the renormalised self energy, and the next subsection takes the spatial Fourier transform of it. \ref{statis} finds the one loop perturbative solution for the statistical propagator, phase space area, and the von Neumann entropy of the system. Finally, in \ref{Conclusion} we conclude our work.

We shall work in $d = 4 − \epsilon$ dimensional $(\epsilon = 0^+)$ Minkowski and inflationary de Sitter spacetime.

\section{Decoherence and entropy generation in the Minkowski spacetime}\label{11}

\subsection{The basic setup}
\label{section : The model}

We consider a hermitian massive scalar field $\phi(x)$ coupled to fermions by the Yukawa interaction, 
\begin{equation}\label{action:tree1}
S = \int \mathrm{d}^{\scriptscriptstyle{d}}\!x \left[-\frac{1}{2} (\partial_\mu\phi)
(\partial^{\mu} \phi)  - \frac{1}{2} m^{2}\phi^{2}-\frac{i}{2}{\bar{\psi}}\gamma^{\mu}\partial_{\mu}\psi-g {\bar{\psi}}\psi \phi\right]
\end{equation}
where $\gamma^{\mu}$'s satisfy the anti-commutation given in \ref{gamma}
%
%
As we have mentioned earlier, the scalar $\phi(x)$ will play the
 role of the system, interacting with the environment $\psi(x)$, $\overline{\psi}(x)$.  We assume that  the environment is at zero temperature and is in its vacuum state.  

\begin{figure}
     \centering
     \includegraphics[scale=.43]{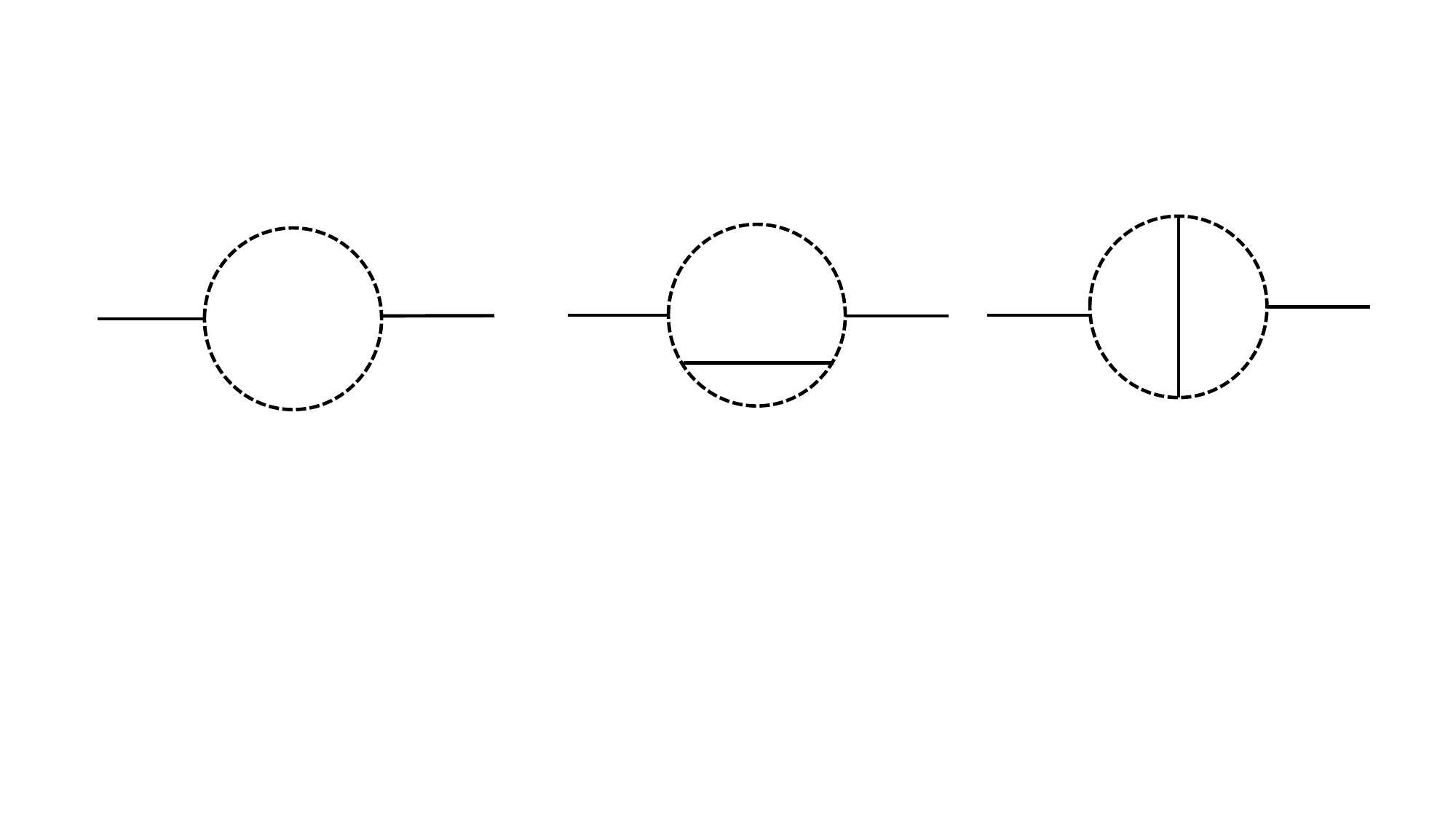}
     \vspace{-40mm}
     \caption{\small \it Self energy diagrams for the scalar field at one and two loop order for the Yukawa coupling.  The broaken lines denote fermions whereas the solid lines denote the scalar. We  shall restrict our computations to  ${\cal O}(g^2)$ only. We refer to main text for discussion. }
     \label{figa}
 \end{figure}

We note that a tree level two point function is essentially  Gaussian. A quantum  corrected such function may or may not be Gaussian.  Higher  correlations containing the effect of interaction, such as the three point correlators, are  non-Gaussian.  We have to  work on perturbative corrections to the self energy and will restrict ourselves to one loop, ${\cal O}(g^2)$. Note also that the two loop ${\cal O}(g^4)$ diagrams of \ref{figa} contains backreaction of the system on the environment,  which will be ignored. Restricting ourselves only to one loop  seems justified at least when the Yukawa coupling constant $g$ is not too strong and second, in particular when there is no secular effect at late times. Note also that since the fermions are taken to be massless, there can be decay of the system (i.e., the scalar) into fermion-anti-fermion pairs starting at ${\cal O}(g^2)$. However, this decay involves three point correlators between the system and the surrounding, with on shell process also going on within the surrounding which is unobserved by the observer.  Hence we shall ignore such decay in our computations. As we have stated in the preceding Section, the lack of observer's ability to measure higher order correlators due to the system-environment interaction essentially leads to lack of information for the system and it decoheres. This lack of information or decoherence will be quantified  by the von Neumann entropy. 

Now we wish to construct the Kadanoff-Baym equations to compute the various two-point functions, using the in-in or the Schwinger-Keldysh formalism outlined in \ref{Propagators in the Schwinger-Keldysh Formalism}.
 
\subsection{Kadanoff-Baym equations in the Minkowski spacetime}
\label{The Kadanoff-Baym equations}

The Kadanoff-Baym equations are integro-differential equations satisfied by the two-point functions in an interacting quantum field theory \cite{KadanoffBaym:1962}. It is well known that the  one particle irreducible effective action, when varied with respect to some background field, gives rise to the quantum corrected field equation. Likewise, the 2PI effective action, when varied with respect to the two point functions, yields equation of motion satisfied by them known as  the Kadanoff-Baym equations. Since the 2PI effective action contains quantum corrections, we obtain extensions of free theory equations like \ref{Feynman propagator}, essentially containing  the effect of loops.  
 One can obtain the 2PI effective action as a double Legendre transform from the generating functional for connected Green functions with respect to the linear source $J$ and also another quadratic source~\cite{NEQFT, Calzetta:1986cq, cornwall, jackiw, 2PIyukawa}. These equations contain the effects of the non-local self energy.  
 
\begin{figure}[t!]
  \centering
   \includegraphics[width=\textwidth]{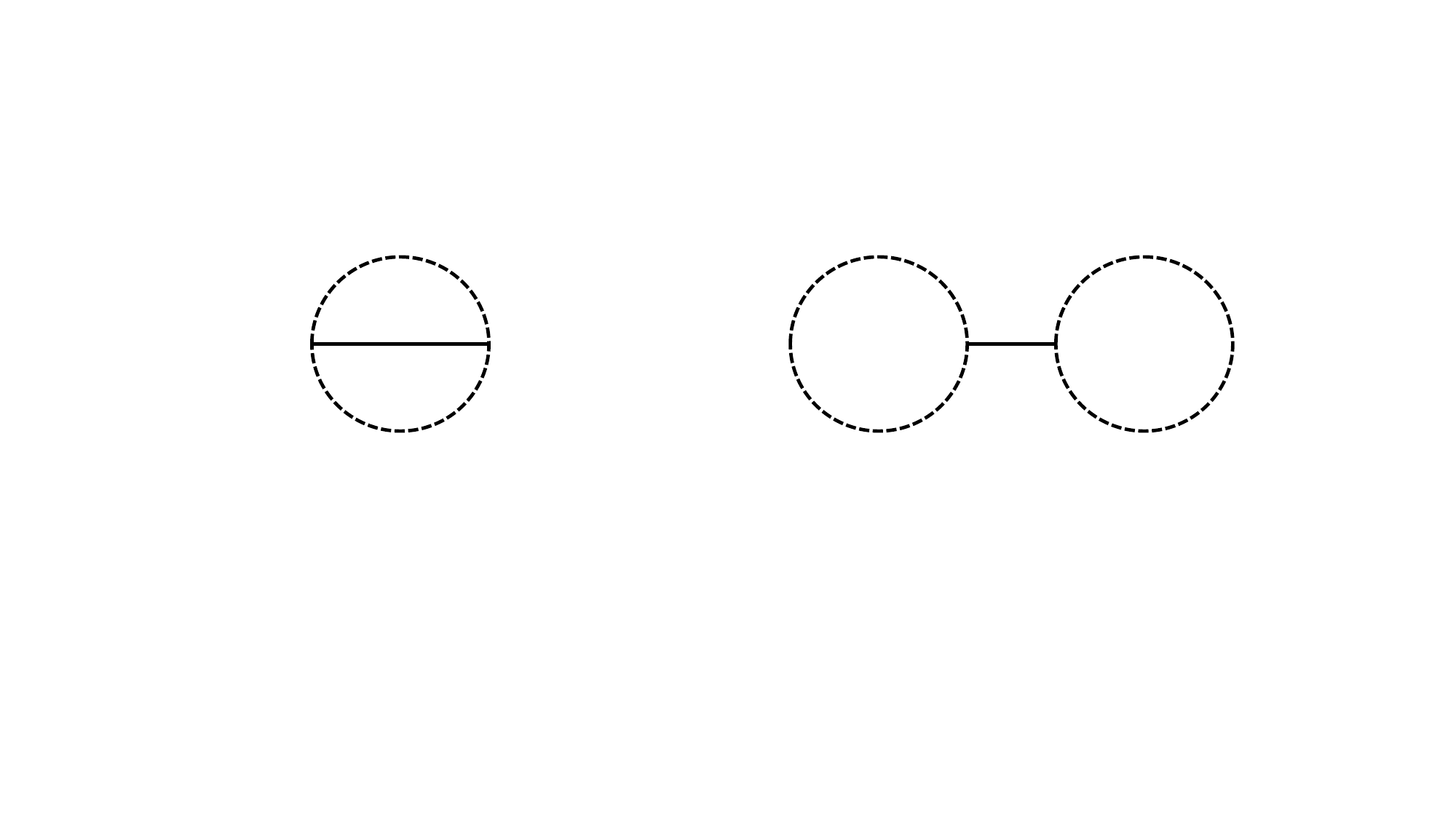}
   \vspace*{-40mm}
   \caption{\small \it Contributions to the 2PI effective action up to two
     loop order. The solid lines denote $\phi$-propagators, whereas the dashed lines correspond
     to $\psi$-propagators. The tadpoles will not make any contribution. We refer to main text for discussion.   \label{fig:2PIEfAction}}
 \end{figure}

 We expand the effective action corresponding to \ref{action:tree1} up to two loop order as
\begin{multline}\label{effectiveaction}
\Gamma[{\phi}^{s},{\psi}^{s},{\bar \psi}^s, \imath\Delta_{\phi}^{ss^{\prime}},\imath{}S^{ss^\prime}_{\psi}]
= S[{\phi}^{s},{\psi}^{s}] + \frac{\imath}{2}
\mathrm{Tr} \ln [ (\imath\Delta_{\phi}^{ss^{\prime}})^{-1}] 
{-\imath} \mathrm{Tr} \ln [
(\imath{}S^{ss^\prime}_{\psi})^{-1}] \\
  +\frac{1}{2} \mathrm{Tr} \frac{\delta^{2}\!
S[{\phi}^{s},{\psi}^{s}]}{\delta\!{\phi}^{s} \delta\!
{\phi}^{s^\prime}} \imath\Delta_{\phi}^{ss^{\prime}} -
\mathrm{Tr} \frac{\delta^{2}
S[{\phi}^{s},{\psi}^{s}]}{\delta\! {\bar \psi}^{s} \delta\!
{\psi}^{s^\prime}} \imath {}S^{ss^\prime}_{\psi} +
\Gamma^{(2)}[{\phi}^{s},{\psi}^{s},\imath\Delta_{\phi}^{ss^{\prime}},\imath {}S^{ss^\prime}_{\psi}]\quad
\end{multline}
where $s,s'=\pm$,  and  $\imath {}S^{ss^\prime}_{\psi}$ are the fermion propagators. $\Gamma^{(2)}$ denotes the 2PI contribution to the effective action at two loop, as shown in  \ref{fig:2PIEfAction}. Note however that since we are considering massless fermions, the tadpoles vanish. Even with a massive fermion, the tadpoles can be completely renormalised away in the flat spacetime. Thus we need to consider  only the sunset like diagram of \ref{fig:2PIEfAction}.
We now explicitly write down the parts of \ref{effectiveaction} containing quantum corrections as
\begin{subequations}
\label{Gamma}
\begin{align}
\hspace{-10mm}\Gamma^{(0)}[\imath\Delta_{\phi}^{ss^{\prime}},\imath {}S^{ss^\prime}_{\psi}) ] &=
\phantom{+} \int \mathrm{d}^{\scriptscriptstyle{d}}\!x
\mathrm{d}^{\scriptscriptstyle{d}}\!x' \sum_{s,s^\prime=\pm}
\frac{s}{2}(\partial_{x}^{2} -
m^{2})\delta^{\scriptscriptstyle{d}}\!(x-x') \delta^{ss^\prime}
\imath\Delta^{s^{\prime}s}_{\phi}(x',x)\label{Gamma0} \\
& - \int \mathrm{d}^{\scriptscriptstyle{d}}\! x
\mathrm{d}^{\scriptscriptstyle{d}}\! x' \sum_{s,s^\prime=\pm}
{s}\imath\slashed{\partial}_{x}\delta^{\scriptscriptstyle{d}}\! (x-x') \delta^{ss^\prime}
\imath {}S^{s^{\prime}s}_{\psi} (x,x') \nonumber
\\
\Gamma^{(1)} [\imath\Delta_{\phi}^{ss^{\prime}},\imath{}S^{ss^\prime}_{\psi} ]
&=-\frac{\imath}{2}{\rm Tr} \ln\left[\imath \Delta^{ss}_{\phi}
(x,x)\right]  +{\imath}{\rm Tr} \ln\left[\imath {}S^{ss}_{\psi} (x,x)\right] \label{Gamma1}
\\
\Gamma^{(2)}[\imath\Delta_{\phi}^{ss^{\prime}},\imath {}S^{ss^\prime}_{\psi} ] &= -
 \frac{\imath
g^{2}}{2} \int \mathrm{d}^{\scriptscriptstyle{d}}\! x
\mathrm{d}^{\scriptscriptstyle{d}}\! x' \sum_{s,s^\prime = \pm}ss^\prime
{\rm Tr}[iS^{ss'}_\psi(x,x')iS^{s's}_\psi(x',x)]
\imath\Delta^{ss^{\prime}}_{\phi}(x,x') \label{Gamma2} ,
\end{align}
\end{subequations}
where the last line corresponds to the first of \ref{fig:2PIEfAction}.

We now vary the effective action with respect to the bosonic  and the fermionic propagators, $i\Delta_{\phi}$ and $iS_{\psi}$, to obtain the equations of motion
\begin{eqnarray}
\frac{ \delta \Gamma[\imath\Delta_{\phi}^{ss^{\prime}},
\imath {}S^{ss^\prime}_{\psi}
]}{\delta \imath\Delta_{\phi}^{ss^{\prime}}} = 0, \label{EOM1a} \qquad {\rm and}\qquad 
\frac{\delta
\Gamma[\imath\Delta_{\phi}^{ss^{\prime}},\imath {}S^{ss^\prime}_{\psi} ]}{\delta
\imath {}S^{ss^\prime}_{\psi}} = 0 \label{EOM1}
\end{eqnarray}
explicitly giving respectively
\begin{subequations}
\label{EOM2}
\begin{align}
&s(\partial_{x}^{2} -
m^{2})\delta^{\scriptscriptstyle{d}}\!(x-x^\prime)
\delta^{ss^\prime}
-\imath\left[\imath\Delta^{ss^\prime}_{\phi}(x,x^\prime)\right]^{-1}
- \imath g^{2}ss^\prime
{\rm Tr}[iS^{ss'}_\psi(x,x')iS^{s's}_\psi(x',x)] = 0
\label{EOM2a} \\
&s\imath\slashed{\partial}_{x}\delta^{\scriptscriptstyle{d}}\!(x-x^\prime)\delta^{ss^\prime}
-\imath\left[\imath {}S^{ss^\prime}_{\psi}(x,x^\prime)\right]^{-1}
 -\imath
g^{2}ss^\prime \, \imath {}S^{s^\prime s}_{\psi}(x',x)
\imath\Delta^{ss^\prime}_{\phi}(x,x') = 0 \label{EOM2b} 
\end{align}
\end{subequations}

We next multiply \ref{EOM2a} and \ref{EOM2b} respectively by $s
\imath\Delta^{s^{\prime}s^{\prime\prime}}_{\phi}(x^\prime,x^{\prime\prime})$ and \\$s
\imath {}S^{s^{\prime}s^{\prime\prime}}_{\psi}(x^\prime,x^{\prime\prime})$ (from the right),
and then integrate over $x^\prime$ and sum
over $s^\prime=\pm$. Recalling that the integration over the product of the propagator and its inverse gives a $\delta$-function,   we obtain the one-loop Kadanoff-Baym equations 
\begin{subequations}
\label{EOM3}
\begin{align}
&(\partial_{x}^{2}-m^{2})\imath
\Delta^{ss^\prime}_{\phi}(x,x^\prime) -\sum_{s^{\prime\prime}=\pm}s^{\prime\prime}\int
\mathrm{d}^{\scriptscriptstyle{d}}\! x''
i M^{ss^{\prime\prime}}_{\phi}(x,x'')\imath \Delta^{s^{\prime\prime}s^{\prime}}_{\phi}(x'',x^\prime) =
\imath s\delta^{ss^\prime}\delta^{\scriptscriptstyle{d}}\!(x-x^\prime)
\label{EOM3a} \\
&\imath\slashed{\partial} \imath{}S^{ss^\prime}_{\psi}(x,x^\prime) -\sum_{s^{\prime\prime}=\pm}s^{\prime\prime}\int
\mathrm{d}^{\scriptscriptstyle{d}}\! x''
iM^{ss^{\prime\prime}}_{\psi}(x,x'')\,\imath {}S^{s^{\prime\prime}s^{\prime}}_{\psi}(x'',x^\prime) =
\imath s\delta^{ss^\prime}\delta^{\scriptscriptstyle{d}}\!(x-x^\prime)
\label{EOM3b} 
\end{align}
\end{subequations}
where  we have abbreviated the one loop self-energies as
\begin{subequations}
\label{selfMass}
\begin{eqnarray}
\imath M^{ss^{\prime\prime}}_{\phi}(x,x') &=&
 \imath g^{2} {\rm Tr}[iS^{ss'}_\psi(x,x')iS^{s's}_\psi(x',x)] =2ss^{\prime\prime}\frac{\delta\Gamma^{(2)}[\imath\Delta_{\phi}^{ss^{\prime}},
\imath {}S^{ss^\prime}_{\psi} ]}{\delta\imath\Delta^{s^{\prime\prime}s}_{\phi}}
\label{selfMassa} \\
 \imath M^{ss^{\prime\prime}}_{\psi}(x,x') &=&
   \imath g^{2} \imath {}S^{s^{\prime\prime}s}_{\psi}(x',x) \imath \Delta^{ss^{\prime\prime}}_{\phi}(x,x')=ss^{\prime\prime}\frac{\delta\Gamma^{(2)}[\imath\Delta_{\phi}^{ss^{\prime}},
\imath {}S^{ss^\prime}_{\psi} ]}{\delta\imath {}S^{s^{\prime\prime}s}_{\psi}}
\label{selfMassb} 
\end{eqnarray}
\end{subequations}
Since we are considering the scalar to be our system which is observed, we shall consider only the scalar self energy, \ref{selfMassa}. The corresponding Feynman diagram is given by the first of~\ref{figa}. Accordingly, by expanding the summations, we rewrite \ref{EOM3a} as
\begin{multline}
\label{EOM3aExtended}
(\partial_{x}^{2}-m^{2}) \imath
\Delta^{++}_{\phi}(x,x^\prime) - \int
\mathrm{d}^{\scriptscriptstyle{d}}\! y \left[\imath
M^{++}_{\phi}(x,y)\imath \Delta^{++}_{\phi}(y,x^\prime) - \imath
M^{+-}_{\phi}(x,y)\imath \Delta^{-+}_{\phi}(y,x^\prime)\right] \\ \qquad\qquad\qquad  \qquad\qquad\qquad  \qquad\qquad\qquad  \qquad\qquad\qquad = \imath\delta^{\scriptscriptstyle{d}}(x-x^\prime)
\\
(\partial_{x}^{2} - m^{2})\imath
\Delta^{+-}_{\phi}(x,x^\prime)- \int
\mathrm{d}^{\scriptscriptstyle{d}}\!y \left[\imath
M^{++}_{\phi}(x,y)\imath \Delta^{+-}_{\phi}(y,x^\prime) - \imath
M^{+-}_{\phi}(x,y)\imath \Delta^{--}_{\phi}(y,x^\prime)\right] =\,
0
\\ (\partial_{x}^{2} - m^{2})\imath
\Delta^{-+}_{\phi}(x,x^\prime) - \int
\mathrm{d}^{\scriptscriptstyle{d}}\!y \left[\imath
M^{-+}_{\phi}(x,y)\imath \Delta^{++}_{\phi}(y,x^\prime) - \imath
M^{--}_{\phi}(x,y)\imath \Delta^{-+}_{\phi}(y,x^\prime)\right] =\,
0
\\ (\partial_{x}^{2}-m^{2})\imath
\Delta^{--}_{\phi}(x,x^\prime)  - \int
\mathrm{d}^{\scriptscriptstyle{d}}\!y \left[\imath
M^{-+}_{\phi}(x,y)\imath \Delta^{+-}_{\phi}(y,x^\prime) - \imath
M^{--}_{\phi}(x,y)\imath \Delta^{--}_{\phi} (y,x^\prime)\right]
 \\=-\imath \delta^{\scriptscriptstyle{d}}(x-x^\prime)
\end{multline}
Setting $g=0$ above makes the self energies vanishing, thereby reproducing the free theory results of~\ref{Feynman propagator}.  Note also that even though the self energies appearing in the above equations are of one loop order, we may integrate these equations to find out the propagators, eventually {\it non-perturbative} in the coupling constant. In other words, the Kadanoff-Baym equations gives a framework to resum the self energies. This seems to be in particular useful in the context of the primordial cosmic inflation, where late time {\it secular effects} may be present, necessitating resummation, e.g.~\cite{Tsamis:2005hd, Cabrer:2007xm, Brunier:2004sb}.

We shall solve \ref{EOM3aExtended} by going to the momentum space. We define the Fourier transform
\begin{eqnarray}
\imath \Delta_{\phi}^{ss^{\prime}}(x,x') &=& \int
\frac{\mathrm{d}^{\scriptscriptstyle{d}}k}{(2\pi)^{\scriptscriptstyle{d}}}
 \imath \Delta_{\phi}^{ss^{\prime}}(k){\rm e}^{\imath k \cdot (x-x')}
\label{Fouriertransformdef2a} 
\end{eqnarray}
in terms of which  \ref{EOM3aExtended} become
\begin{subequations}
\label{EOM4Fourier}
\begin{eqnarray}
(-k^2-m^{2}- \imath M^{++}_{\phi}(k))
\imath \Delta^{++}_{\phi}(k) + \imath M^{+-}_{\phi}(k)
\imath \Delta^{-+}_{\phi}(k) &=& \, \imath
\label{EOM4Fourier++} \\
(-k^2 - m^{2} -\imath M^{++}_{\phi}(k)
)\imath \Delta^{+-}_{\phi}(k) + \imath
M^{+-}_{\phi}(k)\imath \Delta^{--}_{\phi}(k) &=& 0
\label{EOM4Fourier+-} \\
(-k^2 - m^{2} + \imath M^{--}_{\phi}(k)
)\imath \Delta^{-+}_{\phi}(k) - \imath
M^{-+}_{\phi}(k)\imath \Delta^{++}_{\phi}(k) &=& 0
\label{EOM4Fourier-+} \\
(-k^2-m^{2} + \imath
M^{--}_{\phi}(k))\imath \Delta^{--}_{\phi}(k) - \imath
M^{-+}_{\phi}(k)\imath \Delta^{+-}_{\phi}(k) &=&
-\imath \label{EOM4Fourier--}
\end{eqnarray}
\end{subequations}
where  $k^2=k_{\mu}k^{\mu}=-k_{0}^{2}+|\vec{k}|^{2}$. Solving these coupled algebraic equations, we can find out various two point functions. For example, on subtracting \ref{EOM4Fourier++} from \ref{EOM4Fourier-+}, we have the momentum space expression for the advanced propagator \ref{Delta:adv}, 
\begin{equation}\
\imath \Delta^{\mathrm{a}}_{\phi}(k) = \imath \Delta^{++}_{\phi}(k)-\imath \Delta^{-+}_{\phi}(k)=  \frac{-\imath}{
k^2+m^{2}+\imath
M^{\mathrm{a}}_{\phi}(k)}
 \label{FourierAdvanced1} 
\end{equation}
where the advanced self-energy $\imath M^{\mathrm{a}}_{\phi}(k)$ is given by
$$\imath M^{\mathrm{a}}_{\phi}(k) = \imath
M^{++}_{\phi}(k) - \imath
M^{-+}_{\phi}(k) =  \imath M^{+-}_{\phi}(k) - \imath
M^{--}_{\phi}(k)$$
Likewise we find the Wightman functions
\begin{subequations}
\label{FourierWightman1}
\begin{eqnarray}
\imath \Delta^{-+}_{\phi}(k) &=& \frac{ - \imath
M^{-+}_{\phi}(k) \imath
\Delta^{\mathrm{a}}_{\phi}(k)}{k^2+m^{2}+\imath
M^{\mathrm{r}}_{\phi}(k)}
 \label{FourierWightmana1} \\
\imath \Delta^{+-}_{\phi}(k) &=& \frac{ - \imath
M^{+-}_{\phi}(k) \imath
\Delta^{\mathrm{a}}_{\phi}(k)}{k^2+m^{2}+\imath
M^{\mathrm{r}}_{\phi}(k)}
\label{FourierWightmanb1} 
\end{eqnarray}
\end{subequations}
where the retarded self-energy $\imath M^{\mathrm{r}}_{\phi}(k)$ is given by
  $$\imath M^{\mathrm{r}}_{\phi}(k) = \imath
M^{++}_{\phi}(k) - \imath
M^{+-}_{\phi}(k) =  \imath
M^{-+}_{\phi}(k) - \imath M^{--}_{\phi}(k)$$
Substituting \ref{FourierAdvanced1} into 
\ref{FourierWightmana1} and \ref{FourierWightmanb1}, we finally obtain the momentum space expression of the statistical propagator \ref{Delta:causal},
\begin{equation}
\begin{split}
\label{statprop1}
F_{\phi}(k)&=-\frac{ \imath
\Delta^{\mathrm{a}}_{\phi}(k) (\imath
M^{-+}_{\phi}(k)+\imath
M^{+-}_{\phi}(k))}{2(k^2+m^{2}+\imath
M^{\mathrm{r}}_{\phi}(k))} \\ & =\frac{\imath(\imath
M^{-+}_{\phi}(k)+\imath
M^{+-}_{\phi}(k))}{2(\imath
M^{\mathrm{r}}_{\phi}(k)-\imath
M^{\mathrm{a}}_{\phi}(k))}\Bigg(\frac{1}{k^2+m^2+\imath
M^{\mathrm{a}}_{\phi}(k)}-\frac{1}{k^2+m^2+\imath
M^{\mathrm{r}}_{\phi}(k)}\Bigg)
\end{split}
\end{equation}
Recall that the statistical propagator will yield the expression of the phase space area and entropy, \ref{deltaareainphasespace}, \ref{entropy}. Thus in order to compute the statistical propagator, we need to determine various  self-energies, as appearing in \ref{statprop1}. However, note that the above expressions are not renormalised. Hence we shall use the renormalised self energies in \ref{statprop1}, in order to compute the entropy. Also in particular, note that since the self energies are ${\cal O}(g^2)$, the only 
coupling constant dependence of the above expression comes in the denominator of the terms within the parenthesis. Thus the expression for the statistical propagator is actually non-perturbative and contains  the resummed self-energy corresponding to the series of one-loop diagrams  (i.e., the first of \ref{figa}), owing to the Kadanoff-Baym equations.

\subsubsection{The retarded self-energy and its renormalisation}
\label{Renormalising the Kadanoff-Baym Equations}

In this section, we compute the self-energy $\imath M_{\phi}^{ss^\prime}(x,x')$ and find out the renormalised retarded self-energy,  $\imath M_{\phi,\mathrm{ren}}^{\mathrm{r}}(x,x')= \imath
M_{\phi,\mathrm{ren}}^{++}(x,x') - \imath M_{\phi}^{+-}(x,x')$, to be useful for our future purpose. Due to the subtraction of the two propagators, it is easily done in coordinate space. 
Necessary techniques in order to deal with such coordinate space computations can be seen in, e.g.~\cite{Miao:2006pn, woodard}. \\

The fermion propagator  $\imath {}S^{ss^\prime}_{\psi}(x,x')$ is obtained  by acting $i\slashed{\partial}$ on the massless scalar field propagator,
\begin{equation}\label{Feynmanpropposition2}
\begin{aligned}
\imath {}S^{ss^\prime}_{\psi}(x,x') = i\slashed{\partial}\left[\frac{\Gamma\left(\frac{d}{2}-1\right)}{4 \pi^{\frac{d}{2}}}\left[\Delta x_{ss^\prime}^{2}\left(x, x^{\prime}\right)\right]^{1-\frac{d}{2}}\right]
=-\frac{i\Gamma\left(\frac{d}{2}\right)}{2 \pi^{\frac{d}{2}}} \frac{\gamma^{\mu} (\Delta x_{\mu})_{ss'}}{\left[\Delta x_{ss^\prime}^{2}\left(x, x^{\prime}\right)\right]^{\frac{d}{2}}}
\end{aligned}
\end{equation}
The Poincarr\'e  invariant biscalar distance  functions with appropriate $i\epsilon$ prescription, $\Delta x_{ss^\prime}^{2}(x,x')$, necessary for the in-in formalism are defined as
\begin{subequations}
\label{x}
\begin{eqnarray}
\Delta x_{++}^{2}(x,x') &=& - \left(\left|t - t' \right| - i
\epsilon \right)^{2} + | \vec{x} -
\vec{x}'|^{2} \label{x++}  \\
\Delta x_{+-}^{2}(x,x') &=& - \left( \phantom{|}t - t'\phantom{|}
+ i \epsilon \right)^{2} + | \vec{x} - \vec{x}'|^{2}
\label{x+-} \\
\Delta x_{-+}^{2}(x,x') &=& - \left( \phantom{|} t - t'\phantom{|}
 - i \epsilon \right)^{2} + | \vec{x} - \vec{x}'|^{2}
\label{x-+} \\
\Delta x_{--}^{2}(x,x') &=& - \left(\left|t - t' \right| + i
\epsilon \right)^{2} + | \vec{x} - \vec{x}'|^{2} \label{x--}
\end{eqnarray}
\end{subequations}
The one loop scalar self-energy $\imath M_{\phi}^{++}(x,x')$ is readily found from  \ref{selfMassa} and \ref{Feynmanpropposition2}
\begin{equation}\label{SelfMassPosspace}
\imath M_{\phi}^{++}(x,x')= - 
\frac{\imath g^{2}\Gamma^{2}(\frac{d}{2})}{ \pi^{\scriptscriptstyle{d}}}
\frac{1}{ \Delta x_{++}^{2\scriptscriptstyle{d}-2}(x,x')} 
\end{equation}
Similarly, we can find out the other self-energies $\imath M_{\phi}^{--}(x,x')$, $\imath M_{\phi}^{+-}(x,x')$ and $\imath M_{\phi}^{-+}(x,x')$ using
the suitable $i\epsilon$ prescriptions as given in \ref{x}. We will now identify the divergence of \ref{SelfMassPosspace}. Note first that for an arbitrary exponent $\alpha$, we have
\begin{equation}\label{SelfMassPosspace2}
\frac{1}{\Delta x_{++}^{2 \alpha}(x,x')} = \frac{1}{4(\alpha-1)(\alpha -
\frac{d}{2})}
\partial^{2} \frac{1}{\Delta x_{++}^{2(\alpha-1)}(x,x')}
\end{equation}
Furthermore, we can write
\begin{subequations}
\label{SelfMassPosspace3}
\begin{equation}
\partial^{2} \frac{1}{\Delta x_{++}^{\scriptscriptstyle{d}-2}(x,x')} = \frac{4
\pi^{\frac{d}{2}}}{\Gamma(\frac{d-2}{2})} \imath
\delta^{\scriptscriptstyle{d}} (x-x') \label{SelfMassPosspace3a}
\end{equation}
For the other distance functions of \ref{x}, we have
\begin{eqnarray}
\partial^{2} \frac{1}{\Delta x_{--}^{\scriptscriptstyle{d}-2}(x,x')} &=& - \frac{4
\pi^{\frac{d}{2}}}{\Gamma(\frac{d-2}{2})} \imath
\delta^{\scriptscriptstyle{d}} (x-x')  \label{SelfMassPosspace3b} \\
\partial^{2} \frac{1}{\Delta x_{+-}^{\scriptscriptstyle{d}-2}(x,x')} &=& \,0=\,
\partial^{2} \frac{1}{\Delta x_{-+}^{\scriptscriptstyle{d}-2}(x,x')}  \label{SelfMassPosspace3d}
\end{eqnarray}
\end{subequations}
We now rewrite \ref{SelfMassPosspace} using
\ref{SelfMassPosspace2} and \ref{SelfMassPosspace3a} as
\begin{equation}\label{SelfMassPosspace4}
\begin{split}
\imath M_{\phi}^{++}(x,x') =& -  \frac{\imath g^{2}
\Gamma^{2}(\frac{d}{2}-1) }{16 \pi^{\scriptscriptstyle{d}}}
\frac{1}{(d-3)(d-4)} \Biggr[ \partial^{4}\left\{ \frac{1}{ \Delta
x_{++}^{2\scriptscriptstyle{d}-6}(x,x')} -
\frac{\mu^{\scriptscriptstyle{d}-4}}{ \Delta
x_{++}^{\scriptscriptstyle{d}-2}(x,x')} \right\} \\&+ \frac{ 4
\pi^{\frac{d}{2}} \mu^{\scriptscriptstyle{d}-4}
}{\Gamma(\frac{d-2}{2})} \imath \delta^{\scriptscriptstyle{d}}
(x-x') \Biggr]
\end{split}
\end{equation}
where  $\mu$ is an arbitrary mass scale. We now Taylor expand the terms inside the curly brackets around $d=4$ to obtain,
\begin{equation}\label{SelfTaylor}
\imath M_{\phi}^{++}(x,x')= - \frac{\imath g^{2}
\Gamma(\frac{d}{2}-1) \mu^{d-4} }{4
\pi^{\frac{d}{2}} (d-3)(d-4)}\partial^2 \imath\delta^{\scriptscriptstyle{d}}
(x-x') + \frac{\imath g^{2}}{32 \pi^{4}} \partial^{4}\left[
\frac{\ln(\mu^{2}\Delta x_{++}^{2}(x,x'))}{ \Delta
x_{++}^{2}(x,x')}\right] + \mathcal{O}(d-4) 
\end{equation}

 The first term of the above expression contains an ultraviolet divergence around $d=4$ which we have separated and the second term contains
a non-local contribution to the self-energy.
Since the divergence contains a $\partial^2$, we have to add a scalar field  strength renormalisation counterterm ($\phi \to (1+\delta Z)^{1/2}\phi$).  This yields in the action an additional  kinetic term 
$$\int d^d x\left[ \frac12 \delta Z \phi\, \partial^2  \phi\right]$$
where we have ignored a total divergence. The amputated version of the Feynman diagram corresponding to the above term   involves two functional differentiations with respect to the two scalar field operators, yielding two $\delta$-functions. Integrating either of them, and choosing     
\begin{equation}\label{SelfMasscounterterm}
\delta Z=  \frac{
g^{2} \Gamma(\frac{d}{2}-1) \mu^{d-4} }{4
\pi^{\frac{d}{2}} (d-3)(d-4)},
    \end{equation}
we remove the divergence of \ref{SelfTaylor}. Note that such renormalisation could also be performed at the level of the Kadanoff-Baym equations, \ref{EOM3aExtended}. In that case, the counterterm contribution is achieved by replacing the self energies $\imath M_{\phi}^{++}(x,x')$ and $\imath M_{\phi}^{--}(x,x')$, which contain divergent contributions,  by the amputated counterterm contribution corresponding to $\delta Z$ given above.

$\imath M_{\phi}^{--}(x,x')$ is given by just the complex conjugation of \ref{SelfTaylor}
\begin{equation}\label{SelfMassPosspace4Taylor}
\imath M_{\phi}^{--}(x,x')=  \frac{\imath g^{2}
\Gamma(\frac{d}{2}-1) \mu^{d-4}}{4
\pi^{\frac{d}{2}} (d-3)(d-4)}\partial^2\imath \delta^{\scriptscriptstyle{d}}
(x-x') - \frac{\imath g^{2}}{32 \pi^{4}} \partial^{4}\left[
\frac{\ln(\mu^{2}\Delta x_{--}^{2}(x,x'))}{ \Delta
x_{--}^{2}(x,x')}\right] + \mathcal{O}(d-4) 
\end{equation}
     The divergence appearing in the above expression can be tackled as above, with the counterterm \ref{SelfMasscounterterm}. Thus the  renormalised expressions for the above two self-energies are given by 
\label{SelfMassPosspace5}
\begin{eqnarray}
\imath M_{\phi,\mathrm{ren}}^{++}(x,x')=  \frac{\imath g^{2}}{32 \pi^{4}} \partial^{4}\left[
\frac{\ln(\mu^{2}\Delta x_{++}^{2}(x,x'))}{ \Delta
x_{++}^{2}(x,x')}\right] = (\imath M_{\phi,\mathrm{ren}}^{--}(x,x'))^{\star}
 \label{selfmassa} 
\end{eqnarray}
Also, choosing the suitable pole prescription from \ref{x}, $\imath M_{\phi}^{+-}(x,x')$ and $\imath M_{\phi}^{-+}(x,x')$ are determined by
\begin{eqnarray}
\label{Self5}
\imath M_{\phi}^{+-}(x,x')=   \frac{\imath g^{2}}{32 \pi^{4}} \partial^{4}\left[
\frac{\ln(\mu^{2}\Delta x_{+-}^{2}(x,x'))}{ \Delta
x_{+-}^{2}(x,x')}\right]  = (\imath M_{\phi}^{-+}(x,x'))^{\star}
\end{eqnarray}
Note that $\imath M_{\phi}^{+-}(x,x')$ and $\imath M_{\phi}^{-+}(x,x')$ do not need any renormalisation as they do not 
contain any divergence around $d = 4$, as can be readily verified  from   \ref{SelfMassPosspace3d}.
From \ref{selfmassa} and \ref{Self5},  we now have the renormalised retarded self-energy  
\begin{equation}
\begin{split}
\imath M_{\phi,\mathrm{ren}}^{\mathrm{r}}(x,x') &= \imath
M_{\phi,\mathrm{ren}}^{++}(x,x') - \imath M_{\phi}^{+-}(x,x') \\ &= 
\frac{\imath g^{2}}{32 \pi^{4}}
\partial^{4}\left[ \frac{\ln(\mu^{2}\Delta x_{++}^{2}(x,x'))}{
\Delta x_{++}^{2}(x,x')} - \frac{\ln(\mu^{2}\Delta
x_{+-}^{2}(x,x'))}{ \Delta x_{+-}^{2}(x,x')}\right]
\label{R1}
\end{split}
\end{equation}
Using now
\begin{equation}
    \label{prop1}
    \frac{\ln \Delta x^2}{\Delta x^2}=\frac{1}{8} \partial^2 (\ln^2 \Delta x^2-2\ln \Delta x^2)
\end{equation}
we can write the nonlocal terms of \ref{R1} as,
\begin{equation}\label{prop2}
\begin{split}
\frac{\ln \left(\mu^{2} \Delta x_{++}^{2}\right)}{\Delta x_{++}^{2}}-\frac{\ln \left(\mu^{2} \Delta x_{+-}^{2}\right)}{\Delta x_{+-}^{2}} & =\frac{\partial^{2}}{8}\Biggr[\ln ^{2}\left(\mu^{2} \Delta x_{++}^{2}\right)-2 \ln \left(\mu^{2} \Delta x_{++}^{2}\right)
\\& -\ln ^{2}\left(\mu^{2} \Delta x_{+-}^{2}\right)+2 \ln \left(\mu^{2} \Delta x_{+-}^{2}\right)\Biggr]
\end{split}
\end{equation}
Breaking now the logarithms into real and complex parts, we have
\begin{equation}
\begin{split}
\label{prop3}
    \ln ^{2}\left(\mu^{2} \Delta x_{++}^{2}\right)-2 \ln \left(\mu^{2} \Delta x_{++}^{2}\right)&
-\ln ^{2}\left(\mu^{2} \Delta x_{+-}^{2}\right)+2 \ln \left(\mu^{2} \Delta x_{+-}^{2}\right)\\&=4 \pi i (\ln \mu^2 (\Delta t^2 - \Delta x^2)-1)\theta(\Delta t^2-\Delta x^2)
\end{split}
\end{equation}
Putting these all in together the renormalised retarded self-energy,  \ref{R1}, takes the form
\begin{eqnarray}\label{Re1} 
\imath M_{\phi,\mathrm{ren}}^{\mathrm{r}}(x,x')=\frac{g^{2}}{64 \pi^{3}} \partial^{6}\left[\theta(\Delta
t^2-\Delta x^2)\theta(\Delta t) \left\{1 - \ln\left(\mu^{2}(\Delta
t^{2}-\Delta x^{2})\right)\right\} \right]  
\end{eqnarray}
The step functions  appearing above ensure that the retarded self-energy is non-vanishing only if $\Delta t>0$ and $\Delta
t^2-\Delta x^2>0$. This ensures the expected causal characteristics of the retarded self-energy. 

As we have stated earlier, we wish to make a momentum space computation of the entropy and hence the statistical propagator, \ref{statprop1}.   The relevant renormalised expressions, including that of various self energies are found in  \ref{A1} and \ref{A2}.   In particular, the  Fourier transform of  \ref{Re1} is done in \ref{A2}. Using these ingredients, we wish to compute the von Neumann entropy in the following Section.

\subsubsection{Phase space area and entropy}
\label{Phase space area and Entropy}
Renormalised expression for the statistical propagator is given by the renormalised version of~\ref{statprop1}
\begin{equation}
\label{statprop1'}
    F_{\phi}(k)=\frac{\imath(\imath
M^{-+}_{\phi}(k)+\imath
M^{+-}_{\phi}(k))}{2(\imath
M^{\mathrm{r}}_{\phi, {\rm ren}}(k)-\imath
M^{\mathrm{a}}_{\phi, {\rm ren}}(k))}\Bigg(\frac{1}{k^2+m^2+\imath
M^{\mathrm{a}}_{\phi,  {\rm ren}}(k)}-\frac{1}{k^2+m^2+\imath
M^{\mathrm{r}}_{\phi, {\rm ren}}(k)}\Bigg)
\end{equation}
The renormalisation of the various self-energies  has been performed in the preceding Section and \ref{A1}, \ref{A2}.
Substituting now \ref{Fourierretarded6},
 \ref{FourierWightman2} and \ref{Fourieradvance6} into the above equation, we find after some algebra
\begin{eqnarray}
F_{\phi}(k) &=& - \frac{\imath}{2} \mathrm{sgn}(k^{0})
\theta(k_{0}^{2}-|\vec{k}|^{2})
\Bigg[\frac{1}{k^2+m^{2}+\frac{g^{2}k^2}{8
\pi^{2}}\ln\left(\frac{k^2}{4\mu^{2}}
\right)- \frac{\imath
g^{2}k^2}{8\pi}\mathrm{sgn}(k^{0})\theta((k^{0})^{2}-|\vec{k}|^{2})} \nonumber \\
&& \qquad\qquad\qquad- \frac{1}{k^2+m^{2}+\frac{g^{2}k^2}{8
\pi^{2}}\ln\left(\frac{k^2}{4\mu^{2}}
\right)+ \frac{\imath
g^{2}k^2}{8\pi}\mathrm{sgn}(k^{0})\theta((k^{0})^{2}-|\vec{k}|^{2})} \Bigg]
\label{FourierStatisticals} 
\end{eqnarray}
\begin{figure}[!tbp]
  \centering
  \begin{minipage}[b]{0.45\textwidth}
    \includegraphics[scale=.7]{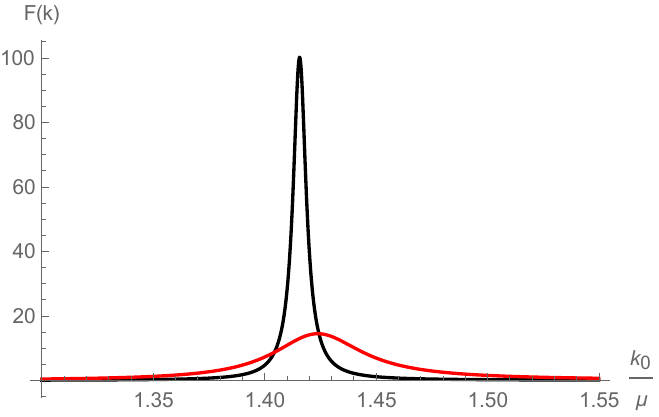}
    \caption{\small \it Variation of the statistical propagator, \ref{FourierStatisticals}, with respect to the dimensionless variable $k^0/\mu$, where we have taken $|\vec{k}|/\mu=1$, $m/\mu=1$ and the Yukawa coupling strengths $g=0.5$ (black curve), $g=1.3$ (red curve).}
     \label{fig:SP}
  \end{minipage}
  \hfill
  \begin{minipage}[b]{0.45\textwidth}
    \includegraphics[scale=.7]{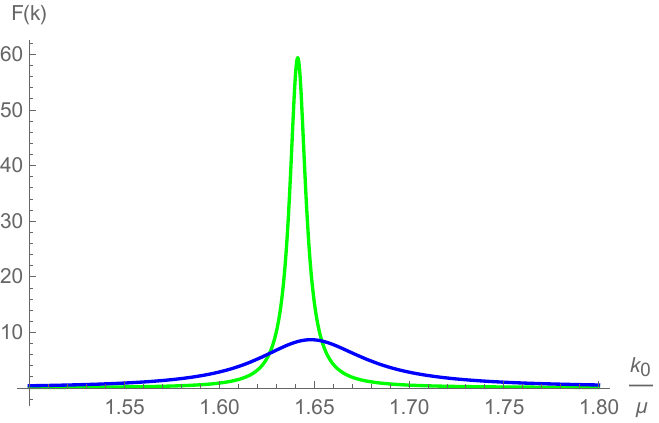}
    \caption{  \small \it Variation of the statistical propagator, \ref{FourierStatisticals}, with respect to the dimensionless variable $k^{0}/\mu$, where we have taken $|\vec{k}|/\mu=1$, $m/\mu=1.3$ and the Yukawa coupling strengths $g=0.5$ (green curve), $g=1.3$ (blue curve). Comparison with \ref{fig:SP} shows lesser numerical value of the statistical propagator with increasing $m/\mu$. We refer to main text for discussion.}
    \label{fig:SP1}
  \end{minipage}
\end{figure}
We have plotted the above statistical propagator with respect to the dimensionless  variable $k^0/\mu$ in~\ref{fig:SP} and \ref{fig:SP1}.
Note that as the Yukawa coupling $g$ gets smaller, $F_{\phi}(k)$ approaches a $\delta$-function dispersion, as expected from \ref{FourierStatisticals}.  However as $g$ increases, the $\delta$-function peak becomes broadened to a quasi-particle peak like that of the  Breit-Wigner kind. Further broadening of the peak with increasing $g$ implies that the resonance becomes broadened and we can no longer sensibly talk about a quasi-particle. Also, these plots show that with the increase in $m/\mu$, the peak gets shifted towards the higher values of $k^0/\mu$ with a decrease in the value of the statistical propagator, which means that the corresponding state becomes less populated.

Now in order to compute the entropy, we would use the definition of the same in terms of the phase space area and the three momentum, \ref{deltaareainphasespace}, \ref{entropy}. We shall assume the  mass of the scalar field to be time independent. However, one can also consider  time-dependent mass as a signature of a non-equilibrium system, as has been  considered in \cite{koksma}. A time dependent mass function will break the time translation invariance, inducing an explicit proper time dependence on the statistical propagator. We also note that in standard quantum field theories, one generally talks about the early or late times when the system is in equilibrium and it is found in some eigenstates of the free Hamiltonian. For a system which is out-of-equilibrium however, an exact distinction between energy states is unclear. A standard approach in such scenario is the adiabatic approximation, in which one specifies a reference  set of approximate states under the assumption of a slowly varying dynamical background. Using the projection of system's evolution onto these approximate states, one may hope to study its dynamics at intermediate times. There are schemes to truncate the adiabatic expansion of the system to form the aforementioned approximate basis set at intermediate times. For example in \cite{koksma, kok}, the Bogoliubov transformations are used to achieve the same.\par 

For turning on the perturbation non-adiabatically which we have not considered here, we consider replacing the coupling constant $g$ with $g\to g\,\theta(t-t_0)$. The step function changes the limit of integration \ref{EOM3aExtended} from $-\infty$ to $t_0$ and $t_0$ to some final time. The self energies vanish in the absence of interaction and would only contribute for the   second time interval.  Thus it is clear that the Schwinger-Keldysh contours presented in \ref{fig:schwingercontour1} are not equivalent to this abrupt switching case. Note that turning on the interaction abruptly may certainly lead to particle creation, which may significantly modify the statistical propagator. Some relevant discussion on this can be seen in eg. \cite{koksma}.

\begin{figure}
    \centering
    \includegraphics[scale=0.65]{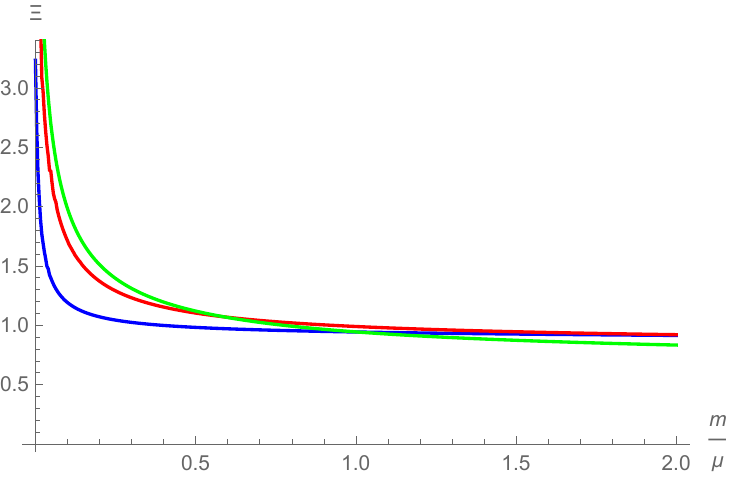}
    \caption{\small \it Variation of the dimensionless phase space area, with respect to the dimensionless mass parameter $m/\mu$ of the system, where we have taken $|\vec{k}|/\mu=1$ and, for $g=0.5$ (blue curve), $g=1$ (red curve) and $g=1.5$ (green curve). We refer to main text for discussion.}
    \label{fig:areamass}
\end{figure}

Using now the Fourier transforms, we have from \ref{FourierStatisticals}
\begin{subequations}
\label{Fconstantmass2}
\begin{eqnarray}
F_{\phi}(|\vec{k}|,0) &=& \int_{-\infty}^{\infty} \frac{\mathrm{d}k^{0}}{2\pi} F_{\phi}(k) \label{Fconstantmass2a} \\
\left.\partial_{t} F_{\phi}(|\vec{k}|, \Delta t) \right|_{\Delta t =0} &=&
- \imath \int_{-\infty}^{\infty} \frac{\mathrm{d}k^{0}}{2\pi}
k^{0}
F_{\phi}(k) \label{Fconstantmass2b} \\
\left.\partial_{t'}\partial_{t} F_{\phi}(|\vec{k}|, \Delta t)
\right|_{\Delta t =0} &=& \int_{-\infty}^{\infty}
\frac{\mathrm{d}k^{0}}{2\pi} (k^{0})^{2} F_{\phi}(k)
\label{Fconstantmass2c}
\end{eqnarray}
\end{subequations}
Substituting now \ref{FourierStatisticals} into the above integrals, we have 
evaluated them numerically. For example, for $|\vec{k}|/\mu=1$, $m/\mu=2$ and $g=0.5$, we
find the numerical value of the phase space area \ref{deltaareainphasespace}, to be
\begin{equation}
\Xi \approx 2.69,  \label{DeltaconstantMass} 
\end{equation}
which is indeed greater than unity and hence indicating a non-vanishing von Neumann entropy, as dictated by~\ref{entropy}.
We have further analysed the variation of the phase space area  with respect to the dimensionless system mass  $m/\mu$, by fixing all the other parameters in \ref{fig:areamass}. Thus the phase space area decreases monotonically with increasing $m/\mu$ and asymptotically reaches unity, indicating very small or almost vanishing entropy. This corresponds to the fact   that with the increasing mass, the system becomes more stable, i.e., it becomes difficult for the surrounding to disturb a heavier system.

From \ref{entropy}, we obtain the non-vanishing von Neumann  entropy corresponding to the phase space area of \ref{DeltaconstantMass} ($|\vec{k}|/\mu=1$, $m/\mu=2$ and $g=0.5$)
\begin{equation}
S \approx 1.27\label{SconstantMass} 
\end{equation}
\begin{figure}[!ht]
    \centering
    \includegraphics[scale=0.53]{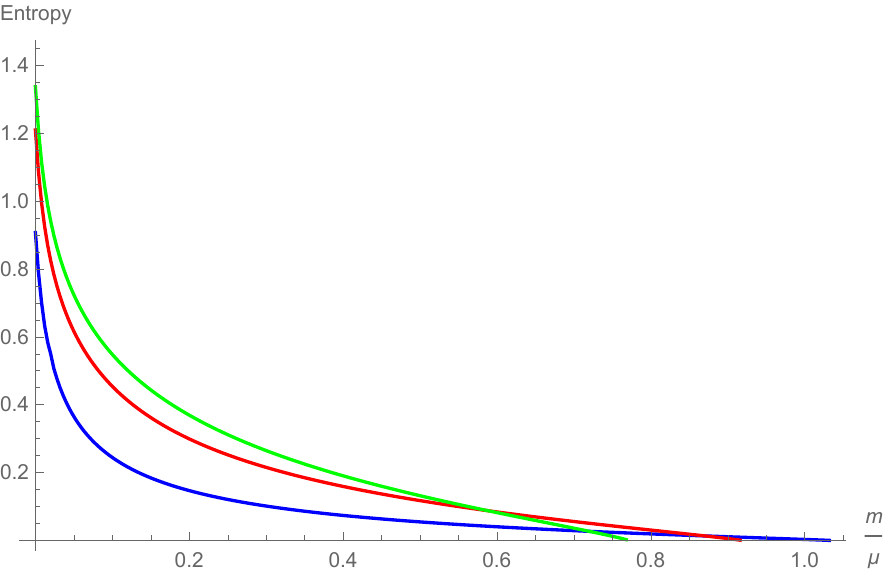}
    \caption{\small \it Variation of the von Neumann entropy with respect to the dimensionless mass parameter $m/\mu$ of the system. We have taken $|\vec{k}|/\mu=1$ and, $g=0.5$ (blue curve), $g=1$ (red curve) and $g=1.5$ (green curve). As of the phase space area, \ref{fig:areamass}, the entropy also decreases monotonically with increasing $m/\mu$. We refer to main text for discussion.   }
    \label{fig:Sconstantmass}
\end{figure}
We have also analysed the variation of the entropy with respect to the system mass by fixing all the other parameters in \ref{fig:Sconstantmass}. Due to the aforementioned reason as that of the phase space area, the entropy is decreasing with the increasing $m/\mu$. 

As we have emphasised earlier, such non-vanishing entropy is due to the
coupling between the system and the environment along with the ignorance of all  kinds of correlations between them for an observer located in the practical world.  The non-trivial statistical  propagator  gives up phase space for the system field that previously was
inaccessible to it. Such increase in the  accessible phase space for the system in turn, implies that less information about the system
field is accessible to the observer, and hence we observe a non-vanishing von Neumann entropy. Due to this reason, as our results show, the increase in the coupling strength increases the von Neumann entropy. 

Let us now proceed to the second part of this Chapter, where we study decoherence and entropy generation in the inflationary de Sitter spacetime.

\section{Decoherence and entropy generation in the inflationary de Sitter spacetime}\label{22}

\subsection{The basic setup} \label{basic}

The metric for the inflationary de Sitter spacetime, which is our interest, is given in \ref{desitterspacetime}. The cosmological time ranges from $0\leq t <\infty$, so that $-H^{-1}\leq \eta < 0^-$.
%

The action for a massless minimally coupled scalar and fermion coupled via the Yukawa interaction  reads, 
\begin{equation}
S =  \int d^d x a^d\left(-\frac12 (\nabla_\mu\phi)(\nabla^\mu\phi)   - i{\bar{\psi}}\gamma^{\mu} D_{\mu}\psi- g \bar{\psi}\psi \phi \right)
\label{action for 2 fields:dS}
\end{equation}
where $D_\mu$ is the spin covariant derivative introduced in \ref{Quantum field theory in curved spacetime}. We note that since we are working with mostly positive signature of the metric, the appropriate anti-commutation relation for the $\gamma$-matrices is given in \ref{gamma1}. 
%
As the Minkowski spacetime case, here also scalar $\phi(x)$ in \ref{action for 2 fields:dS} will play the role of the system, whereas the fermions will be the environment and assume that the environment is in its vacuum state. 

Owing to the spatial translational invariance of the de Sitter spacetime,  one can employ the $3$-momentum space. For the free scalar field satisfying $\Box \phi=0$, the temporal part of the modes then reads
\begin{eqnarray}\label{modfunc}
u(\eta, k) =  \frac{H}{\sqrt{2k^3}} (1+ i k \eta)e^{- i k \eta}    
\end{eqnarray}
along with  its complex conjugation, $u^{\star}$. Here $k=|\vec{k}|$ and $u\,(u^{\star})$ respectively corresponds to the positive (negative) frequency modes in the asymptotic past. The above modes define the Bunch-Davies vacuum state, say $|0\rangle$. 
 
 In terms of the asymptotic positive and negative frequency mode functions \ref{modfunc}, it is easy to write down a  $3$-momentum space version of the relevant propagators or two point functions, reviewed in \ref{Propagators in the Schwinger-Keldysh Formalism}. For instance for the Wightman functions, we have
\begin{align}
i \Delta_{\phi}^{\mp \pm}(\eta, \eta^{\prime},k) =  \frac{H^2}{2 k^3}(1\pm i k \eta) (1 \mp i k \eta^{\prime})e^{\mp i k (\eta-\eta^{\prime} )}  \label{BD2PF}
\end{align}

For our present purpose, apart from the usual propagators we require two more kind of two point functions, i.e.  the spectral or the causal two-point function $ \Delta^c_{\phi}$, and the Hadamard or  the statistical two-point function $F_{\phi }$. Their relationship with the Wightman functions and some other properties has been  reviewed in \ref{Propagators in the Schwinger-Keldysh Formalism}. 

From \ref{modfunc}, \ref{Delta:causal}, it is straightforward to compute
\begin{eqnarray}\label{BDDelta} 
  \Delta^c_{\phi}(\eta, \eta^{\prime},k) =  \frac{H^2}{ k^3} \Big[k(\eta-\eta^{\prime}) \cos  k (\eta - \eta^{\prime}) -(1 + k^2 \eta \eta^{\prime}) \sin  k (\eta - \eta^{\prime} ) \Big]   
\end{eqnarray}
and
\begin{eqnarray}\label{BDF} 
 F_{\phi}(\eta, \eta^{\prime},k) = \frac{H^2}{2 k^3} \Big[ (1 + k^2 \eta \eta^{\prime})\cos  k (\eta - \eta^{\prime}) +k(\eta-\eta^{\prime}) \sin  k (\eta - \eta^{\prime} )  \Big]    
\end{eqnarray}
For our purpose, we shall be computing the one-loop correction to the statistical propagator for the Yukawa interaction in the super-Hubble limit. It is also easy to see from \ref{gaussian invariant} and \ref{BDF} using the Bunch-Davies mode functions that for the free case  $\Xi_{ \phi }(\eta, k)$ becomes unity, making the von Neumann entropy vanishing. This could be attributed to the fact that for the free theory vacuum such as the Bunch-Davies, the uncertainty  is minimal. However, when interactions are introduced, the entropy undergoes perturbative corrections, due to the opening up of new phase space areas. We shall be using \ref{phase} to compute the entropy generation due to the Yukawa interaction. We shall restrict ourselves to one loop order only.
 
%
\subsection{Kadanoff-Baym equation in the inflationary de Sitter spacetime\label{2PF}}
\subsubsection{The $2$PI effective action}\label{effective}

We wish to compute below the one loop correction to the statistical propagator for  the scalar field,  \ref{Delta:causal}, using the in-in or the Schwinger-Keldysh formalism outlined in \ref{Propagators in the Schwinger-Keldysh Formalism}. Certainly, the expressions we find for the effective action here will be formally similar to that of the Minkowski spacetime \ref{11}.


For our theory~\ref{action for 2 fields:dS}, the two loop $2$PI effective action in the Schwinger-Keldysh formalism  reads

%
\begin{equation}
\Gamma[i\Delta^{ss^{\prime}}_{\phi},i S^{ss^{\prime}}_{\psi}]
 =\Gamma^{(0)}[i\Delta_{\phi}^{ss^{\prime}},i S^{ss^\prime}_{\psi}) ]+\Gamma^{(1)} [i\Delta_{\phi}^{ss^{\prime}},iS^{ss^\prime}_{\psi} ]+\Gamma^{(2)}[i\Delta_{\phi}^{ss^{\prime}},i S^{ss^\prime}_{\psi} ], \quad s ,s^{\prime} = \pm,
 \label{effac}
\end{equation}

where $i\Delta^{ss^{\prime}}_{\phi}$ are the four scalar propagators, whereas $iS^{ss^\prime}_{\psi}$ are the four fermion propagators. The three constituent functionals appearing in \ref{effac} are given by
\begin{align}
\Gamma^{(0)}[i\Delta_{\phi}^{ss^{\prime}},i S^{ss^\prime}_{\psi} ] =& \int d^dxd^dx' a^d \bigg(\nonumber
\sum_{s,s^{\prime}=\pm} \Box_x\delta^d(x-x')\frac{s \,\delta^{ss^\prime}}{2}
i\Delta^{s^{\prime}s}_{\phi}(x',x)\\
&-\sum_{s,s^{\prime}=\pm}i {\slashed D_x} \delta^d(x-x')s\delta^{ss^\prime}
i S^{s^{\prime}s}_{\psi} (x,x')\bigg)\label{effac0}\\
\Gamma^{(1)} [i\Delta_{\phi}^{ss^{\prime}},iS^{ss^\prime}_{\psi} ]
=&-\frac{i}{2}{\rm Tr}\ln \Big[i \Delta^{ss}_{\phi}(x;x')\Big]+{i}{\rm Tr} \ln \Big[i S^{ss}_{\psi} (x,x) \Big]\label{effac1}\\
\Gamma^{(2)}[i\Delta_{\phi}^{ss^{\prime}},i S^{ss^\prime}_{\psi} ] =&-\sum_{s,s^{\prime}=\pm}\frac{i ss^{\prime}
g^{2}}{2}\int d^dxd^dx' a^d a^{{\prime}d}{\rm Tr}\Big[i S^{ss'}_\psi(x,x')i S^{s's}_\psi(x',x)\Big]
i\Delta^{ss^{\prime}}_{\phi}(x,x')
\label{effac2}
\end{align}

Applying now the variational principle to \ref{effac0}, \ref{effac1} and \ref{effac2} with respect to the scalar propagators,  we obtain the following four equations of motion, i.e. the one loop Kadanoff-Baym equations, 
\begin{align}
\Box_x i\Delta^{ss^{\prime}}_{\phi}(x,x^{\prime \prime})=\frac{s \,  \delta^{ss^{\prime}}i \delta^d(x-x^{\prime \prime})}{a^d}\label{geneom}
+\sum_{s^{\prime\prime}=\pm} \int d^dx^{\prime } a^{{\prime}d}  s^{\prime\prime} \,i  M^{ss^{\prime\prime}}_{\phi}(x,x^{\prime })i\Delta^{s^{\prime\prime}s^{\prime}}_{\phi}(x^{\prime },x^{\prime \prime})
\end{align}
 Thus  the Kadanoff-Baym equations are basically the second order differential equations satisfied by the two point correlators containing the self energy correction, where the one loop scalar self energies read
\begin{equation}
(aa')^{d}iM^{ss^{\prime}}_{\phi}(x,x^{\prime})=i (aa')^d g^{2}{\rm Tr}\Big[iS^{ss^{\prime}}_\psi(x,x')iS^{s's}_\psi(x',x)\Big]\qquad ({\rm no~sum~on}~s\,{\rm or}\,s')
\label{selfMassIni}
\end{equation}
The corresponding Feynman diagram is given in \ref{figa}. Thus in order to compute the one loop correction to the  statistical propagator defined in \ref{Delta:causal}, we must determine the renormalised self energy first, which we do below. 

\subsubsection{The renormalised scalar self energy}\label{renorm}
The renormalised one loop scalar self energy for the Yukawa interaction, \ref{selfMassIni}, was first computed in \cite{Duffy:2005ue}. Since we shall express our final result in a  little bit different form and as well as eventually use the three momentum space as of \cite{Friedrich:2019hev},   we wish to  briefly keep the relevant computations here.  

Let us first express the propagators for the massless fermion field in the de Sitter spacetime, using the de Sitter  invariant and complexified  length functions
\begin{align}
y_{ss^{\prime}}(x,x') = a(\eta) a(\eta^{\prime} )  H^2  \Delta x_{ss^{\prime}}^2\big(x,x'\big) 
\label{distance}
\end{align}
where $s,s'=\pm$ as earlier and 
\begin{align}
\Delta x^2_{\pm \pm} &= - \big(| \eta \!-\! \eta^{\prime} | \mp i \varepsilon \big)^2 + | \vec{x} \!-\! \vec{x}^{\,\prime} |^2\, \label{distance1} \\
\Delta x^2_{\pm \mp} &= - \big(\eta \!-\! \eta^{\prime}  \pm i \varepsilon \big)^2 + | \vec{x} \!-\! \vec{x}^{\,\prime} |^2\, \qquad \qquad (\varepsilon=0^+)
\label{distance2}
\end{align}
are the invariant distance functions in the Minkowski spacetime, owing to the conformal flatness of the de Sitter spacetime. 

Since a massless fermion is conformally invariant and the de Sitter spacetime is conformally flat, the fermion propagator  $i {}S^{ss^\prime}_{\psi}(x,x')$ in this background is obtained  by simply acting $i\slashed{\partial}$ on the massless scalar field propagator in the flat spacetime, followed by an overall multiplication by an appropriate power of $a(\eta) a(\eta')$,  
\begin{equation}\label{Feynmanpropposition2}
\begin{aligned}
i {}S^{ss^\prime}_{\psi}(x,x') = (aa^{\prime})^{\frac{1-d}{2}} i\slashed{\partial}\left[\frac{\Gamma\left(\frac{d}{2}-1\right)}{4 \pi^{\frac{d}{2}}}\left[\Delta x_{ss^\prime}^{2}\left(x, x^{\prime}\right)\right]^{1-\frac{d}{2}}\right]
= - \frac{i(aa^{\prime})^{\frac{1-d}{2}} \Gamma\left(\frac{d}{2}\right)}{2 \pi^{\frac{d}{2}}} \frac{\gamma^{\mu} \Delta x_{\mu}}{\left[\Delta x_{ss^\prime}^{2}\left(x, x^{\prime}\right)\right]^{\frac{d}{2}}}
\end{aligned}
\end{equation}
where we have abbreviated, $a(\eta)\equiv a$ and $a(\eta')\equiv a'$. Using the anti-commutation relations for the $\gamma$-matrices appropriate for the mostly positive signature of the metric,  the one loop scalar self energy is then readily found from  \ref{selfMassIni}, \ref{Feynmanpropposition2}
\begin{equation}\label{SelfMassPosspace}
i (aa')^d M_{\phi}^{++}(x,x')= - 
(aa')^d\frac{i g^{2}(aa^{\prime})^{{1-d}} \Gamma^{2}(\frac{d}{2})}{ \pi^{\scriptscriptstyle{d}}}
\frac{1}{ \Delta x_{++}^{2d-2}(x,x')}
\end{equation}
Similarly, we can find out the other self-energies $ (aa')^di M_{\phi}^{--}(x,x')$, $(aa')^di M_{\phi}^{+-}(x,x')$ and $(aa')^d i M_{\phi}^{-+}(x,x')$  using
the suitable $i\epsilon$ prescriptions. Using now \ref{distance}, we obtain
\begin{equation}\label{SelfMassPo}
i M_{\phi}^{++}(x,x')= - 
\frac{i g^{2}\Gamma^{2}(\frac{d}{2}) H^{2d-2}}{ \pi^{\scriptscriptstyle{d}}} y_{++}^{1-d}
\end{equation}
Using next the identity for $y^{1-d}$ given in \ref{defAndConv}, we rewrite $i M_{\phi}^{++}(x,x')$ as
\begin{eqnarray}
i M_{\phi}^{++}(x,x')= -
\frac{i g^{2}\Gamma^{2}(\frac{d}{2}) H^{2d-2}}{ 2^{2d-2}\pi^{\scriptscriptstyle{d}}} \Bigg[\frac{2}{(d-2)^2} \frac{\square}{H^2} -\frac{2}{(d-2)}\Bigg] \Big( {\frac{y_{++}}{4}}\Big)^{2-{d}} 
\end{eqnarray}
where the dimensionless d'Alembertian  operator in the de Sitter spacetime reads
\begin{align}
\frac{\square}{H^2} = \eta^2 \Big[- \partial^2_{\eta} + \frac{d-2}{\eta} \partial_{\eta} + \vec{\partial}^2  \Big] 
\end{align}
Substituting the value of $\Big( {\frac{y_{++}}{4}}\Big)^{2-{d}} $ from \ref{defAndConv}, we now obtain
\begin{multline}
i M_{\phi}^{++}(x,x') = \frac{i g^{2}\Gamma^{2}(\frac{d}{2}) H^{2d-2}}{ 2^{2d-2}\pi^{\scriptscriptstyle{d}}} \Bigg[\frac{2}{(d-2)^2} \frac{\square}{H^2} -\frac{2}{(d-2)}\Bigg]  \Bigg[  \frac{\square}{H^2} \Big(\frac{4}{y_{++}}\ln \frac{\mu^2 y_{++}}{H^2}  \Big)  \\ - \frac{4}{y_{++}} \Big(2 \ln \frac{\mu^2 y_{++} }{H^2}  -1\Big) \Bigg]   
 + \mathcal{O} \big( d-4 \big) 
\label{some equation}
\end{multline} 
where $\mu$ is renormalisation scale with mass dimension one. Apart from the ultraviolet finite terms, \ref{some equation} yields a divergent local contribution to the self energy  
\begin{multline}
(aa')^d i M^{ss^{\prime}}_{\phi}(x,x')\vert_{\rm div}
  = -
\frac{i (aa')^dg^{2}\Gamma^{2}(\frac{d}{2}) H^{2d-2}}{ 2^{2d-2}\pi^{\scriptscriptstyle{d}}} \Bigg[\frac{2}{(d-2)^2} \frac{\square}{H^2} -\frac{2}{(d-2)}\Bigg] \\ \Bigg[ \frac{2 (4 \pi )^{d/2} }{(d-3)(d-4)\Gamma\big[\frac{d}{2}-1\big] } \Big(  \frac{ \mu}{H}\Big)^{d-4}  \frac{i \delta^d \big(x-x^{\prime} \big)}{(Ha)^d} \Bigg]
  s \delta^{ss^{\prime}}
\label{counterterm self energy}
\end{multline}
As was shown in~\cite{Duffy:2005ue}, this divergence can be absorbed by the scalar field strength renormalisation  {\it and} a conformal counterterm, such that their combination introduces a term in the Lagrangian density
$$  (aa')^{d/2} \left( \frac{\delta Z}{aa'} \p^2 \delta^d (x-x')\right)  $$
whose contribution, when added to the self energy, leads  us to the choice of the scalar field strength renormalisation counterterm 
\begin{equation}
\delta Z = -\frac{ g^2 \mu^{-\epsilon} \Gamma(1-\epsilon/2) }{2^2\pi^{2-\epsilon/2}\epsilon(1-\epsilon)}
\label{y13}
\end{equation}
The resulting one loop self energy after renormalisation reads
\begin{equation}
(aa')^d i M^{++}_{\phi, \text{ren}}(x,x') =  \frac{i g^{2} H^6 (aa')^d}{ 2^6\pi^4} \Bigg[\frac{\square}{2H^2} -1\Bigg]  \Bigg[  \frac{\square}{H^2} \Big(\frac{4}{y_{++}}\ln \frac{\mu^2 y_{++}}{H^2} \Big) - \frac{4}{y_{++}} \Big(2 \ln \frac{\mu^2 y_{++} }{H^2}  -1\Big) \Bigg]  
\label{renormalized self mass}   
\end{equation}
The other renormalised self energies, $i M^{ss^{\prime}}_{\phi, \text{ren}}(x,x')$ ($s,s^{\prime}=\pm$), can be found in a similar manner. However,  the self-energies of mixed kinds, i.e. those associated with the Wightman functions are free of divergences and hence do not require any renormalisation. 
\subsubsection{Self energy in the  spatial momentum space}

Following \cite{Friedrich:2019hev}, we use the identities \ref{i1}, \ref{i2} derived in \ref{defAndConv}  to rewrite  \ref{renormalized self mass} as,
\begin{multline}
\label{inhomSelfMass}
i M^{++}_{\phi, \text{ren}}(x, x^{\prime}) = \frac{i g^{2} H^6}{ 2^6\pi^4} \Bigg[ \frac{\square}{2H^2} -1\Bigg] \Bigg\lbrace \frac{1}{4}
\frac{\square^2}{H^4} \Bigg[  \frac{1}{2}\ln^2\frac{y_{++}}{4}+   \ln  \frac{4 \mu^2 }{ e H^2}  \ln \frac{y_{++}}{4} \Bigg]	
\\ -  \frac{1}{2}  \frac{\square}{H^2} \Bigg[ \frac{1}{2}\ln^2 \frac{y_{++}}{4}
+  \ln  \frac{4 \mu^2 }{ e^3 H^2}  \ln  \frac{y_{++}}{4} \Bigg] - \frac{3}{2} \ln \frac{ y_{++}}{4} 
\Bigg\rbrace 
\end{multline}
which we rewrite as 
\begin{multline}
\label{inhomSelfMass1}
i M^{++}_{\phi, \text{ren}}(x, x^{\prime}) = \frac{i g^{2} H^6}{ 2^7\pi^4}  \Bigg\lbrace \frac{1}{4}
\Bigg(\frac{\square}{H^2}\Bigg)^3 \Bigg[  \frac{1}{2}\ln^2 \frac{y_{++}}{4} + \ln \frac{4 \mu^2 }{ e H^2}  \ln \frac{y_{++}}{4} \Bigg]	
 -   \Bigg(\frac{\square}{H^2} \Bigg)^2  \Bigg[ \frac{1}{2}\ln^2 \frac{y_{++}}{4}
\\+  \ln  \frac{4 \mu^2 }{ e^2 H^2}  \ln  \frac{y_{++}}{4} \Bigg] 
+ \Bigg(\frac{\square}{H^2} \Bigg) \Bigg[ \frac{1}{2}\ln^2 \frac{y_{++}}{4}
+  \ln  \frac{4 \mu^2 }{ e^{\frac{9}{2}} H^2}  \ln \frac{y_{++}}{4} \Bigg] 
- 3 \ln \frac{ y_{++}}{4} 
\Bigg\rbrace 
\end{multline}
We now take the spatial Fourier transform of \ref{inhomSelfMass1} defined by  
\begin{align}
i M^{++}_{\phi, \text{ren}}\big(\eta, \eta^{\prime} ,k \big) = \int d^{3}\vec{r}\, i M^{++}_{\phi, \text{ren}}\big(x, x^{\prime} \big) e^{-i \vec{k} \cdot\vec{r}}
\end{align}
and use \ref{fTlogsApp}, derived in \ref{fourier}, in order to find  in the momentum space
\begin{multline}
i M^{++}_{\phi, \text{ren}} \big(\eta, \eta^{\prime},k \big)  =  \frac{i g^{2} H^6}{ 2^7\pi^4} \Bigg( -\frac{4 \pi^2}{k^3} \Bigg)\Bigg\lbrace \frac{1}{4}\Bigg(\frac{\square_k}{H^2}\Bigg)^3 \Bigg( \Bigg[2+ \big[1+ i k |\Delta \eta | \big] \Big(  \ln \frac{ 2 | \Delta \eta| \mu^2}{e k \eta \eta^{\prime} H^2 }+  \frac{i \pi}{2}- \gamma_E \Big) \Bigg] \\ \times e^{-i k |\Delta \eta|} 
- \big(1 - i k |\Delta \eta| \big)\Bigg[ \text{ci} \big[ 2 k| \Delta \eta|  \big]  -i \,  \text{si} \big[ 2 k |\Delta \eta|  \big]  \Bigg] e^{+i k |\Delta \eta|} \Bigg) \\
 - \Bigg(\frac{\square_k}{H^2}\Bigg)^2 \Bigg(\Bigg[ 2+ \big[1+ i k |\Delta \eta | \big] \Big(  \ln \frac{ | \Delta \eta|H^2}{ e^2 k \eta \eta^{\prime} \mu^2} +  \frac{i \pi}{2}- \gamma_E \Big)  \Bigg]   e^{-i k |\Delta \eta|} \\
- \big(1 - i k |\Delta \eta| \big)\Bigg[ \text{ci} \big[ 2 k| \Delta \eta|  \big]  -i \,  \text{si} \big[ 2 k |\Delta \eta|  \big]  \Bigg] e^{+i k |\Delta \eta|}  \Bigg)\\
+ \Bigg(\frac{\square_k}{H^2}\Bigg) \Bigg(\Bigg[ 2+ \big[1+ i k |\Delta \eta | \big] \Big(  \ln \frac{ | \Delta \eta|H^2}{ e^{\frac{9}{2}} k \eta \eta^{\prime} \mu^2} +  \frac{ i\pi}{2}- \gamma_E \Big)  \Bigg]   e^{-i k |\Delta \eta|} \\
- \big(1 - i k |\Delta \eta| \big)\Bigg[ \text{ci} \big[ 2 k| \Delta \eta|  \big]  -i \,  \text{si} \big[ 2 k |\Delta \eta|  \big]  \Bigg] e^{+i k |\Delta \eta|}  \Bigg)
\\+3 \Bigg[\big[1+ i k |\Delta \eta | \big]  e^{-i k |\Delta \eta|} \Bigg] \Bigg\rbrace
  \label{M++Full}
\end{multline}
where 
$\square_k$  is the d'Alembertian in the momentum space,
$$\square_k = \eta^2 \Big[- \partial^2_{\eta} + \frac{d-2}{\eta} \partial_{\eta} -\vec{k}^2  \Big] $$
and  $\rm ci$, $\rm si$ are the cosine and sine integral functions respectively, given at the beginning of  \ref{defAndConv}.
 \ref{M++Full} seems to be involved enough to be handled  in the Kadanoff-Baym equations, \ref{geneom}. However, since our chief interest is to look into the dynamics of the quantum field at late times or towards the end of inflation, we shall take its super-Hubble or the infrared limit, i.e., $\eta, \eta' \to 0$. In this limit \ref{M++Full} simplifies to  
\begin{equation}
i M^{++}_{\phi, \text{ren}} (\eta, \eta^{\prime},k)_{ k | \Delta \eta| \ll1}  \approx  \frac{i g^{2} H^6}{ 2^5\pi^2 k^3} \Bigg[ \frac{1}{4}\Bigg(\frac{\square}{H^2}\Bigg)^3 
 -\Bigg(\frac{\square}{H^2}\Bigg)^2 
+\Bigg(\frac{\square}{H^2}\Bigg) \Bigg] \Bigg( \ln \frac{H^2 k^2 \eta \eta^{\prime}}{\mu^2} + 2 i k  |\Delta \eta |  \Bigg)
  \label{M++Full15}
\end{equation}
Note that unlike the case of a massless minimal scalar field \cite{nitin, Bhattacharya:2022aqi}, the above self energy is devoid of any secular logarithm  of the scale factor \cite{Miao:2006pn}, $a=-1/H\eta$. This is not surprising, as the loop we have computed consists of fermion propagators only. 
Using \ref{M++Full15}, we shall now  compute the statistical propagator and the von Neumann entropy in the next section. 

\subsubsection{Perturbative solution for the statistical propagator and the von Neumann entropy} \label{statis}

Let us look at the renormalised version of  equations of motion \ref{geneom} for the propagators $i \Delta_{\phi}^{ss^{\prime}}(x, x^{\prime }) $.
By rewriting the two-point functions in terms of real and imaginary parts, we obtain after using \ref{reduction:F+Deltac} into \ref{geneom}
\begin{multline}
\square_{x} F_{\phi} (x,x^{\prime \prime})  = \frac{i}{2} \int  d^4 x^{\prime }\big(\eta^{\prime} H \big)^{-4}\, \Big[    M^{++}_{\phi,\text{ren}}(x, x^{\prime })  -   M^{--}_{\phi,\text{ren}}(x, x^{\prime }) +   M^{-+}_{\phi,\text{ren}}(x, x^{\prime }) -\\   M^{+-}_{\phi,\text{ren}}(x, x^{\prime })   \Big] F_{\phi} (x^{\prime }, x^{\prime\prime} )  - \frac{1}{4} \int   d^4 x^{\prime }\big(\eta^{\prime} H \big)^{-4}\, \Big[\text{sign} (\eta^{\prime} - \eta^{\prime \prime} ) \Big(    M^{++}_{\phi,\text{ren}}(x, x^{\prime })  \\+    M^{--}_{\phi,\text{ren}}(x, x^{\prime })\Big) -   M^{-+}_{\phi,\text{ren}}(x, x^{\prime }) -   M^{+-}_{\phi,\text{ren}}(x, x^{\prime })   \Big]  \Delta_{\phi}^c (x^{\prime }, x^{\prime\prime} )  \label{statEQX}
\end{multline}

\begin{figure}[!ht]
    \centering
    \includegraphics[scale=0.58] {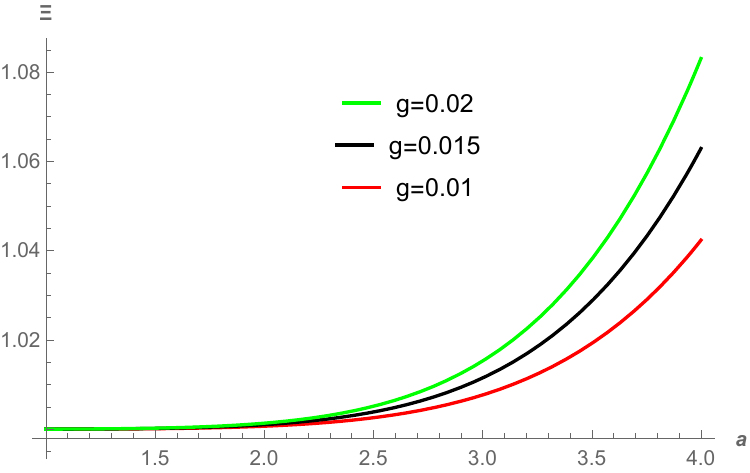}\hspace{1.0cm}
     \includegraphics[scale=0.58] {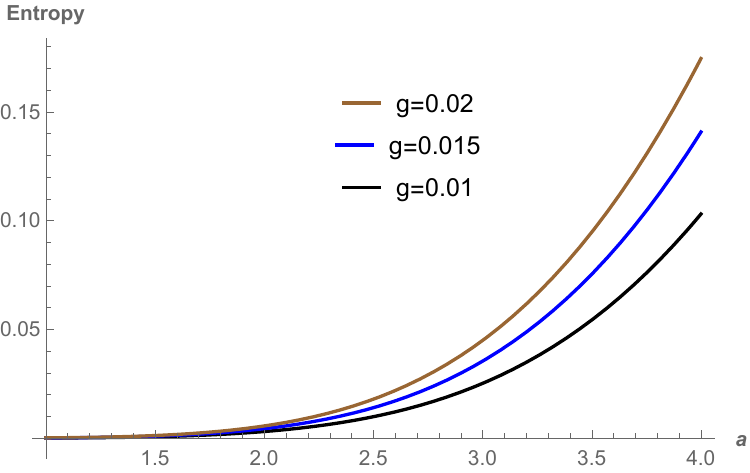}
    \caption{\small \it Variation of the phase space area (left) and the von Neumann entropy (right) with respect to the scale factor. Both  increase monotonically with increasing $a$ and the Yukawa coupling, as expected. }
    \label{fig:phasespace}
\end{figure}

 We wish to solve the above equation perturbatively for the statistical propagator at ${\cal O}(g^2)$, using the one loop self energy.  
In order to do this, we will  substitute the expressions from the free theories given by \ref{BDDelta}, \ref{BDF} for $\Delta_{\phi}^c$ and $ F_{\phi}$, on the right-hand side of \ref{statEQX}. We have  in the momentum space
\begin{multline}
\square_k F_{\phi} (\eta,\eta^{\prime \prime},k)\approx  \int_{- 1/H}^{\eta} d \eta^{\prime} \big(\eta^{\prime} H \big)^{-4}\, M^c(\eta, \eta^{\prime} ,k) F_{\phi} (\eta^{\prime }, \eta^{\prime\prime},k )  \\ + \int_{-1/H}^{\eta^{\prime \prime}} d \eta^{\prime} \big(\eta^{\prime} H \big)^{-4}\, M^F(\eta, \eta^{\prime },k)  \Delta_{\phi}^c (\eta^{\prime }, \eta^{\prime\prime},k )
\end{multline}
where we have used \ref{energies} and \ref{energies1} given in \ref{defAndConv}. We note that while utilising \ref{energies1} into \ref{statEQX}, it becomes evident that $\eta \gtrsim \eta^{\prime}$ and $\eta^{\prime} \lesssim \eta^{\prime \prime}$, consequently determining the upper limits of the above integration. The lower limit is set to be $-1/H$, as explained in \ref{basic}. Using now \ref{newenergies} and \ref{newenergies1} in above equation, we obtain
\begin{multline}
\square_k F_{\phi} (\eta,\eta^{\prime \prime},k)  \approx -2  \int_{- 1/H}^{\eta} d \eta^{\prime} \big(\eta^{\prime} H \big)^{-4} \text{Im} M^{++}(\eta, \eta^{\prime} ,k) F_{\phi} (\eta^{\prime }, \eta^{\prime\prime},k )\\  - \int^{\eta^{\prime \prime} }_{-1/H} d \eta^{\prime} \big(\eta^{\prime} H \big)^{-4}\,  \text{Re} \,  M^{++}(\eta, \eta^{\prime} ,k) \Delta_{\phi}^c (\eta^{\prime }, \eta^{\prime\prime},k ) \label{eqnF}
 \end{multline}
Since we shall use the free theory results for the propagators appearing on the right hand side, we may solve the above equation by using a Green function. The causal result corresponds to the retarded  Green function $G_{\text{ret}}(\eta , \eta^{\prime} , k)$, satisfying in the momentum space  
\begin{align}
\square_k G_{\text{ret}}(\eta , \eta^{\prime} , k) = H^2 \eta^2 \Big[  - \partial_{\eta}^2   + \frac{d-2}{\eta} \partial_{\eta} - k^2 \Big]G_{\text{ret}}(\eta , \eta^{\prime} , k) =  a^{-4}(\eta^{\prime}) \delta(\eta-\eta^{\prime}) 
\end{align}
which can be easily solved by using the free theory mode functions, e.g.~\cite{Friedrich:2019hev},
\begin{eqnarray}
G_{\text{ret}}(\eta , \eta^{\prime} , k) = \theta(\eta - \eta^{\prime}) \frac{H^2}{k^3} \Big[ k(\eta - \eta^{\prime}) \cos  k(\eta - \eta^{\prime} )  -(1 + k^2 \eta \eta^{\prime} ) \sin k(\eta - \eta^{\prime}) \Big],
\end{eqnarray}
which, in the late time or super-Hubble limit we are interested in  becomes
\begin{eqnarray}
G_{\text{ret}}(\eta , \eta^{\prime} , k)_{ k | \Delta \eta| \ll1} = -\theta(\eta-\eta^\prime) \frac{H^2 | \Delta \eta|^3}{3}  
\end{eqnarray}
After substituting the real and imaginary parts of $i M^{++}_{\phi, \text{ren}}$ from \ref{M++Full15} into \ref{eqnF}, we find the one loop corrected statistical propagator in the supper-Hubble limit
\begin{multline}
F_{\phi} (\eta,\eta^{\prime \prime},k)_{ k | \Delta \eta| \ll 1}   =  \frac{1}{H^2}\left[-\frac{1}{4} \Bigg(\frac{\square_k}{H^2} \Bigg)^2 + \Bigg(\frac{\square_k}{H^2} \Bigg)  - 1 \right]\mathcal{O}(\eta, \eta^{\prime},k)_{ k | \Delta \eta| \ll 1} \\ + H \int_{-\infty}^{\infty}d\tau \frac{G_{\text{ret}}(\eta, \tau,k)_{ k | \Delta \eta| \ll 1} }{(\tau H)^4}+ F_{\text {free}} (\eta ,\eta^{\prime \prime},k)   \label{eqnFInv}
\end{multline}
where we have abbreviated 
\begin{multline}
\mathcal{O}(\eta, \eta^{\prime \prime},k)_{ k | \Delta \eta| \ll 1}  = - \frac{ g^{2} H^6}{ 2^3\pi^2 k^3}   \int_{- 1/H}^{\eta} d \eta^{\prime} \big(\eta^{\prime} H \big)^{-4} k  |\Delta \eta | F_{\phi} (\eta^{\prime }, \eta^{\prime\prime},k )_{{\rm free}, k | \Delta \eta| \ll 1}  \\- \frac{ g^{2} H^6}{ 2^5\pi^2 k^3} \int^{\eta^{\prime \prime} }_{-1/H} d \eta^{\prime} \big(\eta^{\prime} H \big)^{-4}\ln \frac{H^2 k^2 \eta \eta^{\prime}}{\mu^2}  \Delta_{\phi}^c (\eta^{\prime}, \eta^{\prime\prime},k )_{{\rm free}, k | \Delta \eta| \ll 1}   \label{BLog}
\end{multline}
and
$
\square_k F_{\text{free}} (\eta ,\eta^{\prime \prime},k)=0$ is the free theory part of the statistical propagator, \ref{BDF}. The expressions for  $F_{\phi} (\eta^{\prime }, \eta^{\prime\prime},k )_{ {\rm free}, k | \Delta \eta| \ll 1}$ and $\Delta_{\phi}^c (\eta^{\prime }, \eta^{\prime\prime},k )_{{\rm free}, k | \Delta \eta| \ll 1}$  in the infrared or super-Hubble  limit are explicitly given in \ref{nc4}, \ref{corr1}. Substituting everything now into \ref{eqnFInv} and integrating, we have the one loop result
\begin{multline}
F_{\phi} (\eta,\eta^{\prime \prime},k)_{ k | \Delta \eta|\ll 1}\vert_{\rm 1\,loop} \approx \frac{g^{2} H^2}{768 \pi ^2 k^3}\Bigg[ \ln \frac{\eta  H^2 k^2 \eta ''}{\mu ^2} \Bigg(\big(\eta ^4 k^4+2 \eta ^2 k^2+4\big) \ln \frac{\eta  H^2 k^2 \eta ''}{\mu ^2}\\ -4
   \big(\eta ^2 k^2+6\big)\Bigg)
   +4 \left(\eta ^2 k^2+6\right) \ln \left(-\frac{\eta  H k^2}{\mu ^2}\right)-\left(\eta ^4 k^4+2 \eta ^2 k^2+4\right) \ln
   ^2\left(-\frac{\eta  H k^2}{\mu ^2}\right)\Bigg] \\ + F_{\text {free}} (\eta ,\eta^{\prime \prime},k)
\end{multline}
We now substitute it into \ref{phase} in order to obtain the increase in the phase space area 
\begin{eqnarray}
\delta \left(\frac{ \Xi_\phi ^2}{4a^4} \right) \approx \frac{g^{2} H^4 \left(\eta ^2 k^2 \ln \frac{\eta ^2 H^2 k^2}{\mu ^2} \left(\eta ^2 k^2 \ln \frac{\eta ^2 H^2 k^2}{\mu ^2}-4\right)+2 \eta ^2
   k^2+2\right)}{384 \pi ^2 \eta ^2 k^6}
\end{eqnarray}
Converting $\eta$ now in terms of the scale factor $a$,  and keeping the dominant terms only, we have
\begin{eqnarray}
\delta \Xi_\phi ^2\approx \frac{ g^{2} H^6 a^6}{48 \pi ^2 k^6}  
\end{eqnarray}
By taking (for example) $k/H\sim1$, we have  plotted the above variation  in \ref{fig:phasespace} for different values of the Yukawa coupling. Similar behaviour was reported in \cite{Friedrich:2019hev} for cubic interactions in inflation in the super-Hubble limit.
Now we use the \ref{entropy} to compute the von Neumann entropy, which is also plotted in \ref{fig:phasespace}. Expectedly, the von Neumann entropy also increases monotonically with increasing scale factor and the coupling constant,  similar to that of the phase space area.

\section{Discussion}
\label{Conclusion}



In this Chapter, we have explored the phenomenon of decoherence in a quantum field theory consisting of scalar and fermions coupled via the Yukawa interaction, and using the correlator approach proposed in ~\cite{JFKTPMGS, koksma, kok}. We begin started with the simplest scenario in the Minkowski spacetime in \ref{11}, and then subsequently extended our analysis to the inflationary de Sitter spacetime in \ref{22}. We have considered the scalar field as the system, while the fermions as the environment in both cases. The scalar field is considered to be massive in the Minkowski spacetime and massless in de Sitter spacetime, while the fermions are assumed to be massless in both cases. We assumed that the observer only measures the Gaussian 2-point correlator of the scalar field. As mentioned in \ref{deco}, the lack of knowledge about the higher order correlators, both Gaussian and non-Gaussian, of the system and its surroundings is the main cause of decoherence and the subsequent generation of entropy. To explore this decoherence process, we have employed the 2-loop 2-particle irreducible effective action and derived the renormalised Kadanoff-Baym equations, which describes the dynamics of the 2-point correlators in the Schwinger-Keldysh formalism, in both spacetimes. These equations include self-energy corrections. We then compute the statistical propagator in terms of the 2-point functions. Using the relationship of the statistical propagator with the phase space area, introduced in \ref{Motivation and Overview}, we have calculated the von Neumann entropy. We note that, in the Minkowski spacetime we have been able to resum the one loop scalar self energy contributions and found a non-perturbative result whereas in the inflationary de Sitter spacetime our results are perturbative in the coupling constant. Our observations for both spacetimes are as follows:\\

\noindent \textbf{In the Minkowski spacetime:} for a fixed value of all the relevant parameters such as mass, coupling, momentum and energy, we observe a higher numerical value of phase space area and entropy in (cf., \ref{Phase space area and Entropy}), in contrast to the scenario when both the system and the surrounding are scalars~\cite{koksma}. This is due to the fact that the Yukawa coupling is dimensionless whereas the cubic coupling considered in~\cite{koksma} is dimensionfull, leading to different momentum dependence in the statistical propagator~\ref{FourierStatisticals}. Although we note that the qualitative aspects of the variation  of the entropy (e.g.~\ref{fig:Sconstantmass}) for both the cases are similar.

\noindent \textbf{In the inflationary de Sitter spcetime:} the phase space area and entropy turns out to be monotonically increasing with respect to the time (the scale factor), as well as the Yukawa coupling. This is in the qualitative agreement with the case when both the system and the surrounding are scalar fields. Our result is also in the qualitative agreement with
the result where such entropy generation was computed using the Feynman-Vernon influence functional technique \cite{Boyanovsky:2018soy}. The increase in the entropy with the scale factor corresponds to the fact that the phase space area must increase with time in an expanding spacetime. The increase of the entropy with respect to the Yukawa coupling must correspond to the opening up of new phase space areas due to the interaction. This feature is qualitatively similar to that of the Minkowski spacetime, even though this result in the Minkowski spacetime was non-perturbative.

As we have previously indicated, our analysis has been limited to \({\cal{O}}(g^2)\), for the de Sitter. At this particular order, we did not encounter any {\it secular effects}. However, if we attempt to extend our investigation to higher loop orders, for example considering the second and third of \ref{figa}, we expect to surely encounter the emergence of {\it secular} logarithm terms, necessitating a non-trivial resummation procedure. This seems to be an important task, which we reserve for future.

 

\chapter{Entanglement degradation in a cosmological black hole spacetime}
\label{chapter4}

In this Chapter we study the phenomenon of entanglement degradation or survival in the Schwarzschild-de Sitter (SdS) black hole spacetime. As we mentioned in \ref{Motivation and Overview}, this spacetime represents a static and spherically symmetric black hole sitting in the de Sitter universe. We will work here in (1 + 1)-dimensions for simplicity, focusing on a free, massless and minimally coupled scalar field. Earlier such investigations on entanglement chiefly involve the Rindler or non-extremal black holes and Friedmann-Lemaitre-Robertson-Walker cosmological spacetimes e.g.~\cite{Fuentes:2010dt, Bhattacharya:2019zno, mann, Pan:2008yi, martinez, martin, Montero:2011sx, richter, Asghari:2018} and references therein. In the Rindler or a maximally extended non-extremal black hole background, particle creation occurs in causally disconnected spacetime wedges. Since the created particles are thermal, the associated randomness destroys the entanglement or quantum correlation between entangled states as the black hole evaporates and gets hotter, first shown in~\cite{mann} and subsequently further explored in e.g.~\cite{Fuentes:2010dt, mann, Pan:2008yi, martinez, martin, Montero:2011sx, richter, Asghari:2018}.  

To the best of our knowledge, all such earlier studies were made in  asymptotically flat spacetimes. However, keeping in mind the observed accelerated expansion or the possible dark energy domination of our current universe, it is physically important to ask: how does such degradation get affected in black hole spacetimes endowed with a positive cosmological constant, $\Lambda$? We note that such spacetimes can also model primordial black holes formed in the early inflationary universe, e.g.~\cite{Gibbons}. The chief qualitative difference of these  black holes with that of $\Lambda\leq 0$ is the existence of the cosmological event horizon for the former, an additional event horizon serving as the outer causal boundary of our universe. For the empty de Sitter universe, we have mentioned the cosmological event horizon in \ref{The de Sitter spacetime} (\ref{deSitter-penrose}). The non zero surface gravities of the black hole event horizon (BEH) and cosmological event horizon (CEH) give rise to the individual particle creation. Usually these two surface gravities are different thereby making the Schwarzschild-de Sitter spacetime a two temperature system. In other words, these two-event horizon spacetimes admit two-temperature thermodynamics and hence are qualitatively much different compared to the single horizon $\Lambda \leq 0$ cases, e.g.~\cite{Gibbons, Kastor, Bousso, Bousso2, JHT, Choudhury:2004ph, bhatta, Bhattacharya:2018ltm, qiu, Goheer, Park, Dolan1, Dolan2, Maeda, Davies, Urano, Saida1, Saida2,  Bhattacharya:2015mja, Zhang, Kanti1, Kanti2, Kanti3, pappas, Das:2000zs}. With this motivation,  we wish to investigate in this Chapter the role of this two temperature particle creation in the entanglement degradation. Our chief goal here is to see whether in this physically well motivated spacetime, the multi-horizon structure brings in any {\it qualitatively new} feature compared to that of the single horizon, i.e. the $\Lambda \leq 0$ cases.

The rest of this Chapter is organised as follows. In the next section we outline very briefly the causal structure of the Schwarzschild-de Sitter spacetime, a static and spherically symmetric black hole  located in the de Sitter universe. In \ref{S3}, we discuss the entanglement degradation in the thermodynamical setup proposed in~\cite{Gibbons},  where an observer can be in thermal equilibrium with either of the horizons and show that the results qualitatively resemble with that of the single horizon  spacetimes. We use  the mutual information and logarithmic negativity for a maximally entangled, bipartite Kruskal-like state corresponding to massless minimal scalar fields as appropriate measures. In \ref{s4}, we adopt the so called total entropy-effective equilibrium temperature description to treat both the horizons combined, e.g.~\cite{Maeda, Bhattacharya:2015mja}. We first present a field theoretic derivation of the effective temperature and show that  unlike the previous cases, the entangled pair creation in this scenario {\it does not} occur in causally disconnected wedges in the extended spacetime. Most importantly, we demonstrate that   the entanglement here actually {\it increases} with the increase in the black hole Hawking temperature, no matter how hot the black hole becomes or how small the cosmological constant is. We emphasise that this phenomenon  is purely an outcome of the two-event horizon geometry and hence has no $\Lambda \leq 0$ analogue.

\section{The basic setup}\label{S2}
\noindent
We consider the Schwarzschild-de Sitter (SdS) spacetime, 
\begin{eqnarray}
ds^2=-\left(1-\frac{2M}{r}-\frac{\Lambda r^2}{3}\right)dt^2+\left(1-\frac{2M}{r}-\frac{\Lambda r^2}{3}\right)^{-1}dr^2+r^2 \left(d\theta^2 +\sin^2\theta d\phi^2 \right)
\label{l1}
\end{eqnarray}
which admits three event or Killing horizons for $0<3M \sqrt{\Lambda} < 1$, e.g.~\cite{Gibbons, JHT}, 
\begin{equation}
r_{H}=\frac{2}{{\sqrt \Lambda} }\cos\frac{\pi+\cos^{-1}(3M\sqrt{\Lambda})}{3},~
r_{C}=\frac{2}{{\sqrt \Lambda} }\cos\frac{\pi-\cos^{-1}(3M \sqrt{\Lambda})}{3},~r_{U}=-(r_H+r_C)
\label{l2}
\end{equation}
$r_H < r_C$ are respectively the black hole and the cosmological event horizon (BEH and CEH), whereas $r_{U}<0$ is unphysical. As $3M\sqrt{\Lambda} \to 1$ we have $r_H \to r_C$, known as the Nariai limit whereas for $3M\sqrt{\Lambda}>1$,  the spacetime is naked singular. Thus unlike $\Lambda \leq 0$, a black hole cannot  be arbitrarily massive here, for a given $\Lambda$. If we set $M=0$ in \ref{l1}, we reproduce the empty de Sitter universe discussed in \ref{The de Sitter spacetime}, whereas setting $\Lambda = 0$ reproduces the Schwarzschild spacetime. We also note that the static form of the Schwarzschild-de Sitter metric \ref{l1}, holds only in the region $r_H<r<r_C$.

The surface gravities of BEH and CEH are respectively given by,
\begin{eqnarray}
\kappa_H=    \frac{\Lambda (2r_H+r_C)(r_C-r_H)}{6 r_H}, \quad 
-\kappa_C=\frac{\Lambda (2r_C+r_H)(r_H-r_C)}{6 r_C}
\label{l3}
\end{eqnarray}
Due to the repulsive effects generated by a  positive $\Lambda$, the surface gravity of CEH is negative.

Since $r=r_H,\,r_C$ are two coordinate singularities of \ref{l1}, we need two Kruskal-like coordinates in order to  extend the spacetime beyond them, given by 
\begin{eqnarray}
ds^2=-\frac{2M}{r}\left\vert1-\frac{r}{r_C}\right\vert^{1+\frac{\kappa_H}{\kappa_C}} \left(1+\frac{r}{r_H+r_C}\right)^{1-\frac{\kappa_H}{\kappa_U}}\, d{\overline u}_H d {\overline v}_H+r^2(d\theta^2+\sin^2\theta d\phi^2)
\label{ds16}
\end{eqnarray}
and
\begin{eqnarray}
ds^2=-\frac{2M}{r}\left\vert\frac{r}{r_H}-1\right\vert^{1+\frac{\kappa_C}{\kappa_H}} \left(1+\frac{r}{r_H+r_C}\right)^{1+\frac{\kappa_C}{\kappa_U}}\, d {\overline u}_C d {\overline v}_C+r^2(d\theta^2+\sin^2\theta d\phi^2)
\label{ds17}
\end{eqnarray}
where,
\begin{eqnarray}
{\overline u}_H=-\frac{1}{\kappa_H}e^{-\kappa_H u},\quad {\overline v}_H=\frac{1}{\kappa_H}e^{\kappa_H v} \quad {\rm and} \quad
{\overline u}_C=\frac{1}{\kappa_C}e^{\kappa_C u},\quad {\overline v}_C=-\frac{1}{\kappa_C}e^{-\kappa_C v}
\label{ds15}
\end{eqnarray}
are the Kruskal null coordinates whereas $u=t-r_{\star}$ and $v=t+r_{\star}$ are the usual retarded and advanced null coordinates. The radial tortoise coordinate $r_{\star}$ is given by
\begin{eqnarray}
r_{\star}= \int \frac{dr}{(1-\frac{2M}{r}-\frac{\Lambda r^2}{3})} = \frac{1}{2\kappa_H}\ln \left\vert\frac{r}{r_H}-1\right\vert -\frac{1}{2\kappa_C} \ln \left\vert1-\frac{r}{r_C}\right\vert +\frac{1}{2\kappa_U}\ln \left\vert\frac{r}{r_U}-1\right\vert
\label{ds5}
\end{eqnarray}
$\kappa_U$ is the `surface gravity' of the unphysical horizon located at $r_U =-(r_H+r_C)$. We note that \ref{ds16} and \ref{ds17} are  free of coordinate singularities respectively on the BEH and CEH. We also note that none of them are free from the singiularity at $r=0$. This corresponds to the fact that just like the Schwarzschild spacetime, the SdS has also a genuine or curvature singularity at $r=0$, as can be seen easily by computing the curvature invariant 
$$
R_{\mu\nu\lambda\rho}R^{\mu\nu\lambda\rho} = \frac{48M^2}{r^6} + \frac{\Lambda^2}{4}
$$

\begin{figure}[h!]
\centering
  \includegraphics[width=12cm]{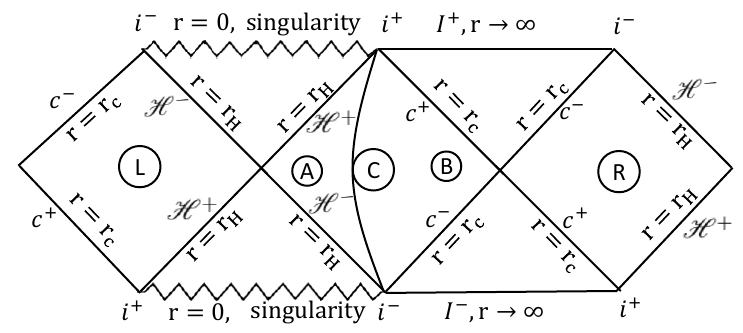}
  \caption{\small \it  The Penrose-Carter diagram of the maximally extended Schwarzschild-de Sitter spacetime. ${\cal H }^{\pm}({\cal C}^{\pm})$ respectively denote the future and past black hole event horizons (cosmological event horizons). 
$i^{\pm}$ respectively represent the future and past timelike infinities, whereas the infinities $ I^{\pm}$, like the de Sitter \ref{deSitter-penrose}, are spacelike. The regions R, L are time reversed with respect to C and all the seven wedges are causally disconnected. The spacetime can further be extended towards both sides indefinitely, but we do not require this for our current purpose. Our region of interest is C $(r_H< r < r_C)$ and hence we shall trace over the states belonging to other regions when it is  relevant. The hyperbola joining $i^{\pm}$ is a thermally opaque membrane separating the region C into two subregions A and B $({\rm C}={\rm A}\cup {\rm B})$, as discussed in \ref{S3}. The hyperbola is drawn with respect to the black hole Kruskal coordinate (the first of \ref{ds5'}), but can be drawn with respect to cosmological Kruskal coordinates as well. In ~\ref{s4} however, such membrane will not be considered.}
  \label{fl1}
\end{figure}

Finally we also note that there is no single Kruskal coordinate for the SdS spacetime that simultaneously removes the coordinate singularities on both the horizons. Defining now the Kruskal timelike and spacelike coordinates as,
$$\overline{u}_H = T_H -R_H,\quad \overline{v}_H = T_H+R_H,\qquad {\rm and} \qquad \overline{u}_C = T_C -R_C,\quad \overline{v}_C = T_C+R_C, $$
and using \ref{ds15}, we respectively have the relations
\begin{eqnarray}
&&-\overline{u}_H \overline{v}_H=R_H^2 -T_H^2= \frac{1}{\kappa_H^2} \left\vert 1-\frac{r}{r_C}\right\vert^{-\kappa_H/\kappa_C} \left\vert \frac{r}{r_U}-1\right\vert^{\kappa_H/\kappa_C}\left(\frac{r}{r_H}-1 \right)\nonumber\\
&&-\overline{u}_C \overline{v}_C=R_C^2 -T_C^2= -\frac{1}{\kappa_C^2} \left\vert \frac{r}{r_U}-1\right\vert^{-\kappa_C/\kappa_U} \left\vert \frac{r}{r_H}-1\right\vert^{-\kappa_C/\kappa_H}\left(1-\frac{r}{r_C} \right)
\label{ds5'}
\end{eqnarray}
Thus with respect to either of the above Kruskal coordinates, an $r={\rm const.}$ line is a hyperbola. \ref{fl1} shows the Penrose-Carter diagram of the maximally extended SdS spacetime. For further discussions on charged and rotating black holes located in a de Sitter universe, we refer our reader to the original work of \cite{Gibbons}. 
%

\section{The Gibbons-Hawking thermal states and entanglement} \label{S3}
\noindent
The two event horizons of the SdS spacetime produce two thermodynamic relationships with temperatures $\kappa_H/2\pi$ and $\kappa_C/2\pi$,
\be
\delta M = \frac{\kappa_H}{2\pi} \frac{\delta A_H}{4}, \qquad \delta M= -\frac{\kappa_C}{2\pi} \frac{\delta A_C}{4} 
\label{l4}
\ee
where $A_H= 4\pi r_H^2$ and  $A_C= 4\pi r_C^2$  are respectively the areas of the BEH and CEH. However, since $r_C\geq r_H$, we have  $\kappa_H \geq \kappa_C$, \ref{l3}, and accordingly one expects that any equilibrium  is not possible. 
One way to tackle this issue is to place a thermally opaque membrane in the region C in \ref{fl1}, thereby splitting it into two thermally isolated subregions, as was suggested in ~\cite{Gibbons}. Thus an observer located at the black hole side detects Hawking radiation at temperature $\kappa_H/2\pi$ and likewise another on the other side detects the same at temperature $\kappa_C/2\pi$. We note in analogy that even in an asymptotically flat black hole spacetime, in order to define the Hartle-Hawking state, which describes thermal equilibrium of the black hole with a blackbody radiation at its Hawking temperature, one also needs  to `encase' the black hole with a perfectly thermally reflecting membrane~\cite{Birrell:1982ix}.

The existence of such states in the SdS spacetime was explicitly demonstrated in~\cite{Gibbons} via the path integral quantisation.  For our purpose, we shall very briefly demonstrate below their existence via the canonical quantisation. Let us consider a  free, massless and minimally coupled  scalar field,
 $\Box \phi(x)=0$, in the (1+1)-dimensional Schwarzschild-de Sitter spacetime. Since the theory is conformally invariant in two dimensions as well since any two dimensional spacetime is conformally flat, the scalar mode functions are simply plane waves. Thus for example for the outgoing modes, the field quantisation will read,

 $$
\phi(x) = \int \frac{d\omega}{\sqrt{2\pi \omega}}[a_\omega e^{-i\omega u} + a_\omega ^\dagger e^{i \omega u}]
 $$
 
where $a_\omega$ and $a_\omega ^\dagger$ are the usual annihilation and creation operators satisfying the canonical commutation relations, $[a_\omega, a_{\omega}] = 0 = [a^\dagger_\omega, a^\dagger_{{\omega}}]$ and $[a_\omega, a_{{\omega^\prime}} ^\dagger ] = \delta (\omega-\omega^{\prime})$.
Let us first consider the side of the  membrane which faces the BEH and call this subregion as A. The field quantisation can be done in a manner similar to that of the Unruh effect~\cite{Unruh:1976db, Crispino:2007eb}, and we shall not go into detail of it here. The local modes correspond to the $t-r_{\star}$ coordinates in A and also in the causally disconnected region L (with the time $t$ reversed) in \ref{fl1}. Note that there are both right and left moving plane wave modes characterised by the retarded and advanced null coordinates $u$ and $v$. The field quantisation can be done with both these kind of positive and negative frequency modes. However, since the left and right moving modes are orthogonal, the creation and annihilation operators associated with these two sectors commute. Accordingly, these two sectors can be treated as independent and without any loss of generality, we may focus on only one sector. This field quantisation yields the local vacuum, $|0_{A}, 0_{L}\rangle $. The global vacuum, $|0\rangle_{\kappa_H}$, in A$\cup$L corresponds to the field quantisation with the Kruskal coordinate of \ref{ds16}, regular on or across the BEH. Likewise, by calling the other subregion of C as B, we use the $t-r_{\star}$ coordinate and \ref{ds17} to make the field quantisation in B$\cup$R. Following the steps similar to the ones described in \ref{chapter2}, we next compute the Bogoliubov relationships and accordingly obtain the squeezed state expansion which similar to that of the Rindler spacetime, 
\be
|0\rangle_{\kappa_H}=\sum_{n=0}^{\infty} \frac{\tanh^n {r}}{\cosh{r}}|n_{A}, n_{L}\rangle \quad {\rm and }\quad  |0\rangle_{\kappa_C}=\sum_{n=0}^{\infty} \frac{\tanh^n {s}}{\cosh{s}}|n_{B}, n_{R}\rangle
\label{l5}
\ee
where $\tanh{r}=e^{-\pi \omega/\kappa_{H}}$ and $\tanh s= e^{-\pi \omega/\kappa_{C}}$. In other words, the Kruskal or the global vacuum states are analogous to that of the Minkowski vacuum, whereas the states appearing on the right hand side of the above equations are analogous to that of the local Rindler states confined to some particular spacetime regions. The squeezed state expansions of \ref{l5} correspond  to  Planck spectra of created particle pairs respectively with temperatures $\kappa_H/2\pi$ and $\kappa_C/2\pi$  in regions A$\cup$L and B$\cup$R given by

$$ n_{BEH}(\omega)= \frac{1}{e^{\frac{2\pi \omega}{\kappa_H}}-1} \qquad n_{CEH}(\omega) = \frac{1}{e^{\frac{2\pi\omega}{\kappa_C} }-1} $$ 

These spectra are detectable respectively by observers located at the black hole and the cosmological horizon side of the thermally opaque membrane.  As we mentioned earlier, this setup was first proposed in~\cite{Gibbons} and accordingly, we shall regard these states as the Gibbons-Hawking thermal states. 

For our purpose of forming entangled states, we shall also require the one particle  excitations $|1\rangle_{\kappa_H}$ and $|1\rangle_{\kappa_C}$, found by applying once the relevant creation operator on $|0\rangle_{\kappa_H}$ and $|0\rangle_{\kappa_C}$.  Using \ref{l5} and the Bogoliubov relations we re-express these one particle states in terms of the local squeezed states,
\begin{equation}
|1\rangle_{\kappa_H}=\sum_{n=0}^{\infty} \frac{\tanh^n {r}}{\cosh^2{r}} \sqrt{n+1}  |(n+1)_{A}, n_L\rangle,\qquad
|1\rangle_{\kappa_C}=\sum_{n=0}^{\infty} \frac{\tanh^n {s}}{\cosh^2{s}} \sqrt{n+1}|(n+1)_{B},n_R \rangle
\label{onep}
\end{equation}

We note that even though the  construction of the Gibbons-Hawking states is mathematically absolutely consistent, we may wonder how one may  practically realise such a thermally opaque membrane between the two horizons. Perhaps one possible way to realise this will be to consider the Klein-Gordon equation in $3+1$-dimensions, with the radial function satisfying,
$$ \left(-\frac{\partial^2}{\partial t^2}+\frac{\partial^2}{\partial r_{\star}^2} \right)R(r)+\left(1-\frac{2M}{r}-\frac{\Lambda r^2}{3} \right)\left(\frac{l(l+1)}{r^2} +\frac{2M}{r^3} -\frac{\Lambda}{3} \right) R(r)=0$$
The effective potential term appearing in the above Schr\"{o}dinger-like equation  vanishes at both the horizons and is positive in between. This bell shaped potential thus will work as a barrier between the two horizons. Modes that cannot penetrate it, will be confined in the regions close to the horizons and hence will be disconnected from each other. The effective potential thus can be thought of as a natural realisation of the thermally opaque membrane mentioned above.

Let us now take a maximally entangled global state,
\begin{eqnarray}
|\psi\rangle=\frac{1}{\sqrt{2}}\left[|0_{\kappa_H}, 0_{\kappa_C}\rangle+|1_{\kappa_H},  1_{\kappa_C}\rangle\right],
\label{x1}
\end{eqnarray}
and imagine that the $\kappa_H$- and $\kappa_C$-type states are located respectively in subregions A and B of C, defined above. Due to this placement, we can use \ref{l5}, \ref{onep} to consistently re-express $|\psi \rangle$ in terms of the local states in regions L, A and  B, R. Tracing out now the states belonging to the  causally disconnected regions R and L, \ref{fl1},   the reduced density operator for $|\psi \rangle $ becomes,
\begin{equation}
\begin{split}
\rho_{AB}&=\frac{1}{{2}} \sum_{n,m=0}^{\infty} \frac{\tanh^{2n} {r}\tanh^{2m} {s}}{\cosh^2{r} \cosh^2{s}}  \left[|n,m\rangle\langle n,m|+\frac{\sqrt{(n+1)(m+1)}}{{{\cosh{r}}\,{\cosh{s}}}}|n,m\rangle\langle n+1, m+1|\right.\\&\left.+\frac{\sqrt{(n+1)(m+1)}}{{{\cosh{r}}\,{\cosh{s}}}} |n+1, m+1\rangle\langle n,m| +\frac{{(n+1)}{(m+1)}}{{{\cosh^2{r}}{\cosh^2{s}}}} |n+1, m+1\rangle\langle n+1, m+1|\right]
\label{x2}
\end{split}
\end{equation}
where in any ket or bra, the first and second entries respectively belong to  A and B. With the help of this reduced, bipartite and mixed density matrix, we shall compute two appropriate measures of quantum entanglement, the mutual information and the logarithmic negativity, reviewed in \ref{Motivation and Overview}. 

\begin{figure}[h!]
\begin{center}
  \includegraphics[width=7.8cm]{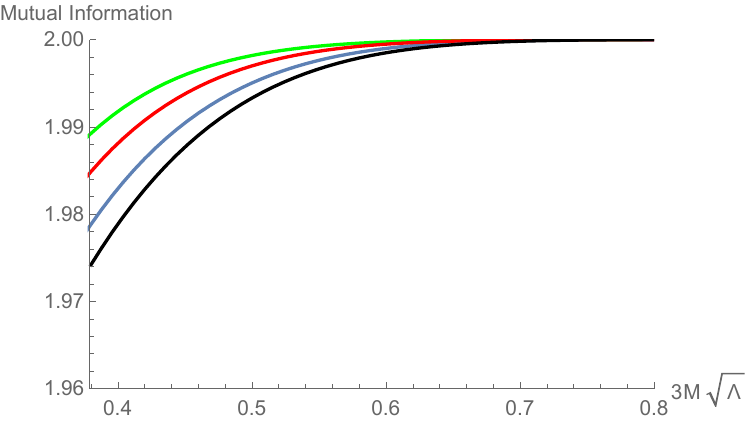}
  \includegraphics[width=7.8cm]{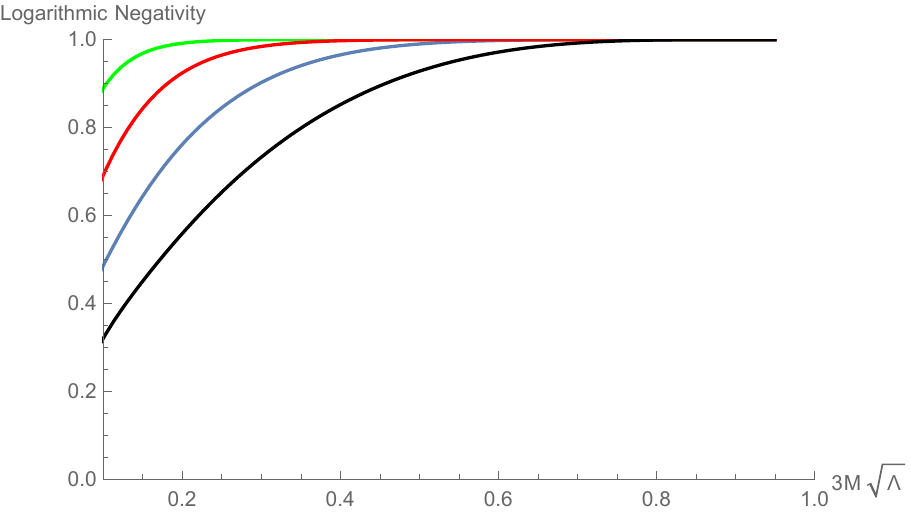}
  \caption{\small \it  (Left) The mutual information for the maximally entangled state
 ${|\psi \rangle }$, \ref{x1}, vs. the dimensionless parameter $3M\sqrt{\Lambda}$.  The black,  blue, red, green curves correspond respectively to $\omega/ \sqrt{\Lambda}  = 0.94, 1,\,1.1,\,1.2$. (Right) The logarithmic negativity vs. $3M\sqrt{\Lambda}$. The black. blue, red, green curves correspond respectively to $\omega/ \sqrt{\Lambda}  = 0.6, 1,\,1.7,\,3$. For smaller values of $\omega/ \sqrt{\Lambda}$, the  curves will show further smaller values of the information quantities, for a given $3M\sqrt{\Lambda}$. This is because such lesser values (for a fixed $\omega$) would correspond to higher $\Lambda$ values, leading to higher rate of cosmological particle creation, eventually degrading the entanglement more. We refer to main text for discussion.}

  \label{fig2}
\end{center}
\end{figure}

The quantum mutual information of A and B is defined as
\begin{eqnarray}
{\cal{I}}_{AB}=S(\rho_{A})+S(\rho_{B})-S(\rho_{AB}),
\label{c1}
\end{eqnarray}
where $S= - {\rm Tr} (\rho \ln \rho)$ is the von Neumann entropy. Tracing out further the states belonging to  the subregions B, and A, we respectively have, 
\begin{equation}
\rho_{A}
=\frac{1}{{2}} \sum_{n=0}^{\infty} \frac{\tanh^{2n} {r}   }{\cosh^2{r}  } \left(1+\frac{n}{\sinh^2{r}}\right) |n\rangle\langle n|\quad {\rm and} \quad
\rho_{B}
=\frac{1}{{2}} \sum_{m=0}^{\infty} \frac{\tanh^{2m} s}{\cosh^2s}\left(1+\frac{m}{\sinh^2{s}}\right) |m\rangle\langle m|
\label{cc3}
\end{equation}
Using \ref{x2}, \ref{cc3}, we now compute ${\cal{I}}_{AB}$ numerically in Mathematica.  Implicitly assuming $\Lambda$ to be fixed, the variation of ${\cal{I}}_{AB}$ with respect to the dimensionless parameter $3M{\sqrt \Lambda}$  is depicted   in the first of \ref{fig2} for three different values of the dimensionless parameter $\omega/\sqrt{\Lambda}$. Thus the mutual information increases monotonically with increasing $M$ and  saturates to two in the Nariai limit,  $3M\sqrt{\Lambda}\to 1$. \ref{l2}, \ref{l3} shows that the increase in this parameter corresponds to  the decrease in both the surface gravities, eventually becoming vanishing in the Nariai limit.
On the other hand as $3M\sqrt{\Lambda}\to 0$, we have $\kappa_H \sim M^{-1}$ and $\kappa_C \sim \sqrt{\Lambda}$. Thus for a fixed value of $\Lambda$,  black hole's Hawking  temperature becomes very large in this limit, resulting in an extremely high rate of particle creation.

Intuitively, the increase in the Hawking temperature increases the degree of randomness of the created thermal particles, which  degrades or destroys quantum correlation, as has been reflected in \ref{fig2}, analogous to that of the single horizon spacetimes reported earlier~~\cite{Fuentes:2010dt, mann, Pan:2008yi, martinez, martin, Montero:2011sx, richter, Asghari:2018}. We also note that for a given value of $\omega$, the mutual information   degrades more with increasing $\Lambda$. This  is due to the increase in the   Hawking temperature of the CEH  with increasing $\Lambda$,  resulting in  degradation  in the  correlation further, compared to the $\Lambda=0$ case.  

The logarithmic negativity  is related to the eigenvalues of \ref{x2}, but after transposing one of sectors (say A), $(\rho_{AB})^{T_{A}}$. It is defined as ${\cal{L_{N}}}=\log\sum_i |\lambda_{i}|$,  where $\lambda_{i}$'s are the eigenvalues $(\rho_{AB})^{T_{A}}$, e.g.~\cite{Nielsen:2000}. 
We have also computed it numerically in Mathematica and have plotted in the second of ~\ref{fig2}. The qualitative conclusions remain the same as that of the mutual information. 

 While the entanglement degradation with the Gibbons-Hawking states thus seems to be intuitively well acceptable, we wish to present below another viable description of the SdS thermodynamics where such intuitions seems to fail.

\section{The total entropy and the effective temperature}\label{s4} 
\noindent
The Gibbons-Hawking framework discussed in the preceding section allows us to treat the two event horizons of the Schwarzschild-de Sitter spacetime independently, and provides a thermal description of them in terms of their individual temperatures $\kappa_H/2\pi$ and $\kappa_C/2\pi$. As we mentioned earlier, these individual temperatures correspond to individual Bekenstein-Hawking entropies $A_H/4$ and $A_C/4$ respectively. Since the entropy is a measure of lack of information to an observer, for an observer located in region C in \ref{fl1}, one can also define a total entropy $S= (A_{H}+A_C)/4$. Thus for a fixed $\Lambda$,
$$\delta S= \frac14 \left(\frac{\delta A_H}{\delta M}+\frac{\delta A_C}{\delta M} \right)\delta M  $$
Using \ref{l2}, \ref{l3}, one then obtains after a little algebra
 a thermodynamic relationship with an  {\it effective equilibrium} temperature~\cite{Maeda, Davies, Urano, Saida1, Saida2, Bhattacharya:2015mja, Zhang, Kanti1, Kanti2, Kanti3, pappas},
\begin{eqnarray}
\delta M =- \frac{\kappa_H\kappa_C}{2\pi(\kappa_H-\kappa_C)}\delta S=-T_{\rm eff}\delta S
\label{eff}   
\end{eqnarray}
Even though  various computations including that of phase transition has been done using $T_{\rm eff}$, e.g.~\cite{Zhang, Kanti1, Kanti2, Kanti3, pappas}, a clear   understanding of it in terms of  field quantisation and  explicit quantum states seems to be missing.  

Let us first try to understand the emergence of this effective temperature intuitively. We note that in this picture where the two horizons are combined, we must consider emission as well as absorption of Hawking radiations, both of which change the horizon areas. For example ignoring the greybody effects, a particle emitted from the BEH will propagate towards CEH and will eventually get absorbed. Likewise a particle emitted from the CEH will propagate inward and will be absorbed by the BEH.  Since $\kappa_H >\kappa_C$, \ref{l3}, the flux of outgoing particles emitted from the BEH at any point $r_H < r<r_C$ will be greater than the flux of particles propagating inward emitted from the CEH, resulting in an effective outward flux and evaporation of the black hole. The existence of such effective temperature can then be intuitively understood as follows. Let {\small $P_E^H(P_A^H$)} and {\small $P_E^C(P_A^C)$} respectively be the single particle emission (absorption) probabilities for the BEH and CEH, so that
 {\small $P^{H}_{E}=P^{H}_{A}e^{-2\pi \omega/\kappa_H}$}  and {\small $P^{C}_{E}=P^{C}_{A}e^{-2\pi \omega/\kappa_C }$}.
By treating the above probabilities as independent, we may define an {\it effective emission probability}, {\small $P^E_{\rm eff}: = P^{C}_{E}P^{H}_{A}= P^{C}_{A}P^{H}_{E}e^{-\omega/T_{\rm eff}}$},  corresponding to the effective inward flux of the cosmological Hawking radiation on the BEH. Likewise the effective absorption probability {\small $P^A_{\rm eff}=P^{C}_{A}P^{H}_{E}$}, corresponds to the   effective outward flux of  black hole's Hawking radiation on the CEH.  Since {\small $P^A_{\rm eff} > P^E_{\rm eff}$}, the black hole gets evaporated.

From \ref{l2}, \ref{l3}, we also have
\begin{eqnarray}
 \lim_{3M\sqrt{\Lambda} \to 0} T_{\rm eff} \to \frac{\kappa_C}{2\pi} \approx \frac{1}{2\pi}\sqrt{\frac{\Lambda}{3}}, \qquad \lim_{3M\sqrt{\Lambda} \to 1} T_{\rm eff} \approx \frac{1}{2\pi}\frac{3\sqrt{\Lambda}}{4}
\label{}   
\end{eqnarray}
Thus  even though individually the surface gravities $\kappa_H$ and $\kappa_C$ become vanishing in the Nariai limit ($3M\sqrt{\Lambda} \to 1$), $T_{\rm eff}$ is non-vanishing. Moreover, the effective temperature in the Nariai limit is greater than that of when the black hole is extremely hot ($3M\sqrt{\Lambda} \to 0$, with $\Lambda$ fixed). This corresponds to the fact that $T_{\rm eff}$ is related to an emission probability which corresponds to the inward particle flux on the BEH created by CEH, as described above. As the black hole Hawking temperature increases due to decrease in $3M \sqrt{\Lambda}$, the black hole radiates more resulting in larger outward flux on CEH, corresponding to  reduced  effective inward flux or reduced effective temperature. 

However, a thermodynamic relationship such as \ref{eff} always needs to be proven via the explicit demonstration of particle creation. 
Accordingly, we now wish to explicitly find out  the   quantum states corresponding to the above description. Since we are treating both the horizons together, it is natural to ask, could there be a global vacuum in L$\cup$C$\cup$R in \ref{fl1}, which plays a role here? The answer is no,  for there exists no analytic Feynman propagator that connects both the horizons~\cite{Gibbons}. This implies that (unlike the single horizon cases) one cannot construct any single global mode which is analytic on or across both the horizons and hence in
L$\cup$C$\cup$R. Such non-existence should be attributed to the fact that   there exists no single Kruskal-like coordinates which remove the coordinate singularities of both the horizons. 

We recall  that  once we relax the idea of Lorentz invariance such as in a curved spacetime, we have the liberty to choose any coordinate system to describe a given phenomenon, as each such coordinate system represents a viable observer. For example in the Schwarzschild spacetime, one chooses different null coordinates to construct various vacuum states, e.g.~\cite{Choudhury:2004ph} and references therein. Thus we shall now introduce a new coordinate system to address the issue of this effective temperature, as follows.

We note that the `surface gravity' $\kappa_U$, \ref{ds5},  of the unphysical horizon at $r_U=-(r_H+r_C)$   is given by\, $-\partial_r g_{tt} (r_U)/2$. From \ref{l3}, it is easy to see that
$$\frac{1}{\kappa_U}= \frac{1}{\kappa_C}-\frac{1}{\kappa_H}$$
Using \ref{ds5}, let us now try to remove the `singularity' of the metric at $r=r_U$. Accordingly, we rewrite the $t-r$ part of \ref{l1} as  
\begin{eqnarray}
ds^2=-\frac{2M}{r}\left\vert1-\frac{r}{r_C}\right\vert^{1+\frac{\kappa_U}{\kappa_C}} \left\vert\frac{r}{r_H}-1\right\vert^{1-\frac{\kappa_U}{\kappa_H}}\, d{\overline u} d {\overline v}
\label{c4}
\end{eqnarray}
where we have defined,
\begin{eqnarray}
{\overline u}=-\frac{1}{\kappa_U}e^{-\kappa_U u},\quad {\overline v}=\frac{1}{\kappa_U}e^{\kappa_U v} \quad 
\label{c5}
\end{eqnarray}
Apparently it might appear that we are analytically extending the spacetime metric  at $r_U$. However due to the singularity at $r=0$, the spacetime cannot be extended to negative radial values.  We note also that the metric in \ref{c4} is {\it not} well behaved on or across any of the event horizons. Thus \ref{c4}, \ref{c5} {\it do not} correspond to the beyond horizon extensions of \ref{fl1} and hence they represent a coordinate system only in region C, $r_H < r <r_C$, which is our region of interest anyway. By considering incoming and outgoing null geodesics, it is easy to check that,
$$ u_{\rm in}(r_C) \to  -\infty,~u_{\rm in}(r_H) \to  \infty,\quad v_{\rm out}(r_H) \to -\infty,~v_{\rm out}(r_C) \to \infty $$
yielding $-\infty < \overline{u} \leq 0$ and $0 \leq  \overline{v} < \infty$. It is also easy to see the hyperbolic locus of the $r={\rm const.}$ curves with respect to $\overline{u}$ and $\overline{v}$, as of the previous cases, \ref{ds5'}.

We now define a field quantisation in terms of  $(\overline{u},\overline{v})$  and an alternative one in terms of the usual  $(u,v)$ coordinates as earlier. Using the ranges of coordinates and following the standard procedure, e.g.~\cite{JHT}, we can compute the Bogoliubov coefficients  by choosing the integration surface infinitesimally close to, eg. the BEH. Denoting the vacuum defined by the  $(\overline{u},\overline{v})$ modes by $| \overline{0}\rangle$, we have the squeezed state relationship between the two kind of states corresponding to the above two field quantisations, 
\begin{eqnarray}
| \overline 0\rangle =\sum_{n=0}^{\infty} \frac{\tanh^n {w}}{\cosh{w}}|n, n\rangle,\quad {\rm with }~~\tanh w = e^{-\pi \omega/\kappa_U}
\label{c6}
\end{eqnarray}
the above corresponds to a pair creation with temperature $T_{\rm eff}=\kappa_U/2\pi$.  We emphasise once again that the (entangled) pair creation is occurring in this case only in $r_H < r < r_C$ and {\it not} in the causally disconnected wedges as of the preceding Section. Accordingly, $|\overline{0}\rangle$ should not be regarded as any analogue of the global or Minkowski vacuum.  We also emphasise that the appearance of $\kappa_U$ (instead of $\kappa_H$ or $\kappa_C$) in \ref{c5} has guaranteed the emergence of the temperature $T_{\rm eff.}$. Due to this reason, the coordinate system in \ref{c4}, \ref{c5} seems to be unique, as far as this effective description is concerned. Since we are quantising the field using two different coordinatisations ($u,v$ and $\overline{u}, \overline{v}$), the associated vacua are energetically different, giving rise to the  Bogoliubov relationship and particle creation.  Since the global and the local vacua are confined in the single region $r_H < r< r_C$, the particle creation here qualitatively rather resembles with that of in the cosmological spacetimes, e.g.~\cite{Bhattacharya:2020sjr}. 

\begin{figure}[h!]
\begin{center}
  \includegraphics[width=7.8cm]{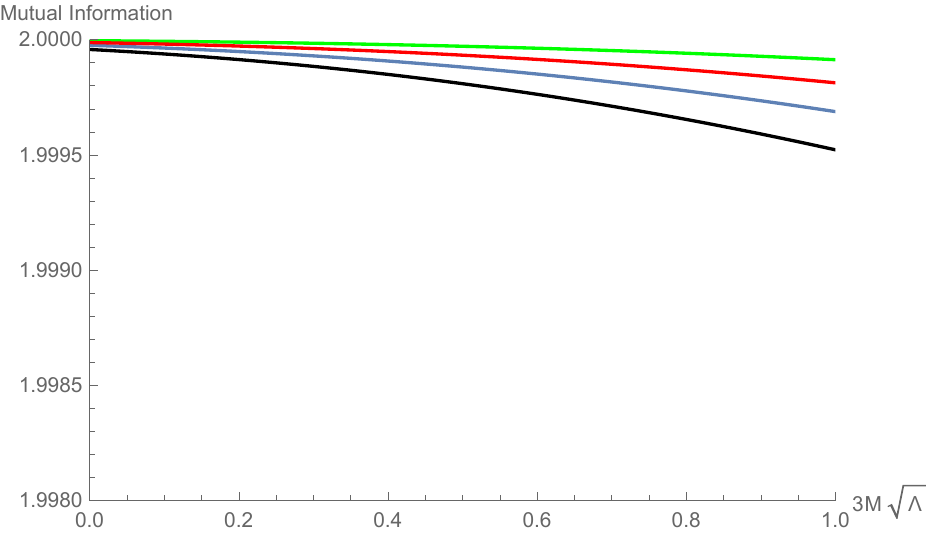}
  \includegraphics[width=7.8cm]{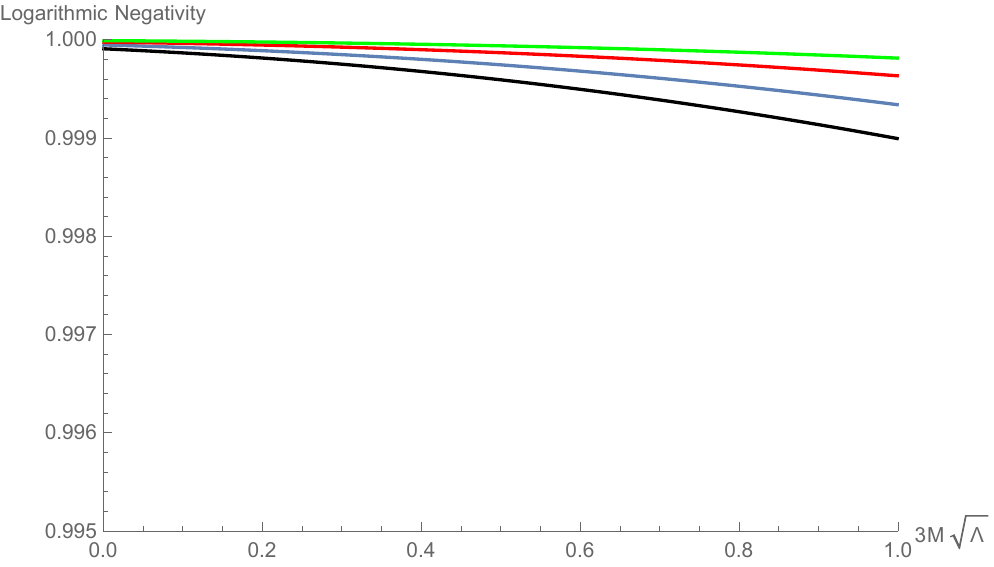}
  \caption{\small \it (Left) The mutual information corresponding to the maximally entangled state $|\chi\rangle$, \ref{c7}, vs. $3M\sqrt{\Lambda}$.  (Right) The logarithmic negativity vs. $3M\sqrt{\Lambda}$. The black,  blue, red, green curves correspond respectively to $\omega/ \sqrt{\Lambda}  = 0.95, 1,\,1.07,\,1.15$. Note the complete qualitatively opposite behaviour with respect to the Gibbons-Hawking states, \ref{fig2}. We refer to main text for discussion.}

  \label{fig3}
\end{center}
\end{figure}

We now take a maximally entangled state analogous to \ref{x1},
\begin{eqnarray}
|\chi \rangle = \frac{1}{\sqrt{2}}\left[|\overline{0}, \overline{0}\rangle+|\overline{1},  \overline{1}\rangle\right],
\label{c7}
\end{eqnarray}
and  expand it using \ref{c6}. We then trace out parts of it in order  to form a mixed bipartite system and compute as earlier the mutual information and the logarithmic negativity, plotted in \ref{fig3}. As expected from the characteristics of $T_{\rm eff}$   discussed above, the entanglement does not degrade even when the black hole is extremely hot $(3M \sqrt{\Lambda} \to 0$, for a fixed $\Lambda$), but actually it is maximum in this limit. Moreover, the entanglement degrades as we approach the Nariai limit.  This is completely contrary to what was obtained with the Gibbons-Hawking states in Sec.~3, or to the best of our knowledge, what has been reported in the literature so far, ~\cite{Fuentes:2010dt, mann, Pan:2008yi, martinez, martin, Montero:2011sx, richter, Asghari:2018}. We note also from the figure that (for a fixed $\omega$) the entanglement degrades with increasing $\Lambda$ value, corresponds to the increasing temperature and particle creation by the CEH. In the next Section we have explained that $T_{\rm eff.}$ can have {\it no} analogue in single horizon (i.e. $\Lambda \leq 0$) spacetimes.  
\section{Discussion}
\noindent
In this Chapter, we have analysed the entanglement degradation for maximally entanglement Kruskal-like states in the Schwarzschild-de Sitter spacetime, by exploring  two viable descriptions of thermodynamics and particle creation in this background. In \ref{S3}, we have taken the Gibbons-Hawking proposition~\cite{Gibbons}, where an observer can be in thermal equilibrium with either of the horizons, by the means of placing a thermally opaque membrane in between the two horizons. We have shown that the entanglement degrades in this case with increasing Hawking temperature of either of the horizons. This is qualitatively similar to the earlier results found for single horizon spacetimes~~\cite{Fuentes:2010dt, mann, Pan:2008yi, martinez, martin, Montero:2011sx, richter, Asghari:2018}. In \ref{s4}, we have addressed the total entropy-effective temperature formalism~\cite{Maeda, Davies, Urano}, in order to treat both the horizons in an equal footing. By introducing a suitable coordinate system, we have found the vacuum states necessary for such description and have shown that  entanglement degradation never happens in this scenario, no matter how hot the black hole becomes or how small the cosmological constant is. We note from \ref{l2} that the total entropy is minimum in the Nariai limit ($3M\sqrt{\Lambda} \to 1$) and maximum as $3M\sqrt{\Lambda} \to 0$. Thus if we begin from the Nariai limit (where the Hawking temperature of both horizons are vanishingly small), assuming $\Lambda$ is fixed the black hole will evaporate and the spacetime will evolve towards $3M\sqrt{\Lambda} \to 0$. In this course, we shall keep recovering the entanglement or correlation, \ref{fig3}. This is in complete qualitative contrast with the existing cases~~\cite{Fuentes:2010dt, mann, Pan:2008yi, martinez, martin, Montero:2011sx, richter, Asghari:2018}.  Finally,   note from \ref{l3}, \ref{eff} that $T_{\rm eff} \to 0$ as $\Lambda \to0$. Also, there can be no cosmological event horizon for $\Lambda \leq 0$. This means that $T_{\rm eff}$ and the associated entanglement phenomenon  can have {\it no} $\Lambda \leq 0$ analogue.   Since a black hole spacetime endowed with a positive  $\Lambda$  furnishes a nice toy model for the global structure of a black hole spacetime in the current as well as in the early inflationary universes, these results seem to be physically interesting in its own right. 

Although we have simply worked in $1+1$-dimensional Schwarzschild-de Sitter spacetime, the qualitative features of entanglement we have found here would certainly remain the same in higher dimensions.  This is because the Bogoliubov coefficients and hence the particle creation does not depend  upon the angular eigenvalues for a spherically symmetric spacetime, e.g.~\cite{Birrell:1982ix}.

We note that the `entanglement degradation'  here (and also in the earlier relevant works~~\cite{Fuentes:2010dt, mann, Pan:2008yi, martinez, martin, Montero:2011sx, richter, Asghari:2018}) does not correspond to some dynamical decoherence procedure such as in \ref{chapter3} but rather to some observers who use different coordinate systems and hence different physically well motivated  vacuum states. On the other hand, the decoherence mechanism is likely to involve a more realistic collapse scenario, in order to see the time evolution of a given initial vacuum state. Nevertheless,  since particle creation in an eternal black hole spacetime may effectively model that of in a collapsing geometry~\cite{Birrell:1982ix}, one may expect that the above demonstrations indeed have connections to the Hawking radiation and entanglement degradation/decoherence in a collapsing geometry. However, we are unaware of any such explicit computations and certainly this warrants further attention. 
\chapter{Summary and Outlook}\label{Summary&discussion}

Finally, in this concluding Chapter we provide a summary of the findings presented in the preceding main chapters of the thesis. We started with an introduction on the foundational motivation behind this study, aiming to shed light on the physics of the early universe. Next, it proceeds with an overview of quantum correlations, encompassing diverse measures like entanglement entropy, Bell's inequality, logarithmic negativity, and quantum discord. Further, decoherence within the context of quantum field theory and a brief review of de Sitter spacetime geometry is discussed.

In \ref{chapter2}, we explored two measures of quantum correlations and entanglement, specifically the violation of BMK inequalities and the quantum discord, in the context of Dirac fermions in a cosmological de Sitter background. We have focused on the vacuum state for the two and four mode squeezed states to examine the extent of BMK violation, and demonstrated the maximum violation achievable whereas for the quantum discord we worked with maximally entangled initial Bell state. Our motivation was to see how the results differ subject to different coordinatisations of the de Sitter as well as the spin of the field. We thus first note the qualitative similarities of the variations with respect to parametrisation angles with a scalar field theory in the context of transition from de Sitter to radiation dominated era~\cite{bell:2017} and the hyperbolic de Sitter spacetime~\cite{Kanno:2016gas}. However, the chief qualitative difference for the cosmological de Sitter from them or the non-inertial frame, or even the static de Sitter spacetime is that the Bogoliubov coefficients for the `in' and `out' vacua in this case is independent of the spatial momentum or the total energy. Similar feature is seen for a scalar field theory in the global or the cosmological de Sitter spacetime. The variation of the Bogoliubov coefficients correspond to the variation of the rest mass only. We have also noted that there exists no extreme squeezing limit ($|\beta_k| \to 1$) for the states constructed for Dirac fermions in a cosmological de Sitter background and we always have $|\beta_k| \lesssim 0.707$. Due to this reason, the logarithmic negativity is never vanishing in our case. This is qualitatively different from a scalar field theory where the logarithmic negativity can indeed vanish, indicating the complete decay of quantum entanglement due to particle creation but the discord, being a measure of all correlations, survive.

In \ref{chapter3}, We have explored decoherence in a quantum field theory comprising fermions and scalars coupled with the Yukawa interaction. We started with the simplest scenario in the Minkowski spacetime and subsequently extending our analysis to the more complex inflationary de Sitter spacetime. Although the analytical techniques required for both spacetimes are identical, we adopted perturbative methods for the latter due to its intricacies. In both cases, we treat the scalar field as the system of interest, while fermions act as the environment. The scalar field is assumed to be massive in the Minkowski spacetime and massless in the de Sitter spacetime, while fermions are considered massless in both scenarios. In the Minkowski spacetime, we successfully resummed one loop scalar self-energy contributions, yielding a non-perturbative outcome. However, in the inflationary de Sitter spacetime, our results remain perturbative in the coupling constant. In the Minkowski spacetime the phase space area decreases steadily as system mass increases, eventually approaching unity, indicating minimal entropy. This reflects increased system stability with greater mass, making it harder for the surroundings to disrupt it. Furthermore, higher coupling values increase both the phase space area and entropy, surpassing the numerical values observed in the scenario involving both the system and surroundings as scalar fields. In the inflationary de Sitter spcetime the phase space area and entropy consistently increase as the scale factor and Yukawa coupling grow. This aligns qualitatively with scenarios involving both the system and surroundings as scalar fields. Our findings also concur qualitatively with results obtained through the Feynman-Vernon influence functional method for entropy generation \cite{Boyanovsky:2018soy}. The entropy rise with the scale factor corresponds to the fact that the phase space area must increase with time in an expanding spacetime. Likewise, the rise in entropy associated with the Yukawa coupling signifies the emergence of new phase space regions resulting from interactions. This similarity is reminiscent of the behavior in the Minkowski spacetime, even though the result in the Minkowski spacetime was non-perturbative.

In \ref{chapter4}, We examined the quantum entanglement within the Schwarzschild-de Sitter black hole spacetime, focusing on mutual information and logarithmic negativity of entangled bipartite states for massless minimal scalar fields. This spacetime involves both a black hole and a cosmological event horizon, leading to particle creation at distinct temperatures. We have explored two thermodynamic perspectives. The first considers thermal equilibrium for observers near either horizon, revealing that entanglement degrades as Hawking temperatures increase, similar to asymptotically flat/anti-de Sitter black holes. The second perspective combines both horizons to define total entropy and an effective equilibrium temperature. Surprisingly, in this scenario, entanglement do not degrade but actually increases as the black hole temperature rises. Importantly, this remains consistent regardless of the black hole's temperature or the magnitude of the cosmological constant. This phenomenon is distinct from asymptotically flat/anti-de Sitter black hole spacetimes.

A logical progression of above scenario is to explore how entanglement degradation manifests in multi-event horizon spacetime for fermions, with a focus on contrasting their behavior with scalars. Additionally, we would like to see if entanglement persists increasing as the black hole temperature rises and when the cosmological constant is small in this case also. The rotating black hole spacetimes, due to  the existence of various exotic vacuum states for a massless scalar~\cite{Kay:1988mu} also seems to be interesting in this context. However, perhaps it may be more interesting in this context to consider the acoustic analogue gravity phenomenon~\cite{Barcelo:2005fc}, where the propagation of the perturbation in a fluid is associated with an internal acoustic geometry endowed with a sonic causal structure and horizon. Accordingly, there is creation of phonons with Hawking like spectra. For a multi-component fluid, perhaps  one may naively then  expect a multi-sonic horizon structure analogous to the SdS. For such an acoustic analogue  system, we may look for analogous entanglement properties as that of the the SdS. These construction, like the other analogue gravity phenomenon, might make some interesting predictions testable in the laboratory.  We hope to return to these issues in future works. 

As mentioned in \ref{chapter3}, an obvious extension of decoherence analysis would be to go to the two loop self energies, the second and third of \ref{figa}. Interestingly, at this order we might also expect some effect of system’s backreaction onto the surrounding. Perhaps more importantly, one should also consider the three-point scalar-fermion-anti-fermion correlators containing the effect of the decay of the scalar. As previously stated, when dealing with a massless minimal scalar in an inflationary de Sitter background, we anticipate the emergence of late-time secular effects. Here, we have limited our analysis to the one loop self-energy. Expanding to higher orders is likely to introduce these secular effects. Resumming them non-perturbatively using the Kadanoff Baym equations or some other method, for e.g. Schwinger–Dyson framework, could be an interesting and challenging task. The computation of dynamical mass generation via these framework also seems to be interesting. It would also be interesting to extend the decoherence analysis for the finite temperature field theory where distinction between the system and the environment is obvious. It would be interesting to explore whether finite temperature field theory reveals some more insights compared to the zero temperature scenario. We would also like to compute the decoherence rate for the these cases, which we have not considered here, quantifying how quickly quantum superpositions and interference effects fade away, making the system appear classical. It will help us to understand the stability and robustness of the quantum systems. 



\appendix
\chapter{}

\section{Derivation of phase space area and entropy}\label{Derivation of phase space area and entropy}

In order to derive the expression for phase space area and entropy briefly, we shall take the approach motivated in \cite{Koksma:2010zi}. We start by examining the free Hamiltonian of a time-dependent harmonic oscillator
\begin{equation}\label{dd1}
H(p, x) = \frac{p^2}{2m} + \frac{1}{2}m\omega^2 x^2 + H_{\mathrm{s}}, \quad H_{\mathrm{s}} = xj.
\end{equation}

where, $H_{\mathrm{s}}$ represents the source Hamiltonian, and $j = j(t)$ corresponds to the associated current. We may examine Gaussian density matrices centered around the origin, whose temporal dynamics are dictated by a Hamiltonian, \ref{dd1}, with $j(t) \rightarrow 0$. When expressed in position space representation, the density operator characterizing a quantum Gaussian state centered at the origin generally takes the following form
\begin{equation}\label{dd2}
\hat{\rho}_{\mathrm{g}}(t)=\int_{-\infty}^{\infty} \mathrm{d} x \int_{-\infty}^{\infty} \mathrm{d} y|x\rangle \rho_{\mathrm{g}}(x, y ; t)\langle y|,
\end{equation}
where
\begin{equation}\label{dd3}
\rho_{\mathrm{g}}(x, y ; t)=\mathcal{N}(t) \exp \left[-a x^2-b y^2+2 c x y\right],
\end{equation}
where, $a=a(t)$, $b=b(t)$, and $c=c(t)$ are determined from the von Neumann equation. 
From $\hat{\rho}_{\mathrm{g}}^{\dagger}=\hat{\rho}_{\mathrm{g}}$, we deduce that $b^*=a$ and $c^*=c$. If $c=0$, we recover a pure state with vanishing entropy but when $c \neq 0$, the density matrix is mixed, then it can not be expressed in the simple diagonal form. The normalization $\mathcal{N}(t)$ is obtained as 
\begin{equation}
\operatorname{Tr}\left[\hat{\rho}_{\mathrm{g}}\right]=\int_{-\infty}^{\infty} \mathrm{d} \tilde{x}\left\langle\bar{x}\left|\hat{\rho}_{\mathrm{g}}\right| \tilde{x}\right\rangle=\int_{-\infty}^{\infty} \mathrm{d} x \rho_{\mathrm{g}}(x, x ; t)=\mathcal{N}(t) \sqrt{\frac{\pi}{2\left(a_{\mathrm{R}}-c\right)}}=1
\end{equation}
so
\begin{equation}
\mathcal{N}(t) = \sqrt{\frac{2\left(a_{\mathrm{R}}-c\right)}{\pi}},
\end{equation}

provided that $c<a_{\mathrm{R}}$, where $a_{\mathrm{R}}= \rm{Re}[a]$. The functions $a(t)$, $b(t)$, and $c(t)$ of the density matrix \ref{dd3} can easily be obtained from the von Neumann equation, $i \hbar\frac{d\hat{\rho}}{dt} = [\hat{H}, \hat{\rho}]
$, which in this Gaussian case, reads

\begin{equation}\label{dd4}
\begin{aligned}
\imath \hbar \partial_t \rho_{\mathrm{g}}(x, y ; t) & = -\frac{\hbar^2}{2 m}\left(\partial_x^2-\partial_y^2\right) \rho_{\mathrm{g}}(x, y ; t) + \frac{1}{2} m \omega^2\left(x^2-y^2\right) \rho_{\mathrm{g}}(x, y ; t)
\end{aligned}
\end{equation}

If we insert equation \ref{dd3} into the equation above, we find

\begin{equation}
\begin{aligned}
\frac{\mathrm{d} a_{\mathrm{R}}}{\mathrm{d} t} & = \frac{4 \hbar}{m} a_{\mathrm{I}} a_{\mathrm{R}} \\
\frac{\mathrm{d} c}{\mathrm{d} t} & = \frac{4 \hbar}{m} a_{\mathrm{I}} c \\
\frac{\mathrm{d}}{\mathrm{d} t} \ln (\mathcal{N}) & = \frac{2 \hbar}{m} a_{\mathrm{I}} \\
\frac{\mathrm{d} a_{\mathrm{I}}}{\mathrm{d} t} & = \frac{2 \hbar}{m}\left(a_{\mathrm{I}}^2-a_{\mathrm{R}}^2+c^2\right)+\frac{m \omega^2}{2 \hbar},
\end{aligned}
\end{equation}

where $a_{\mathrm{I}}=\rm{Im}[a]$. The Wigner function is defined as a Wigner transform of the density matrix

\begin{equation}\label{dd5}
W(q, p ; t)=\int_{-\infty}^{\infty} \mathrm{d} r \mathrm{e}^{-i p r / \hbar} \rho_{\mathrm{g}}(q+r / 2, q-r / 2 ; t),
\end{equation}

where we have defined, $q=(x+y) / 2$ and $r=x-y$, respectively. A Wigner transform can be thought of as an ordinary Fourier transform with respect to the relative coordinate $r$ of the density matrix. We obtain Wigner function by performing the integral above

\begin{equation}\label{wg1}
W(q, p ; t)  = \mathcal{M}(t) \exp \left[-\alpha(t) q^2-\beta(t)\left(p(t)+q p(t)\right)^2\right],
\end{equation}
\begin{equation}
\alpha(t)  = 2\left(a_{\mathrm{R}}-c\right), \qquad
\beta(t)  = \frac{1}{2 \hbar^2\left(a_{\mathrm{R}}+c\right)}, \qquad
p(t)  = 2 \hbar a_{\mathrm{I}}, \qquad
\mathcal{M}(t)  = \sqrt{\frac{4\left(a_{\mathrm{R}}-c\right)}{a_{\mathrm{R}}+c}}  
\end{equation}

Using the expressions \ref{dd4} and \ref{dd5}, we get

\begin{equation}\label{dd6}
\left(\partial_t+\frac{p}{m} \partial_q-m \omega^2 q \partial_p\right) W(q, p ; t) = 0
\end{equation}

where we have made use of

\begin{equation}
\begin{aligned}
\left(\partial_x^2-\partial_y^2\right) \rho_{\mathrm{g}}(x, y ; t) & = 2 \partial_q \partial_r \rho_{\mathrm{g}}(q+r / 2, q-r / 2 ; t) \\
r \mathrm{e}^{\imath p r / \hbar} & = -\imath \hbar \partial_p \mathrm{e}^{i p r / \hbar}
\end{aligned}
\end{equation}

Note that, with $p / m=\dot{x}=\partial_p H$, and $-m \omega^2 x=\dot{p}=-\partial_x H$, we see equation \ref{dd6} is nothing but the Liouville equation for the Boltzmann's distribution function $f(x, p ; t)$

\begin{equation}\label{dd7}
\partial_t f(x, p ; t)+\dot{x} \partial_x f(x, p ; t)+\dot{p} \partial_p f(x, p ; t) = 0
\end{equation}

A plausible way to interpret the Wigner function probabilistically is $W(q, p, t) \Leftrightarrow f(x, p ; t)$.

Note that, the equations \ref{dd6} and \ref{dd7} are identical only for a free harmonic oscillator theory, when interactions are included, differences arise between the von Neumann equation for the density operator $\hat{\rho}_{\mathrm{g}}(t)$ or the equation for $W(q, p ; t)$, and also the Boltzmann equation for $f(x, p ; t)$. In this approach, the entropy is approximated by
\begin{equation}\label{dd8}
S \approx S_{W} \equiv -\operatorname{Tr}[W(q, p ; t) \ln (W(q, p ; t))]
\end{equation}
where the trace $\operatorname{Tr} \rightarrow \int \mathrm{d} p \mathrm{~d} q /[2 \pi \hbar]$ should be interpreted as an integration over the phase space volume. We now insert equation \ref{wg1} into equation \ref{dd8} to find
\begin{equation}
S_{W}=\frac{1}{2} \ln \left(\frac{a_{\mathrm{R}}+c}{a_{\mathrm{R}}-c}\right)+1-\ln (2)=\ln \left(\frac{\Xi(t)}{2}\right)+1,
\end{equation}
where we have
\begin{equation}
\Xi^2(t)=\frac{a_{\mathrm{R}}+c}{a_{\mathrm{R}}-c} 
\end{equation}

We can extract the following correlators from the Gaussian state
\begin{align}
&\left\langle\hat{x}^2\right\rangle =\operatorname{Tr}\left[\hat{\rho}_{\mathrm{g}} \hat{x}^2\right]=\int_{-\infty}^{\infty} \mathrm{d} \tilde{x}\left\langle\tilde{x}\left|\hat{\rho}_{\mathrm{g}} \hat{x}^2\right| \tilde{x}\right\rangle=\frac{1}{4\left(a_{\mathrm{R}}-c\right)} \\
&\left\langle\frac{1}{2}\{\hat{x}, \hat{p}\}\right\rangle  =-\hbar \frac{a_{\mathrm{I}}}{2\left(a_{\mathrm{R}}-c\right)}, \qquad \left\langle\hat{p}^2\right\rangle =\hbar^2 \frac{|a|^2-c^2}{a_{\mathrm{R}}-c}
\end{align}

This enables us to find the equivalent inverse relations
\begin{equation}
a_{\mathrm{I}}  =-\frac{\left\langle\frac{1}{2}\{\hat{x}, \hat{p}\}\right\rangle}{2 \hbar\left\langle\hat{x}^2\right\rangle}, \qquad
a_{\mathrm{R}}  =\frac{\Xi^2(t)+1}{8\left\langle\hat{x}^2\right\rangle}, \qquad
c  =\frac{\Xi^2(t)-1}{8\left\langle\hat{x}^2\right\rangle},
\end{equation}
and we express $\Xi(t)$ in terms of the correlators as
\begin{equation}
\Xi^2(t)=\frac{4}{\hbar^2}\left[\left\langle\hat{x}^2\right\rangle\left\langle\hat{p}^2\right\rangle-\left\langle\frac{1}{2}\{\hat{x}, \hat{p}\}\right\rangle^2\right] .
\end{equation}

Hence, we can define the uncertainty relation for a Gaussian state, generalizing Heisenberg's uncertainty relation, as
\begin{equation}
\frac{(\hbar \Xi(t))^2}{4}=\left\langle\hat{x}^2\right\rangle\left\langle\hat{p}^2\right\rangle-\left\langle\frac{1}{2}\{\hat{x}, \hat{p}\}\right\rangle^2 \geq \frac{\hbar^2}{4} .
\end{equation}

Furthermore, it is natural to define the statistical particle number density in terms of phase space area as \cite{Koksma:2010zi}
\begin{equation}
n(t) = \frac{\Xi(t)-1}{2}
\end{equation}

The entropy of $n$-Bose particles per quantum state, following the Boltzmann’s distribution function, is given by
\begin{equation}
S = (1 + n) \ln(1 + n) - n \ln(n)
\end{equation}

Substituting particle number density in terms of $\Xi(t)$ we get,

\begin{equation}\label{entropy1}
S = \frac{ \Xi(t)+1}{2}
\ln\left(\frac{\Xi(t)+1}{2}\right) - \frac{
\Xi(t)-1}{2} \ln\left(\frac{\Xi(t)-1}{2}\right) 
\end{equation}

The above derivation is straightforward within the framework of quantum mechanics. By extending the concept of density matrices in terms of fields, we can similarly apply this analysis within the domain of quantum field theory but we shall not go into further detail of this. For more detail we refer our reader to \cite{Koksma:2010zi}.

\chapter{}

\section{The Dirac mode functions and Bogoliubov coefficients}\label{The Dirac mode functions and Bogoliubov coefficients}

In this appendix, we will solve the Dirac equation \ref{DQ} in the cosmological de Sitter background in order to compute the Bogoliubov coefficients and the particle creation.  We shall also require for our purpose the standard cosmological time $t$, given by $ t = -\frac{1}{H} \ln  (-H \eta)$. With the choice of the tetrad,

\begin{eqnarray}
e^{\mu}{}_a  \equiv H\eta \, {\rm diag}\, \left(1, \, 1,\, 1,\, 1 \right),
\label{c2}
\end{eqnarray}
 the Dirac equation \ref{DQ} in the de Sitter  background \ref{desitterspacetime} becomes,  
\begin{eqnarray}
\left[i\gamma^0\eta \partial_{\eta} -\frac{3i \gamma^0}{2}+i\eta\, {\vec \gamma}\cdot{\vec \partial} -\frac{m}{H}\right]\psi=0,
\end{eqnarray}

$\psi(x)$ can be quantised  in terms of the conformal time as,
\begin{eqnarray}
\psi(x)= \int \frac{d^{3} \vec{ k}}{(2\pi)^{3/2}}\,  \sum_{s= 1}^2 \left[a_{\rm in}(\vec{k},s) u_{\vec{k}, {\rm in }}^{(s)}(\eta) e^{i {\vec{k}\cdot \vec{x}}} + b_{\rm in}^{\dagger}(\vec{k},s) v_{\vec{k}, {\rm in }}^{(s)}(\eta)e^{-i \vec{k}\cdot \vec{x}} \right],
\label{f1}
\end{eqnarray}
where the temporal part of the Bunch-Davies mode functions are given by, e.g.~\cite{Collins:2004wj},
\begin{equation}
\begin{split}
u_{\vec{k}, {\rm in }}^{(s)}(\eta)=\frac{\sqrt{\pi k}\, e^{\frac{m\pi}{2H}}}{2} \, \eta^2  \begin{pmatrix} H_{\nu}^{(2)}(k\eta) \\ i \beta_s  H_{\nu-1}^{(2)}(k\eta) \end{pmatrix} \phi_{ \hat{{k}}}^{(s)}, \quad
v_{\vec{k}, {\rm in }}^{(s)}(\eta)=\frac{ \sqrt{\pi k}\, e^{\frac{m\pi}{2H}}}{2}\,  \eta^2  \begin{pmatrix} \beta_s H_{\nu}^{(1)}(k\eta) \\ -i   H_{1-\nu}^{(1)}(k\eta) \end{pmatrix} \chi_{ \hat{k}}^{(s)}
\label{inmodes}
\end{split}
\end{equation}
where $\beta_s=\pm1$ is the helicity of the mode corresponding respectively to $s=1,2$. $\phi_{ \hat{k}}^{(s)}$ and $\chi_{\hat{k}}^{(s)}$ are eigenvectors of the helicity operator $\hat{k}\cdot {\vec\gamma}$, with eigenvalues $\beta_s$. They are given by, 
\begin{eqnarray}
\begin{split}
 &\phi_{\hat{k} }^{(1)}=- \chi_{\hat{k} }^{(2)}=\frac{e^{-i\delta}}{\sqrt 2} \begin{pmatrix} \frac{{\hat{k}_{x}}-i{\hat{k}_y}}{\sqrt{1-{\hat{k}_{x}}}} \\  \sqrt{1-{\hat{k}_{z}}} \end{pmatrix}, \qquad
\phi_{\hat{k} }^{(2)}= \chi_{\hat{k} }^{(1)}=\frac{e^{i\delta}}{\sqrt 2} \begin{pmatrix} -\sqrt{1-{\hat{k}_{z}}} \\   \frac{{\hat{k}_{x}}+i{\hat{k}_{y}}}{\sqrt{1-{\hat{k}_{x}}}} \end{pmatrix}\\&
\qquad \qquad \quad \quad  \phi_{-\hat{ {k}} }^{(s)}= \chi_{\hat{k} }^{(s)}, \qquad  e^{2i\delta}=- \frac {\left(\hat{k}_y+i \hat{k}_x \right)}{\sqrt{\hat{k}_x^2+\hat{k}_y^2}}   \label{spinors}
\end{split}
\end{eqnarray}
where $\hat{k}_x , \,\hat{k}_y$ are  momentum unit vectors.  Also, $\nu$ is a constant given by
$$\nu=\frac12+ \frac{im}{H}$$
$H_{\nu}^{(1)}(k\eta)$, $H_{\nu}^{(2)}(k\eta)$ are the Hankel functions of the first and second kind respectively. 
It is easy to check using their asymptotic properties~\cite{AS} that $u(\eta \to -\infty)\sim e^{-ik\eta}$ and $v(\eta \to -\infty) \sim e^{ik\eta}$ in \ref{inmodes} and thus respectively behave as positive and negative frequency solutions in the asymptotic past. The orthonormality of the mode functions appearing in \ref{f1} can be checked using \ref{spinors} by computing their inner products, say on an $\eta \to -\infty$ hypersurafce. 
\ref{inmodes} should be regarded as the fermionic analogue of the Bunch-Davies mode functions~\cite{Collins:2004wj, Bunch:1978yq, Spindel}.

The `out' modes are those which have definite positive and negative frequency characteristics in the asymptotic future, $\eta\to 0^-$. It is easy to see from the asymptotic expansion of the Hankel function  that the conformal time cannot be a good coordinate for  this purpose and accordingly we  resort to the cosmological time, $t$.  In terms of $t$,  we have from \ref{inmodes} in the asymptotic future, 
\begin{eqnarray}
\begin{split}
u_{\vec{k}, {\rm in }}^{(s)}(t\to \infty)&=&\frac{\sqrt{\pi k} \,e^{\frac{m\pi}{2 H}}}{2\pi H^2}  e^{-\frac{3Ht}{2}}  \begin{pmatrix} \Gamma(\nu) \left(-\frac{k}{2H}\right)^{-\nu}\,e^{iHmt} \\  \beta_s e^{-\frac{m\pi}{H}}\Gamma{(1-\nu)} \left({-\frac{k}{2H}}\right)^{\nu-1}e^{-iHmt}\end{pmatrix} \phi_{\hat{k} }^{(s)}
\\
v_{\vec{k}, {\rm in }}^{(s)}(t \to \infty)&=&\frac{\sqrt{\pi k} \,e^{\frac{m\pi}{2H}}}{2\pi H^2}  e^{-\frac{3Ht}{2}}  \begin{pmatrix} - \beta_s e^{-\frac{m\pi}{H}}\Gamma(\nu) \,\left({-\frac{k}{2H}}\right)^{-\nu}e^{iHmt} \\  \Gamma{(1-\nu)} \left({-\frac{k}{2H}}\right)^{\nu-1}e^{-iHmt}\end{pmatrix} \chi_{\hat{k}}^{(s)}
\label{out1}
\end{split}
\end{eqnarray}
Accordingly, we choose the temporal part of the `out' modes  as,
\begin{eqnarray}
\begin{split}
&u_{\vec{k}, {\rm out }}^{(s)}(t)= e^{-\frac{3Ht}{2}}\, e^{iHmt} \begin{pmatrix} 1 \\  0  \end{pmatrix} \phi_{\hat{k} }^{(s)}, \qquad
v_{\vec{k}, {\rm out }}^{(s)}(t)={e^{-\frac{3Ht}{2}}}  e^{-iHmt}\begin{pmatrix} 0 \\ 1  \end{pmatrix} \chi_{\hat{k} }^{(s)}
\label{out2}
\end{split}
\end{eqnarray}
Thus we have the field quantisation,
\begin{eqnarray}
\psi(x)= \int \frac{d^{3} \vec{k}}{(2\pi)^{3/2}}\,  \sum_{s= 1}^2 \left[a_{\rm out}(\vec{k},s) u_{\vec{k}, {\rm out }}^{(s)}(t) e^{i {\vec{k}\cdot  \vec{x}}} + b_{\rm out}^{\dagger}(\vec{k},s) v_{\vec{k}, {\rm out }}^{(s)}(t)e^{-i \vec{k}\cdot \vec{ x}} \right],
\label{f2}
\end{eqnarray}

Equating \ref{f1} and \ref{f2} and using \ref{spinors}, \ref{out1}, we find the following Bogoliubov relations, 
\begin{eqnarray}
\begin{split}
a_{\rm out}(\vec{k},s)&=&\frac{e^{\frac{m \pi}{2H}}}{\sqrt{2\cosh \frac{m \pi }{H}}} \left[a_{\rm in}(\vec{k},s) -e^{-\frac{m\pi}{H}} b^{\dagger}_{\rm in}(-\vec{k},s)\right],\\
b^{\dagger}_{\rm out}(\vec{k},s)&=&\frac{e^{\frac{m \pi}{2H}}}{\sqrt{2\cosh \frac{m \pi }{H}}}  \left[b^{\dagger}_{\rm in}(\vec{k},s) +e^{-\frac{m\pi}{H}} a_{\rm in}(-\vec{k},s)\right]
\label{bglv3}
\end{split}
\end{eqnarray}
For any such  $(\vec{k}, s)$ mode we have the fermionic black body distribution with temperature $H/2\pi$,
\begin{eqnarray}
\langle0_{\rm in}|a^{\dagger}_{\rm out} a_{\rm out}|0_{\rm in}\rangle=|\beta_{{k}}|^2 =\frac1 {e^{\frac{2\pi m}{H}}+1}
\label{crp}
\end{eqnarray}
where 
$$|\alpha_{k}| =\frac{e^{\frac{m \pi}{2H}}}{\sqrt{2\cosh \frac{m \pi }{H}}}, \qquad |\beta_k|=  \frac{e^{-\frac{m \pi}{2H}}}{\sqrt{2\cosh \frac{m \pi }{H}}} \qquad ({\rm for~ both~ }s=1,2)$$
where the absolute values indicate that the Bogoliubov relations can be defined up to some global phase factors.
Note that \ref{crp} is independent of $k$. This originates from the fact that the frequencies of the `out' modes, \ref{out2} are solely determined by the rest mass $m$ of the field. Similar thing occurs for  scalar field theory in the global and the cosmological de Sitter backgrounds~\cite{Mottola:1984ar, Markkanen:2016aes}. Although we are certain that this result for fermions must have been reported earlier, we were unable to locate any particular reference stating the same.

Recall also that a conformally invariant field like a massless fermion cannot create particles in the conformal vacuum of a conformally flat spacetime such as the de Sitter, e.g.~\cite{Parker:2009uva, L. Parker, L. Parker1}. Thus we have an apparent ambiguity with setting $m= 0$ in \ref{crp}. However, we note that in this case the modes \ref{out2} have no positive and negative frequency characteristics at all. Thus it seems that the $m\to 0$ limit for \ref{crp}  is not smooth and one needs to treat the purely massless case separately, retaining explicitly its conformal symmetry by using the conformal vacuum. However, this case will not be relevant for our current purpose. Finally, we note the range of the Bogoliubov coefficient,
$$ 0 \leq |\beta_k| \lesssim 0.707$$
where the upper bound correspond to $m/H \to 0$. Note that, as the limit $H \to 0$ is approached, the number density described by \ref{crp} tends to zero, resembling the outcome of flat spacetime.

 \chapter{}

\section{Propagators in the in-in formalism}
\label{Propagators in the Schwinger-Keldysh Formalism}

\begin{figure}
        \begin{center}
\includegraphics[scale=.37]{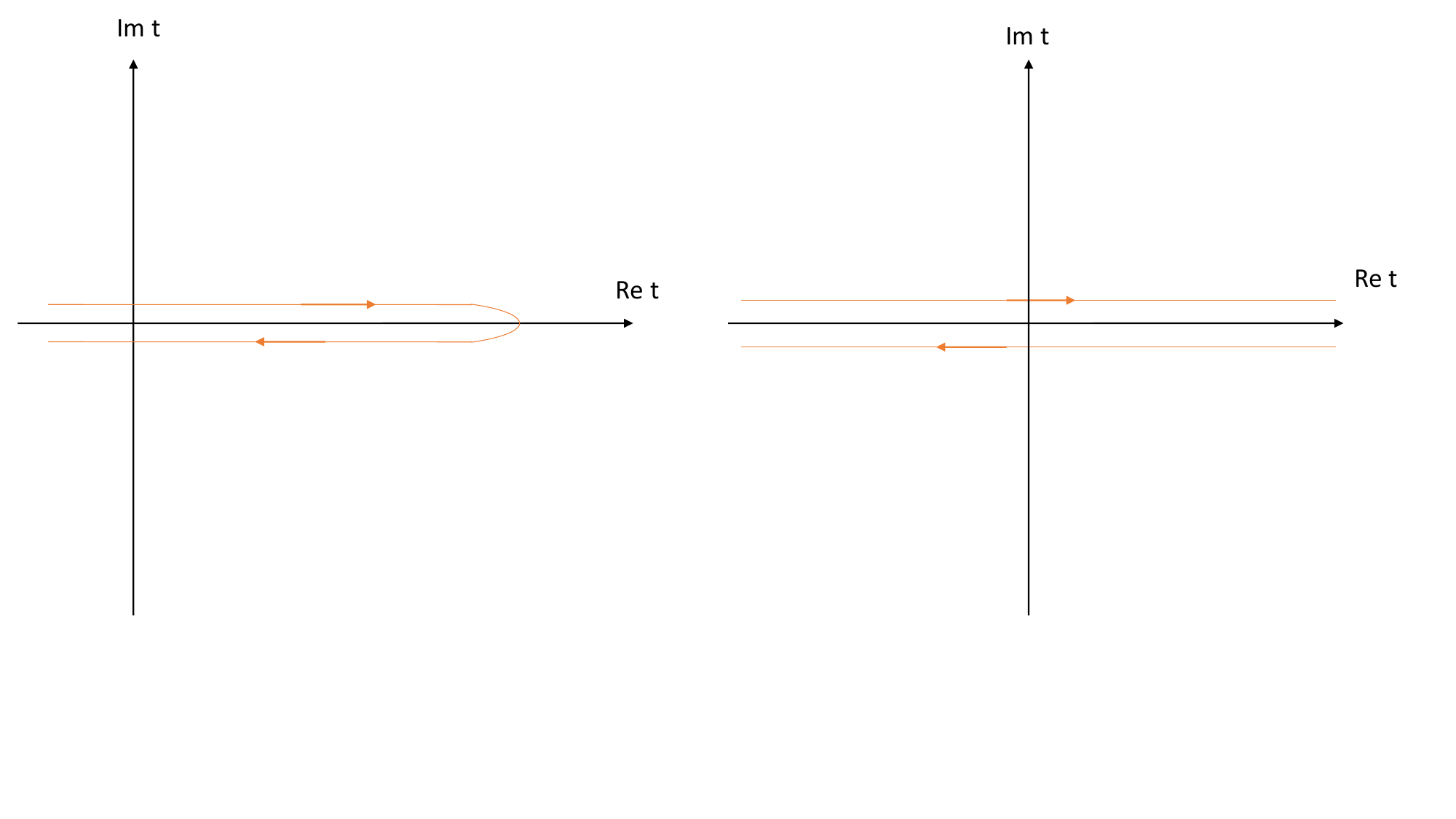}
\vspace*{-20mm}
   {\em \caption{\small \it Schwinger-Keldysh contours with finite and infinite initial and  final times.
   \label{fig:schwingercontour1} }}
        \end{center}
\end{figure}

The conventional in-out formalism for the S-matrix in quantum field theory becomes less applicable in dynamic or nonequilibrium scenarios, particularly when the initial vacuum state can undergo decay through particle pair creation. This situation is exemplified in the cosmological spacetimes we are interested in. The Schwinger-Keldysh formalism, also known as the in-in formalism, proves invaluable when delving into the quantum dynamics of a system within a non-equilibrium context~\cite{Schwinger:1960qe, Keldysh:1964ud}. This framework enables the meaningful computation of a causal expectation value for an operator relative to a suitable initial state. Essentially, it involves tracing the evolution of an operator from an initial state at time $t = t_0$, even without possessing explicit knowledge of the appropriate states at later times. This evolution comprises a time-ordered process from $t_0$ to $t$, followed by an anti-time-ordered process from $t$ back to $t_0$, illustrated in \ref{fig:schwingercontour1}.
 
Thus in the in-in formalism or Schwinger-Keldysh formalism, the expectation value of an operator  $O(t)$ with respect to some initial density operator $\rho(t_{0})$, defined in the Heisenberg picture, is given by,
\begin{equation} \label{expectationvalues}
\resizebox{1.0\hsize}{!}{$ \langle O(t) \rangle =
\mathrm{Tr}\left[\rho(t_{0})O(t)\right] =
\mathrm{Tr}\left[ \rho(t_{0}) \left\{ \overline{T}
\exp\left(\imath \int_{t_0}^t \mathrm{d}t^\prime 
H(t^\prime)\right) \right\} O(t_{0}) \left\{ T \exp
\left(-\imath \int_{t_0}^t \mathrm{d} t'' H(t'')
\right) \right\}\right] $}
\end{equation}
 where $\overline{T}$ stands for the anti-time  ordering,
and $ H(t)$ is the Hamiltonian. The  generating functional for the corresponding path integral  for \ref{action:tree1}, subject to \ref{fig:schwingercontour1} is given by

\begin{multline}\label{Z:inin}
{\cal Z}[J_{+}^{\phi}, J_{-}^{\phi}, J_{+}^{\psi},
J_{-}^{\psi}, \rho(t_0)]
 \quad  \!=\! \int \! {\cal D}\phi^{+}_{0}{\cal D}\phi^{-}_{0}
[{\cal D}\psi^{+}_{0}] [{\cal D}\psi^{-}_{0}] \langle\phi^{+}_{0},
\psi^{+}_{0}| { \rho}(t_{0})|\phi^{-}_{0}, \psi^{-}_{0}
\rangle \\
\times\!\int_{\phi_{0}^{+}}^{\phi_{0}^{-}}\! {\cal
D}\phi^{+}{\cal D}\phi^{-}
\delta[\phi^{+}(t_{f}\!)-\phi^{-}(t_{f}\!)]\!
\int_{\psi_{0}^{+}}^{\psi_{0}^{-}} \![{\cal D}\psi^{+}][{\cal
D}\psi^{-}] [\delta[\psi^{+}(t_{f}\!)-\psi^{-}(t_{f}\!)]]
 \\
\times {\rm exp}\Big[\imath \int
\mathrm{d}^{d-1} \vec{x}\int_{t_{0}}^{t_{f}} \mathrm{d}t^\prime
\Big({\cal L}[\phi^{+},\psi^{+},t']-{\cal
L}[\phi^{-},\psi^{-},t'] +J_{+}^{\phi}\phi^{+} +
J_{-}^{\phi}\phi^{-}\\ + [J_{+}^{\psi} \psi^{+}] +
[J_{-}^{\psi}\psi^{-}] \Big)\Big]
\end{multline}
where for notational convenience we have written $[{\cal D}\psi]={\cal D}\bar{\psi}{\cal D}\psi$, $[J_{+}^{\psi} \psi^{+}]= J_{+}^{\psi} \psi^{+}+\bar{\psi}^{+}\bar{J_+}^{\bar\psi^+} $  etc. 
   The fields $\phi,\, \psi,\,\bar{\psi} $ have their sources $J^{\phi}$,
 $J^{\psi}$ $\bar{J}^{\bar\psi}$ respectively. Note that there are two species of fields and sources for each category. The $+$ sign denotes forward evolution in time whereas the $-$ sign denotes backward evolution. The $\delta$-function ensures the field configurations are the same on the final hypersurface at $t=t_f$. The expectation values of $n$-point functions can be found by taking functional differentiation of the generating functional~\ref{Z:inin} with respect to the sources $J$'s, for instance
\begin{equation} \label{npointfunctions}
\left. \mathrm{Tr}\left[ {\rho}(t_{0})
\overline{T}[\phi(x_1)\dots \phi(x_n)]
T[\phi(y_1)\dots \phi(y_k)] \right] =
\frac{\delta^{n+k}{\cal Z}[J, \rho(t_0)] } {\imath\delta \!
J_{-}^{\phi}(x_{1})\cdots \imath\delta\! J_{-}^{\phi}(x_{n})
\imath\delta\! J_{+}^{\phi}(y_{1})\cdots \imath\delta\!
J_{+}^{\phi}(y_{k})} \right|_{J = 0}  
\end{equation}

With the help of the above, we now define the following propagators for the scalar field,

\begin{multline}\label{propagators}
\imath\Delta^{++}_{\phi}(x,x^\prime) = \mathrm{Tr}\left[{\rho}(t_{0})
T[\phi(x^\prime)\phi(x)] \right]  =
 \mathrm{Tr}\left[{\rho}(t_{0}) \phi^+(x)\phi^+(x^\prime)\right] = 
\left.\frac{\delta^2{\cal Z}[J, \rho(t_0)]} {\imath\delta\!
J_{+}^{\phi}(x) \imath\delta\! J_{+}^{\phi}(x^\prime)}
\right|_{J=0} \\ 
\imath\Delta^{--}_{\phi}(x,x^\prime)  =
 \mathrm{Tr}\left[{\rho}(t_{0}) \overline{T} [ \phi(x^\prime)\phi(x)]
\right] = \mathrm{Tr}\left[{\rho}(t_{0}) \phi^-(x^\prime)\phi^-(x)\right] =
\left. \frac{\delta^2{\cal Z} [J, \rho(t_0)]} {\imath\delta\!
J_{-}^{\phi}(x)\imath\delta\! J_{-}^{\phi}(x^\prime)}
\right|_{J=0} \\
\imath\Delta^{-+}_{\phi}(x,x^\prime) =
 \mathrm{Tr}\left[{\rho}(t_{0})  \phi(x)\phi(x^\prime)\right] = \mathrm{Tr}\left[{\rho}(t_{0}) \phi^-(x)\phi^+(x^\prime)\right] =
\left.\frac{\delta^2{\cal Z} [J, \rho(t_0)] } {\imath\delta\!
J_{-}^{\phi}(x)\imath\delta\! J_{+}^{\phi}(x^\prime)}
\right|_{J=0}\\
\imath\Delta^{+-}_{\phi}(x,x^\prime) =
 \mathrm{Tr}\left[{\rho}(t_{0})\phi(x^\prime)\phi(x)\right] =
 \mathrm{Tr}\left[{\rho}(t_{0})\phi^-(x^\prime)\phi^+(x)\right]=
\left.\frac{\delta^2{\cal Z} [J, \rho(t_0)] } {\imath\delta\!
J_{+}^{\phi}(x)\imath\delta\! J_{-}^{\phi}(x^\prime)}
\right|_{J=0}\\
\end{multline}
where we have taken $t>t'$ above.
The time ordered and the anti-time ordered propagators are respectively the Feynman and anti-Feynman propagators and the rest are the two Wightman functions. One can also write
\begin{eqnarray}
\label{propagatoridentities}
\imath\Delta^{++}_{\phi}(x,x^\prime) &=& \theta(t-t^\prime)\imath
\Delta^{-+}_{\phi}(x,x^\prime) + \theta(t^\prime-t)\imath
\Delta^{+-}_{\phi}(x,x^\prime) \nonumber
\\
\imath\Delta^{--}_{\phi}(x,x^\prime) &=& \theta(t^\prime-t)\imath
\Delta^{-+}_{\phi}(x,x^\prime) + \theta(t-t^\prime)\imath
\Delta^{+-}_{\phi}(x,x^\prime) 
\end{eqnarray}
These propagators and  Wightman functions also satisfy the following properties, 
\begin{eqnarray}
\imath\Delta^{++}_{\phi}(x,x^\prime) +
\imath\Delta^{--}_{\phi}(x,x^\prime) &=& \imath
\Delta^{-+}_{\phi}(x,x^\prime) +\imath
\Delta^{+-}_{\phi}(x,x^\prime) \label{propagatoridentitiesc}
\\
\imath \Delta^{-+}_{\phi}(x,x^\prime)&=&\imath
\Delta^{+-}_{\phi}(x^\prime,x) \label{propagatoridentitiesd}
\end{eqnarray}
The retarded  and the advanced
 propagators, useful for our later purpose, are respectively defined as
\begin{subequations}
\label{propagators2}
\begin{eqnarray}
\imath\Delta^{\mathrm{r}}_{\phi}(x,x^\prime) &=& \imath
\Delta^{++}_{\phi}(x,x^\prime) - \imath
\Delta^{+-}_{\phi}(x,x^\prime)
\\
\imath\Delta^{\mathrm{a}}_{\phi}(x,x^\prime)&=& \imath
\Delta^{++}_{\phi}(x,x^\prime) - \imath
\Delta^{-+}_{\phi}(x,x^\prime)
\label{Delta:adv}
\end{eqnarray}
\end{subequations}
One also defines the spectral two point function and the statistical propagator as (e.g.~\cite{koksma}),
\begin{eqnarray} \label{Delta:causal}
&&\imath\Delta^{c}_{\phi} (x,x^\prime) = \mathrm{Tr}\left(
{\rho}(t_{0})  [\phi(x),\phi(x^\prime)]\right)= \imath
\Delta^{-+}_{\phi}(x,x^\prime) - \imath
\Delta^{+-}_{\phi}(x,x^\prime)\nonumber\\
&&F_{\phi}(x,x') = \frac{1}{2} \mathrm{Tr}\left[ {\rho}(t_{0})
[ \phi(x),\phi(x')]_+ \right]=
\frac{1}{2}\Big(\imath\Delta^{-+}_{\phi}(x,x') +
\imath\Delta^{+-}_{\phi}(x,x')\Big) 
\end{eqnarray}
The spectral function provides insights into the system's states and spectrum, but it doesn't convey information about the occupation of these states. In contrast, the statistical propagator offers valuable information about state population. Consequently, the latter is more pertinent for investigating decoherence and entropy, as discussed in~\cite{NEQFT} also references therein.

In terms of the retarded, advanced, spectral and statistical two-point functions, we have

\begin{subequations}
\label{reduction:F+Deltac}
\begin{eqnarray}
\imath \Delta^{+-}_{\phi}(x,x^\prime) &=& F_{\phi}(x,x^\prime)-
\frac{1}{2}\imath\Delta^{c}_{\phi}(x,x^\prime)
\label{reduction:F+Deltaca}\\
\imath \Delta^{-+}_{\phi}(x,x^\prime) &=& F_{\phi}(x,x^\prime) +
\frac{1}{2}\imath\Delta^{c}_{\phi}(x,x^\prime)
\label{reduction:F+Deltacb}\\
\imath \Delta^{++}_{\phi}(x,x^\prime) &=& F_{\phi}(x,x^\prime) +
\frac{1}{2}\mathrm{sgn}(t- t^\prime)
\imath\Delta^{c}_{\phi}(x,x^\prime)
\label{reduction:F+Deltacc}\\
\imath \Delta^{--}_{\phi}(x,x^\prime) &=& F_{\phi}(x,x^\prime)-
\frac{1}{2}\mathrm{sgn}(t-t^\prime)\imath\Delta^{c}_{\phi}(x,x^\prime)
\,. \label{reduction:F+Deltacd}
\end{eqnarray}
\end{subequations}

The free Feynman, anti-Feynman and the Wightman  functions  satisfy
\begin{equation} \label{Feynman propagator}
(\partial_x^2 - m^2)\imath \Delta^{ss'}_{\phi}(x,x^\prime) =
\imath s \delta_{ss'}\delta^{\scriptscriptstyle{D}}\!(x-x^\prime)\qquad (s,s'=\pm,\,\,\,\,{\rm no~sum~on~}s)
\end{equation}

The fermionic field propagators can be similarly defined, taking into account their anti-commutation properties and replacing the d'Alembertian operator $\partial^2$ with $\slashed{\partial}$. It's important to note that we are considering a zero-temperature field theory, so the various averages mentioned in \ref{propagators} should be interpreted as vacuum expectation values.

We shall be using the Schwinger-Keldysh formalism for decoherence computation in the inflationary de Sitter spacetime also, where cosmological time $t$ will be replaced with the conformal time $\eta$ in above equations. Technicality, for this spacetime will remain the same as the Minkowski spacetime. With these equipments, we shall be computing the loop correction to the two-point functions and to find out the Kadanoff-Baym equations in \ref{chapter3}.


\section{Calculation of renormalised $\imath M_{\phi,{\rm ren}}^{++}(k)$ and $\imath M_{\phi,{\rm ren}}^{--}(k)$}
\label{A1}

In this appendix, we compute the renormalised self-energies $\imath M_{\phi}^{++}$ and $\imath M_{\phi}^{--}$
in Fourier space, to be useful for our future purpose. From \ref{selfMassa}, we have in momentum space
\begin{eqnarray}\label{SelfMassFourier1}
\imath M_{\phi}^{++}(k) 
&=& \imath g^{2} {\rm Tr} \int
\frac{\mathrm{d}^{\scriptscriptstyle{d}}p}{(2\pi)^{\scriptscriptstyle{d}}}\Big[
\frac{ \slashed{p}}{p^2-\imath \epsilon} \times
\frac{(\slashed{p}+\slashed{k})}{(p+k)^2-\imath \epsilon}\Big]\nonumber \\
&=& 4\imath g^{2} \int
\frac{\mathrm{d}^{\scriptscriptstyle{d}}p}{(2\pi)^{\scriptscriptstyle{d}}}
\frac{1}{p^2-\imath \epsilon}
\frac{(p^2+k\cdot p)}{(p+k)^2-\imath \epsilon}\nonumber \\
&=& 4\imath g^{2} \int_{0}^{1} \mathrm{d}x \int
\frac{\mathrm{d}^{\scriptscriptstyle{d}}q}{(2\pi)^{\scriptscriptstyle{d}}}
\frac{q^2-k^2x(1-x)}{(q^2+k^2x(1-x)-\imath
\epsilon)^{2}}\nonumber 
\end{eqnarray}
where we have  used Feynman's trick (e.g.~\cite{Peskin:2018}). Here $p$ and $k$ are the internal and external momentum corresponding to fermion and scalar propagators respectively.  We have also used
$q=p+xk$, as the new integration
variable. Performing now the usual Euclideanisation of $q$ in the last integral of the above equation, we have
\begin{eqnarray}\label{rr}
\imath M_{\phi}^{++}(k)=-4g^{2} \int_{0}^{1} \mathrm{d}x \int\frac{\mathrm{d}^{\scriptscriptstyle{d}}q_E}{(2\pi)^{\scriptscriptstyle{d}}}
\frac{q_{E}^{2}-k^{2}x(1-x)}{(q_{E}^{2}+k^{2}x(1-x)-\imath\epsilon)^{2}}
\end{eqnarray}
The integral can now straightforwardly be performed,
 yielding
\begin{equation}\label{SelfMassFourier2}
\imath M_{\phi}^{++}(k) = -\frac{g^{2}\mu^{d-4}k^2}{4 \pi^{2}}\Big(\frac{1}{d-4}+\frac{\gamma_E}{2}+\ln \sqrt{\pi}\Big)+
\frac{g^{2}k^2}{8\pi^{2}}
\ln\left(\frac{k^2-\imath\epsilon}{4
\mu^{2}}\right) + \mathcal{O}(d-4)
\end{equation}
As of the coordinate space expression \ref{SelfMassPosspace4Taylor}, the divergence can be absorbed in the scalar field strength renormalisation.   Accordingly we have 
\begin{equation}\label{renM++}
\imath M_{\phi,{\rm ren}}^{++}(k)= 
\frac{g^{2}k^2}{8\pi^{2}}
\ln\left(\frac{k^2-\imath\epsilon}{4
\mu^{2}}\right) + \mathcal{O}(d-4)
\end{equation}
$\imath M_{\phi}^{--}(k)$ on the other hand, will be the negative  
of the complex conjugation of the above, follows directly from \ref{selfmassa}, 
\begin{equation}\label{renM--}
\imath M_{\phi,{\rm ren}}^{--}(k)= -
\frac{g^{2}k^2}{8\pi^{2}}
\ln\left(\frac{k^2+\imath\epsilon}{4
\mu^{2}}\right) + \mathcal{O}(d-4)
\end{equation}
We note that the above expression is similar to that  of the case when the environment is also a scalar \cite{koksma}, however in our  case we have one extra factor of $k^2$ multiplied with the logarithm.

\section{Computations for $\imath M_{\phi,\mathrm{ren}}^{\mathrm{r}}(k)$, $\imath M_{\phi,\mathrm{ren}}^{\mathrm{a}}(k)$,  $\imath M^{+-}_{\phi}(k)$ and $\imath M^{-+}_{\phi}(k)$   }\label{A2}

In this appendix, we wish to find out the momentum space renormalised expressions of the retarded and advanced self energies as well as  the self energies corresponding to the Wightman functions $\imath M^{+-}_{\phi}(k)$ and $\imath M^{-+}_{\phi}(k)$, as dictated by \ref{statprop1}.

The renormalised expression for the retarded self-energy in coordinate space was found in \ref{Re1}. We shall now take the Fourier transform of it. We do it in two steps, for the sake of convenience of calculations. We first take the Fourier transformation of \ref{Re1} with respect to its spatial part  only
\begin{eqnarray}\label{RetardedSelfMass2}
\imath M_{\phi,\mathrm{ren}}^{\mathrm{r}}(|\vec{k}|, t, t') &=&
\frac{g^{2}}{64 \pi^{3}} (\partial^{2}_{t}+|\vec{k}|^{2})^{3} \int
\mathrm{d}^{3}\Delta{\vec x}\,\, \theta(\Delta
t^2-\Delta x^2)\,\theta(\Delta t)  \nonumber\\ && \left[1 - \ln\left(\mu^{2}(\Delta
t^{2}-\Delta x^2)\right) \right] e^{-\imath \vec{k}\cdot\Delta\vec{x}}\nonumber\\
&=& \frac{g^{2}}{16 \pi^{2} |\vec{k}| } (\partial^{2}_{t}+|\vec{k}|^{2})^{3}
\theta(\Delta t) \Delta t^{2} \Bigg[  \frac{\sin(|\vec{k}| \Delta t)-|\vec{k}|
\Delta t \cos(|\vec{k}| \Delta t)}{(|\vec{k}| \Delta
t)^{2}} \nonumber \\ && \left(1-\ln(\mu^{2}\Delta
t^{2}) \right) - \int_{0}^{1} \mathrm{d}z
\, z \sin(|\vec{k}| \Delta t z)\ln\left(1-z^{2}\right) \Bigg] 
\end{eqnarray}
 The last integral is given by the special function~\cite{AS},
\begin{eqnarray}\label{eu}
&&\xi(|\vec{k}| \Delta t) = \int_{0}^{1} d z z \sin (|\vec{k}| \Delta t z) \ln \left(1-z^{2}\right)\nonumber\\
&&= \frac{2}{(|\vec{k}| \Delta t)^{2}} \sin (|\vec{k}| \Delta t)-\frac{1}{(|\vec{k}| \Delta t)^{2}}[\cos (|\vec{k}| \Delta t)+|\vec{k}| \Delta t \sin (|\vec{k}| \Delta t)]\left[\operatorname{si}(2 |\vec{k}| \Delta t)+\frac{\pi}{2}\right]\nonumber\\&&+\frac{1}{(|\vec{k}| \Delta t)^{2}}[\sin (|\vec{k}| \Delta t)-(|\vec{k}| \Delta t) \cos(|\vec{k}| \Delta t)]\left[\operatorname{ci}(2 |\vec{k}| \Delta t)-\gamma_E-\ln \left(\frac{(|\vec{k}| \Delta t)}{2}\right)\right]
\end{eqnarray}
where $\operatorname{si}(x)$ and $\operatorname{ci}(x)$ are the sine and cosine integral functions, given by \cite{AS}
\begin{eqnarray}\label{sici}
\begin{aligned}
\operatorname{si}(x) &=-\int_{x}^{\infty} d t \frac{\sin t}{t}=-\frac{\pi}{2}+\int_{0}^{x} d t \frac{\sin t}{t} \\
\operatorname{ci}(x) &=-\int_{x}^{\infty} d t \frac{\cos t}{t}=\gamma_E+\ln x+\int_{0}^{x} d t\left[\frac{\cos t-1}{t}\right]
\end{aligned}
\end{eqnarray}
Using  \ref{eu} into \ref{RetardedSelfMass2}, we arrive at 
\begin{eqnarray}\label{RetardedSelfMass3}
\imath M_{\phi,\mathrm{ren}}^{\mathrm{r}}(|\vec{k}|, t, t') &=&
\frac{g^{2}}{16  \pi^{2} |\vec{k}|^{3}} (\partial^{2}_{t}+|\vec{k}|^{2})^{3}
\theta(\Delta t) \Bigg[  \left( |\vec{k}| \Delta t \cos(|\vec{k}| \Delta t) -
\sin(|\vec{k}| \Delta t)\right) \nonumber \\
&& \left(\mathrm{ci}(2|\vec{k}|\Delta t) - \gamma_{E} - \ln\left(\frac{|\vec{k}|}{2\mu^{2}\Delta t}\right) -1 \right) +\left( \cos(|\vec{k}|\Delta t)+
|\vec{k}|\Delta t \sin(|\vec{k}|\Delta
t)\right) \nonumber \\
&& \qquad\qquad\qquad \qquad \left(\frac{\pi}{2}+\mathrm{si}(2|\vec{k}|\Delta
t)\right)-2\sin(|\vec{k}|\Delta t) \Bigg] \,
\end{eqnarray}
Note that the $\theta(\Delta t)$ appearing above commutes with  one of the $(\partial^{2}_{t}+|\vec{k}|^{2})$ operators because the term in square brackets is proportional to $(\Delta
t)^2$ when $\Delta t \rightarrow 0$.
Accordingly, by taking one $(\partial^{2}_{t}+|\vec{k}|^{2})$, we have
\begin{multline}\label{RetardedSelfMassFinalResult}
\imath M_{\phi,\mathrm{ren}}^{\mathrm{r}}(|\vec{k}|, t, t') =
\frac{g^{2}}{8 \pi^{2} |\vec{k}| } (\partial^{2}_{t}+|\vec{k}|^{2})^2\theta(\Delta
t) \Bigg[  \cos(|\vec{k}| \Delta t)\\
\left(\frac{\pi}{2}+\mathrm{si}(2|\vec{k}|\Delta t)\right) - \sin(|\vec{k}| \Delta
t) \left(\mathrm{ci}(2|\vec{k}|\Delta t) - \gamma_{E} -
\ln\left(\frac{|\vec{k}|}{2\mu^{2}\Delta t}\right)\! \right)\!\Bigg]\!
\end{multline}
We next take the Fourier transform of the above equation with respect to time as well
\begin{multline}
\imath M^{\mathrm{r}}_{\phi,\mathrm{ren}}(k) =
\int_{-\infty}^{\infty}\mathrm{d}\Delta t e^{\imath k^{0}\Delta t}
\imath M^{\mathrm{r}}_{\phi,\mathrm{ren}}(|\vec{k}|,t,t^\prime) \label{Fourierretarded1} = - \frac{g^{2}}{16\pi^{2}|\vec{k}|} (-(k^{0})^{2}+|\vec{k}|^{2})^2
\\ \times \int_{0}^{\infty}\mathrm{d}\Delta t \Bigg[
e^{\imath(k^{0}+|\vec{k}|)\Delta t}
\left\{-\imath\left(\mathrm{ci}(2|\vec{k}|\Delta t)-\ln(2|\vec{k}|\Delta
t)-\gamma_{E}\right) -\frac{\pi}{2}-\mathrm{si}(2|\vec{k}|\Delta t) \right\}\\ +
e^{\imath(k^{0}-|\vec{k}|)\Delta t}
\left\{\imath\left(\mathrm{ci}(2|\vec{k}|\Delta t)-\ln(2|\vec{k}|\Delta
t)-\gamma_{E}\right) -\frac{\pi}{2}-\mathrm{si}(2|\vec{k}| \Delta
t) \right\}\\ -2
\imath \ln\left(2\mu\Delta t
\right)\left(e^{\imath(k^{0}+|\vec{k}|+\imath\epsilon)\Delta t} -
e^{\imath(k^{0}-|\vec{k}|+\imath\epsilon)\Delta t}\right) \Bigg] 
\end{multline}
where we have introduced $\epsilon=0^+$  at necessary places, in order to regularise the integral. In order to evaluate the above equation further, we make use of the following equations \cite{AS}
\begin{subequations}
\begin{equation}
\int_{0}^{\infty}\mathrm{d}z \ln(\beta z) e^{\imath \alpha z} =
-\frac{\imath}{\alpha}\left[\ln\left(\frac{-\imath \alpha +
\epsilon}{\beta}\right)+\gamma_{E}\right]
 \label{Fourierretarded2} 
\end{equation}
\begin{equation}
e^{\imath(k^{0} \pm |\vec{k}|)\Delta t} = \frac{-\imath}{k^{0} \pm
|\vec{k}|}\partial_{t} e^{\imath(k^{0} \pm |\vec{k}|)\Delta t}
\label{Fourierretarded3} 
\end{equation}
\end{subequations}
Using these, \ref{Fourierretarded1} evaluates to
\begin{eqnarray}
\imath M^{\mathrm{r}}_{\phi,\mathrm{ren}}(k)&=&-
\frac{g^{2}}{16|\vec{k}|\pi^{2}}(-k_{0}^{2}+|\vec{k}|^{2}) \Bigg[ 2(k^{0}-|\vec{k}|)\left(
\ln\left(\frac{-\imath(k^{0}+|\vec{k}|)+\epsilon}{2\mu}\right) +
\gamma_{E}\right) \nonumber \\ && - 2(k^{0}+|\vec{k}|)\left(
\ln\left(\frac{-\imath(k^{0}-|\vec{k}|)+\epsilon}{2\mu}\right) +
\gamma_{E}\right) 
 \nonumber \\ && + \int_{0}^{\infty}\mathrm{d}\Delta t \frac{2
k^{0}}{\Delta t}\left(e^{\imath(k^{0}+|\vec{k}|)\Delta t} -
e^{\imath(k^{0}-|\vec{k}|)\Delta t}\right)\Bigg] 
\label{Fourierretarded4}
\end{eqnarray}
To evaluate the remaining integrals, we use, for  real $\alpha , \beta $, \cite{AS}
\begin{equation}
\int_{0^+}^{\infty}\mathrm{d}\Delta t\left[
\frac{\cos(\alpha\Delta t)-1}{\Delta t} - \frac{\cos(\beta\Delta
t)-1}{\Delta t}\right] = \ln\left\vert\frac{\beta}{\alpha}\right\vert
\label{Fourierretarded5} 
\end{equation}
Putting these all in together, we obtain
\begin{equation}
\imath M^{\mathrm{r}}_{\phi,\mathrm{ren}}(k) =
\frac{g^{2}}{8
\pi^{2}}(-k_{0}^{2}+|\vec{k}|^{2})\left[\ln\left(\frac{-k_{0}^{2}+|\vec{k}|^{2}-\imath
\mathrm{sgn}(k^{0})\epsilon}{4\mu^{2}}\right)+2\gamma_{E}\right]
\label{Fourierretarded6}
\end{equation}
The additive constant  $2\gamma_E$ appearing in the above equation  can be further  absorbed in a scalar field strength renormalisation counterterm as of \ref{A1}.
Thus we have the final expression
\begin{equation}
\imath M^{\mathrm{r}}_{\phi,\mathrm{ren}}(k) =
\frac{g^{2} k^2}{8
\pi^{2}}\ln\left(\frac{k^2-\imath
\mathrm{sgn}(k^{0})\epsilon}{4\mu^{2}}\right)
\label{Fourierretarded6}
\end{equation}
%
We next wish to identify the self-energies $\imath
M^{+-}_{\phi}(k)$ and $\imath M^{-+}_{\phi}(k)$, corresponding to the Wightman functions. The simplest way to achieve this is to use the retarded self-energy \ref{Fourierretarded6}, and $\imath
M^{++}_{\phi,\mathrm{ren}}(k)$  derived
in \ref{A1}. The relationship between $\imath
M^{++}_{\phi,\mathrm{ren}}(k)$, $\imath
M^{--}_{\phi,\mathrm{ren}}(k)$, $\imath M^{\mathrm{r}}_{\phi,\mathrm{ren}}(k)$, $\imath
M^{+-}_{\phi}(k)$ and $\imath
M^{-+}_{\phi}(k)$ is given by the renormalised version of the equation appearing below \ref{FourierWightmanb1},  
\begin{equation}\label{kk}
\imath M^{\mathrm{r}}_{\phi,\mathrm{ren}}(k) = \imath
M^{++}_{\phi,\mathrm{ren}}(k) - \imath
M^{+-}_{\phi}(k)=\imath
M^{-+}_{\phi}(k) - \imath M^{--}_{\phi,\mathrm{ren}}(k)
\end{equation} 
 Using now \ref{renM++}, \ref{renM--}, \ref{Fourierretarded6} and \ref{kk}, we  find out the Wightman self-energies
\begin{subequations}
\label{FourierWightman2}
\begin{eqnarray}
\imath M^{+-}_{\phi}(k) &=& - \frac{\imath g^{2}k^2}{4\pi}\theta(-k^{0}-|\vec{k}|) \label{FourierWightman2a} \\
\imath M^{-+}_{\phi}(k) &=& - \frac{\imath
g^{2}k^{2}}{4\pi}\theta(k^{0}-|\vec{k}|) \label{FourierWightman2b} 
\end{eqnarray}
\end{subequations}

Likewise, we can find out the the renormalised advanced self-energy by using the renormalised version of the equation appearing below  \ref{FourierAdvanced1},
\begin{equation}
    \imath M^{\mathrm{a}}_{\phi,\mathrm{ren}}(k)= \imath
M^{++}_{\phi,\mathrm{ren}}(k) - \imath
M^{-+}_{\phi}(k) =  \imath M^{+-}_{\phi}(k) - \imath
M^{--}_{\phi,\mathrm{ren}}(k)
\label{advanced1}
\end{equation}
Using \ref{renM++}, \ref{renM--} and \ref{FourierWightman2}, we have
\begin{equation}
    \label{Fourieradvance6}
    \imath M^{\mathrm{a}}_{\phi,\mathrm{ren}}(k) =
\frac{g^{2}k^2}{8
\pi^{2}}\ln\left(\frac{k^2+\imath
\mathrm{sgn}(k^{0})\epsilon}{4\mu^{2}}\right)
\end{equation}
Being equipped with all these, we have computed the renormalised statistical propagator in  momentum space as quoted in the main text,~\ref{statprop1'}.\\

\noindent
 Finally, as a check of consistency, from \ref{FourierAdvanced1},  \ref{FourierWightman1}, \ref{Fourierretarded6} and \ref{Fourieradvance6}, we compute the Wightman functions,
\begin{subequations}
\label{FourierWightman3}
\begin{eqnarray}
\imath \Delta^{+-}_{\phi}(k) &=& \imath \theta(-k^{0}-|\vec{k}|)
\Bigg[\frac{1}{k^2+m^{2}+\frac{g^{2}k^2}{8
\pi^{2}}\ln\left(\frac{k^2}{4\mu^{2}}
\right)- \frac{\imath
g^{2}k^2}{8\pi}\mathrm{sgn}(k^{0})\theta(k_{0}^{2}-|\vec{k}|^{2})} \nonumber\\
&& \qquad\qquad-
\frac{1}{k^2+m^{2}+\frac{g^{2}k^2}{8
\pi^{2}}\ln\left(\frac{k^2}{4\mu^{2}}\right)
+ \frac{\imath
g^{2}k^2}{8\pi}\mathrm{sgn}(k^{0})\theta(k_{0}^{2}-|\vec{k}|^{2})} \Bigg]  \\ \label{FourierWightman3a} 
\imath \Delta^{-+}_{\phi}(k) &=& - \imath \theta(k^{0}-|\vec{k}|)
\Bigg[\frac{1}{k^2+m^{2}+\frac{g^{2}k^2}{8
\pi^{2}}\ln\left(\frac{k^2}{4\mu^{2}}\right)- \frac{\imath
g^{2}k^2}{8\pi}\mathrm{sgn}(k^{0})\theta(k_{0}^{2}-|\vec{k}|^{2})} \nonumber \\
&& \qquad\qquad-
\frac{1}{k^2+m^{2}+\frac{g^{2}k^2}{8
\pi^{2}}\ln\left(\frac{k^2}{4\mu^{2}}
\right)+ \frac{\imath
g^{2}k^2}{8\pi}\mathrm{sgn}(k^{0})\theta(k_{0}^{2}-|\vec{k}|^{2})} \Bigg] 
\label{FourierWightman3b} 
\end{eqnarray}
\end{subequations}
Using 
\begin{equation}
    \lim_{\epsilon \to 0} \frac{1}{x\pm i \epsilon}= \text{PV} \frac{1}{x} \mp i \pi \delta(x), 
\end{equation}
it is easy to see that in the limit $g \rightarrow 0$, the  above expressions agree
with the free Wightman functions
 \begin{subequations}
 \label{VacuumPropagator}
 \begin{eqnarray}
 \imath\Delta_{\phi}^{+-}(k) &=& 2\pi
 \delta(k^2+m^2) \theta(-k^{0})
 \label{VacuumPropagator+-}
 \\
 \imath\Delta_{\phi}^{-+}(k) &=& 2\pi
 \delta(k^2+m^2) \theta(k^{0}) 
 \label{VacuumPropagator-+}
 \end{eqnarray}
 \end{subequations}
%

\section{Some useful identities \label{defAndConv}}

We have to frequently use the following special functions for our present purpose \cite{AS, Gr},
\begin{eqnarray}\label{ci,si}
\text{si} (z)&=&-\int_{z}^{\infty}\frac{\sin t \,dt}{t}=
\int_{0}^{z}\frac{\sin t \,dt}{t}-\frac{\pi}{2}
\, 
\label{sine integral}\\
\text{ci}(z)&=&-\int_{z}^{\infty}\frac{\cos t\,dt}{t}
\, 
\label{cos integral}
\end{eqnarray}
where  the first and the second  are respectively the sine and cosine  integral functions. 

We shall use some  relations derived in~\cite{Prokopec:2008gw} for the complexified de Sitter invariant $2$-point functions, $y_{\pm\pm}$, defined in \ref{distance}, \ref{distance1} and \ref{distance2}. For example, we have 
\begin{eqnarray}
 \Big( {\frac{y_{\pm \pm}}{4}}\Big)^{1-{d}} = \Big[\frac{2}{(d-2)^2} \frac{\square}{H^2} -\frac{2}{(d-2)}\Big] \Big( {\frac{y_{\pm \pm}}{4}}\Big)^{2-{d}} 
\label{1}   
\end{eqnarray}
We also have 
\begin{multline}
\Big( {\frac{y_{\pm \pm}}{4}}\Big)^{2-{d}} = \Big[\frac{2}{(d-3)(d-4)} \frac{\square}{H^2} -\frac{d(d-2)}{2 (d-3)(d-4)}   + \frac{d-6}{2(d-3)} \Big]\Big( {\frac{y_{\pm \pm}}{4}}\Big)^{3-{d}} \\
-\Big[ \frac{2}{(d-3)(d-4)} \frac{\square}{H^2} - \frac{d(d-2)}{2 (d-3)(d-4)} \Big]\Big( {\frac{y_{\pm \pm}}{4}}\Big)^{1-d/2} \\ \pm \frac{2(4 \pi)^{d/2}}{(d-3)(d-4)\Gamma\big[\frac{d}{2}-1 \big]} \frac{i \delta^d (x-x^{\prime})}{(Ha)^d}\label{deSitterRel}
\end{multline}
as well as
\begin{eqnarray}
 \frac{\square}{H^2}\Big( {\frac{y_{\pm \pm}}{4}}\Big)^{1-d/2} = \pm \frac{(4 \pi)^{d/2}}{\Gamma\big[\frac{d}{2}-1 \big]} \frac{i \delta^d (x-x^{\prime})}{(Ha)^d} + \frac{d(d-2)}{4}\Big( {\frac{y_{\pm \pm}}{4}}\Big)^{1-d/2}
\end{eqnarray}
Introducing a renormalisation scale $\mu$ with mass dimension one, we can modify \ref{deSitterRel} by including a term dependent on $\mu$. This modification allows us to eliminate the divergence in the self energy using appropriate counterterms in the action.
Additionally, we have
\begin{eqnarray}
  \Big( {\frac{y_{\pm \pm}}{4}}\Big)^{3-{d}} = \Big( {\frac{y_{\pm \pm}}{4}}\Big)^{1-d/2}\Big[1- \frac{d-4}{2}\ln   y_{\pm \pm}  + \mathcal{O}(d-4)^2  \Big]  
\end{eqnarray}
Putting things together now, we have from \ref{deSitterRel},
\begin{multline}
\Big( {\frac{y_{\pm \pm}}{4}}\Big)^{2-{d}} =
 \pm \frac{2 (4 \pi )^{d/2} }{(d-3)(d-4)\Gamma\big[\frac{d}{2}-1\big] } \Big(  \frac{ \mu}{H}\Big) ^{d-4}  \frac{i \delta^d \big(x-x^{\prime} \big)}{(Ha)^d}
\\-  \frac{\square}{H^2} \Big(\frac{4}{y_{\pm \pm}}\ln \frac{\mu^2 y_{\pm \pm}}{H^2} \Big) + \frac{4}{y_{\pm \pm}} \Big(2 \ln  \frac{\mu^2 y_{\pm \pm} }{H^2} -1 \Big)  + \mathcal{O}(d-4)
\end{multline}

We also note that for any function $f(y)$, not containing $y^{-1}$, we have 
\begin{align}\label{i1}
\frac{\square}{H^2} f(y) = (4-y) y f^{\prime \prime} (y) + 4(2-y)f^{\prime}(y)
\end{align}
where a prime denotes derivative once with respect to $\eta$. The above gives identities useful for our purpose
\begin{eqnarray}\label{i2}
\frac{1}{y} &=& \frac{1}{4} \frac{\square}{H^2} \ln y + \frac{3}{4}\, \\
\frac{\ln y}{y}   &=&
  \frac{1}{8} \frac{\square}{H^2} \Big[ \ln^2 y - 2  \ln y  \Big]
  + \frac{3}{4} \ln  y  	
- \frac{1}{2}\,
\end{eqnarray}

We also note  from \ref{distance1}, \ref{distance2} that as $\epsilon\to 0$
\begin{eqnarray}
&&\ln \big( \Delta x_{\mp \pm}^2 \big) = \ln  \big|\Delta \eta^2 - r^2 \big| 
 \pm  i \pi \, \text{sign}\big(\eta,\eta^{\prime} \big) \theta \big( \Delta \eta^2 - r^2 \big) \nonumber\\
&& \ln \big( \Delta x_{\pm \pm}^2 \big) = \ln  \big|\Delta \eta^2 -r^2 \big| 
 \pm  i \pi \,  \theta \big( \Delta \eta^2 - r^2 \big) 
\end{eqnarray}\label{log}
where $\text{sign}(\eta,\eta^{\prime})$ is the usual signum function.

For our purpose of solving the Kadanoff-Baym equations, \ref{2PF}, it will be convenient  to define \cite{Friedrich:2019hev},
\begin{eqnarray}\label{newenergies}
M^{F}(\eta, \eta^{\prime}) &:=& \frac{1}{2}\Big[M^{++}(\eta, \eta^{\prime})+M^{--}(\eta, \eta^{\prime}) \Big] = \text{Re} \, M^{++}(\eta, \eta^{\prime})\, , \\ \nonumber
M^{c}(\eta, \eta^{\prime}) &:=&-i\,\text{sign} \big(\Delta\eta \big) \Big[M^{++}(\eta, \eta^{\prime})-M^{--}(\eta, \eta^{\prime})\Big] \\&=&2 \, \text{sign} \big(\Delta\eta \big)\,  \text{Im} \, M^{++}(\eta, \eta^{\prime})\label{newenergies1}
\end{eqnarray}
where $aa'M^{++}$ and $aa'M^{--}$ are respectively the self energies corresponding to the Feynman and anti-Feynman propagators. We also note
\begin{eqnarray}\label{energies}
\Big[M^{++}(\eta, \eta^{\prime})-M^{--}(\eta, \eta^{\prime}) \pm \Big(M^{-+}(\eta, \eta^{\prime})-M^{+-}(\eta, \eta^{\prime})\Big)\Big]  && \nonumber \\ =  \pm 2 \theta(\pm \Delta \eta )
 \, i M^{c}  (\eta, \eta^{\prime}) \\
\Big[M^{++}(\eta, \eta^{\prime})+M^{--}(\eta, \eta^{\prime})\Big] +
  \text{sign} (\eta  \!-\!  \eta^{\prime})  \Big[M^{-+}(\eta, \eta^{\prime})+M^{+-}(\eta, \eta^{\prime})\Big] \nonumber && \\ = 4 \theta(\eta \!-\! \eta^{\prime})\, M^{F}(\eta, \eta^{\prime})\label{energies1}
\end{eqnarray}
where $aa'M^{-+}$ or $aa'M^{+-}$ are the self energies corresponding to the Wightman functions. 

\section{Fourier transform of logarithms \label{fourier}}

In order to take \ref{inhomSelfMass1} to momentum space, \ref{M++Full}, we show in this appendix
\begin{multline}
\int d^{3}\vec{r}\, e^{-i \vec{k} \cdot \vec{r}}\Bigg[\frac{1}{2}\ln^2 \frac{y_{++}}{4} + f\big(\eta, \eta^{\prime} \big)\ln \frac{y_{++}}{4}  \Bigg] 
\\ =
-\frac{4 \pi^2}{k^3} \Bigg[2+ \big[1+ i k |\Delta \eta | \big] \Big(  \ln \frac{a a^{\prime} H^2| \Delta \eta|}{2k} +  \frac{ i\pi}{2}- \gamma_E + f(\eta, \eta^{\prime})\Big) \Bigg] e^{-i k |\Delta \eta|} \\
+\frac{4 \pi^2}{k^3} \big(1 - i k |\Delta \eta| \big)\Bigg[ \text{ci} \big[ 2 k| \Delta \eta|  \big]  -i \,  \text{si} \big[ 2 k |\Delta \eta|  \big]  \Bigg] e^{+i k |\Delta \eta|}\label{fTlogsApp}
\end{multline}
where $\Delta \eta  = \eta - \eta^{\prime}$ and $f(\eta, \eta^{\prime})$ is some $k$-independent function. Above equation was first derived in \cite{Friedrich:2019hev}, we briefly review it here for the sake of completeness. In order to solve above equation, we require integrals of the following type,
\begin{align}
 \mathcal{I}_n (x)  & \equiv x^2 \int_0^{\infty} dz \, z \sin \big[x z  \big] { \ln^n \Big( |1- z^2  |\Big) }
 \\
 &= x^2 \Bigg[ \frac{d^n}{db^n} \int_0^{\infty} dz \, z \sin \big[x z  \big]  |1- z^2  |^{b}  \Bigg]_{{b}=0} 
 \end{align}
 Using 
 \begin{multline}
  \int_0^{\infty} dz \, z \sin (x z)  |1- z^2  |^b =  \frac{\sqrt{\pi}}{2} \Big( \frac{2}{x} \Big)^{b+ \frac{1}{2}} \Gamma \big[ b+1\big] \Big[ J_{b + \frac{3}{2}} \big(x \big) + Y_{-b - \frac{3}{2}} \big(x \big) \Big]\, \,\,\, (x> 0, \quad  -1< b < 0)
 \end{multline}
 where $J_{n}\, , Y_{m}$ are the Bessel functions of the first and second kind respectively, we find 
 \begin{align}
\mathcal{I}_{1} \big(x \big)   =  - \pi \Big[\cos x + x \sin x   \Big]
 \end{align}
 and 
  \begin{multline}
\mathcal{I}_{2}(x )   =   2 \pi \left[-2\cos x +   ( \cos x + x \sin x ) \left(  \text{ci}(2 x)  + \gamma_E - \ln \frac{2}{x} \right) +\left(    \sin x   -x \cos x \right)  \text{si}(2 x)\right]
 \end{multline}
 where the sine and cosine integral functions are defined in \ref{sine integral} and \ref{cos integral} in the preceding appendix and $\gamma_E$ is the Euler constant.
 
Using \ref{log}, we establish
\begin{multline}
\label{log1}
\int d^{3}\vec{r} e^{-i \vec{k} \cdot\vec{r}} \ln  \frac{y_{++}}{4}   =  \frac{4 \pi }{k} \int_0^{\infty} dr \, r \sin k r \ln   \frac{y_{++}}{4}  \\
=   \frac{4 \pi }{k} \int_0^{\infty} dr \, r \sin k r \ln  |1- {r^2}{\Delta \eta^{-2} }|  
+   \frac{4 \pi^2 i}{k} \int_0^{\infty} dr \, r \sin k r\, \theta \big( \Delta \eta^2 - r^2  \big)  
\\
=   \frac{4 \pi }{k^3} ( k\Delta \eta  )^2 \int_0^{\infty} dz \, z \sin ( k | \Delta \eta | z  ) \ln  |z^2 - 1  |  
+   \frac{4 \pi^2 i}{k} \int_0^{| \Delta \eta | } dr \, r \sin k r   
 \\= -\frac{4 \pi^2}{k^3} \big[1+i k |\Delta \eta|  \big]e^{- i k |\Delta \eta| } 
\end{multline}
We also compute
\begin{multline}
\int d^{3}\vec{r} e^{-i \vec{k} \cdot\vec{r}} \ln^2 \frac{y_{++}}{4} 
=  \frac{4 \pi }{k} \int_0^{\infty} dr \, r \sin k r\, \ln^2  |1- {r^2}{\Delta \eta^{-2} }  | 
\\ + \frac{8 \pi }{k} \ln  \frac{a a^{\prime} H^2 \Delta \eta^2}{4}  \int_0^{\infty} dr \, r \sin k r\, \ln  |1- {r^2}{\Delta \eta^{-2} } |  
\\+   \frac{8 \pi^2 i}{k} \int_0^{\infty} dr \, r \sin k r\, \ln  |1- {r^2}{\Delta \eta^{-2} }   |\theta \big( \Delta \eta^2 - r^2  \big) 
\\ -\frac{4 \pi^3 }{k}\int_0^{\infty} dr \, r \sin k r\,\theta \big( \Delta \eta^2 - r^2  \big) 
+ \frac{8 \pi^2 i}{k}\ln\frac{a a^{\prime} H^2 \Delta \eta^2}{4} \int_0^{\infty} dr \, r \sin k r\, \theta \big( \Delta \eta^2 - r^2  \big)  \\
=  \frac{4 \pi }{k^3}( k\Delta \eta )^2 \int_0^{\infty} dz \, z \sin k |\Delta \eta | { \ln^2  |1- z^2  | } \\+ \frac{8 \pi }{k^3} \ln \frac{a a^{\prime} H^2 \Delta \eta^2}{4}  (k\Delta \eta)^2 \int_0^{\infty} dz \, z \,\sin (k |\Delta \eta |z ) \ln |1- z^2  | \\ 
+  \frac{8 \pi^2i }{k^3} ( k\Delta \eta)^2 \int_0^{1} dz \, z\, \sin (k |\Delta \eta | z ) \ln  |1- z^2  |
-\frac{4 \pi^2 }{k} \Bigg[\pi - 2i\ln  \frac{a a^{\prime} H^2 \Delta \eta^2}{4}  \Bigg]\int_0^{|\Delta \eta |} dr \, r \sin k r   \\
= -\frac{8 \pi^2}{k^3} \Bigg[2+ \big[1+ i k |\Delta \eta | \big] \Big(  \ln \frac{a a^{\prime} H^2| \Delta \eta|}{2k} +  \frac{i \pi}{2}- \gamma_E \Big) \Bigg] e^{-i k |\Delta \eta|} 
\\+\frac{8 \pi^2}{k^3} \big(1 - i k |\Delta \eta| \big)\Bigg[ \text{ci} \big[ 2 k| \Delta \eta|  \big]  -i \,  \text{si} \big[ 2 k |\Delta \eta|  \big]  \Bigg] e^{+i k |\Delta \eta|}\label{log2}
\end{multline}
We next combine the results \ref{log1} and \ref{log2} in order to finally obtain \ref{fTlogsApp}. One can similarly find the results  for $y_{--}$ and $y_{+-}$ and $y_{-+}$.

\section{Tree level infrared correlators }\label{corr1}
The infrared (IR) effective field theory in de Sitter pertains to the super-Hubble modes that have been redshifted over time, and is free from any ultraviolet divergences. This field theory is composed of truncated IR modes, expressed as
\begin{eqnarray}
\phi(\eta, \vec{x})  = \int \frac{d^3 \vec{k}}{(2\pi)^{3/2}} \theta (Ha-k) \left[a_{\vec k}\, u(k,\eta) e^{-i\vec{k}\cdot \vec{x}}+a^{\dagger}_{\vec k}\, u^{\star}(k,\eta) e^{i\vec{k}\cdot \vec{x}}\right]
\label{c1}
\end{eqnarray}
where $u(\vec{k},\eta)$, the Bunch-Davies mode function, is given by \ref{modfunc}.
By taking the limit $\eta \to 0^-$, the super-Hubble IR modes (where $k$ is restricted by a cut off) can be expanded as 
\begin{eqnarray}
u(\vec{k},\eta) \vert_{\rm IR} \approx \frac{H}{\sqrt{2} k^{3/2}}\left[1+\frac12 \left(\frac{k}{Ha} \right)^2 + \frac{i}{3}\left(\frac{k}{Ha} \right)^3 +\,{\rm subleading~terms } \right]
\label{c2}
\end{eqnarray}
Thus the leading temporal part of the Bunch-Davies modes becomes nearly a constant in this limit. Substituting this into \ref{c1} and dropping  the suffix `IR' without any loss of generality, we have
\begin{eqnarray}
\phi(\eta, \vec{x})  =\frac{H}{\sqrt{2}} \int \frac{d^3 \vec{k}}{(2\pi)^{3/2}} \frac{\theta (Ha-k)}{k^{3/2}} \left[a_{\vec k}\,  e^{-i\vec{k}\cdot \vec{x}}+a^{\dagger}_{\vec k}\,  e^{i\vec{k}\cdot \vec{x}}\right]+\,{\rm subleading~terms}
\label{c3}
\end{eqnarray}
The step function appearing above ensures that we are essentially dealing with long wavelength modes. 

The IR Wightman functions in terms of these modes are accordingly given by (cf., \ref{Propagators in the Schwinger-Keldysh Formalism})
\begin{eqnarray}
i \Delta^{-+}_{\phi}(x,x^\prime)&&=\frac{H^2}{2} \int \frac{d^3 {\vec k}}{(2\pi)^3 k^3} e^{i{\vec k}\cdot({\vec x}-{\vec y})}\theta(Ha-k)\theta(Ha'-k)(1+ik\eta)(1-ik\eta')e^{-ik(\eta -\eta')}\nonumber\\ &&= \int \frac{d^3 {\vec k}}{(2\pi)^3} e^{i{\vec k}\cdot({\vec x}-{\vec y})}i \Delta_{-+}(k,\eta,\eta')\nonumber\\
i \Delta^{+-}_{\phi}(x,x^\prime) &&=\frac{H^2}{2} \int \frac{d^3 \vec{k}}{(2\pi)^3 k^3} e^{i\vec{k}\cdot(\vec{x}-\vec{y})}\theta(Ha-k)\theta(Ha'-k)(1-ik\eta)(1+ik\eta')e^{ik(\eta -\eta')}\nonumber\\&&= \int \frac{d^3 {\vec k}}{(2\pi)^3} e^{i{\vec k}\cdot({\vec x}-{\vec y})}i \Delta_{+-}(k,\eta,\eta') 
\label{nc2}
\end{eqnarray}
Using now \ref{c2}, we  compute
\begin{eqnarray}
 i \Delta^{+-} _\phi (\eta',\eta'', k)\vert_{ k | \Delta \eta| \ll 1}\approx  \frac{H^2 \theta(Ha'-k)\theta(Ha''-k)}{2k^3} \left(1+\frac{ik^3}{3H^3 a''^3} \right)\quad (\eta' \gtrsim \eta'')
\label{nc4'}
\end{eqnarray}
and  $i \Delta^{+-}=(i \Delta^{-+})^{\star}$.  We also have for our purpose from \ref{Delta:causal}
\begin{multline}
\Delta_{\phi}^c (\eta, \eta^{\prime}, k )_{ k | \Delta \eta| \ll 1} = i \Delta^{-+} _\phi (\eta,\eta',k) - i \Delta^{+-} _\phi (\eta,\eta',k) \approx  -\frac{i \theta(Ha-k)\theta(Ha'-k)}{3H a'^3}\\
F_{\phi} (\eta, \eta^{\prime}, k )_{ k | \Delta \eta| \ll 1}  = \frac12\left(i \Delta^{+-} _\phi (\eta,\eta', k) + i \Delta^{-+} _\phi (\eta,\eta', k)\right)\approx \frac{H^2}{2k^3}\theta(Ha-k)\theta(Ha'-k)
\label{nc4}
\end{multline}
%


\pagestyle{empty}


\bibliographystyle{cas-model2-names}




\end{document}